\documentclass[preprintnumbers,amsmath,amssymb,prd, 
notitlepage,nofootinbib,twocolumn,superscriptaddress]{revtex4-2}

\pdfoutput=1

\usepackage{setspace}
\usepackage{xspace}

\usepackage{graphicx}
\usepackage{epstopdf}
\usepackage{amssymb}
\usepackage{dcolumn}

\usepackage{graphicx}
\usepackage{epstopdf}
\usepackage{amsmath}
\usepackage{amsfonts}
\usepackage{amssymb}
\usepackage[usenames,dvipsnames]{xcolor}
\usepackage{array}
\usepackage{nonfloat}
\usepackage[normalem]{ulem}
\usepackage{bm}

\usepackage{multirow}
\usepackage{psfrag}
\pdfoutput=1

\usepackage[colorlinks,bookmarks]{hyperref}
\definecolor{linkblue}{rgb}{0,0,0.8}
\definecolor{linkgreen}{rgb}{0,0.5,0}

\hypersetup{pdfpagemode=UseNone, pdfstartview=FitH, linkcolor=linkblue, 
	citecolor=linkgreen, urlcolor=linkblue}

\def\beq{\begin{equation}}
\def\eeq{\end{equation}}
\newcommand\numberthis{\addtocounter{equation}{1}\tag{\theequation}}
\newcommand{\lp}{\left(}
\newcommand{\rp}{\right)}
\newcommand{\lb}{\left[}
\newcommand{\rb}{\right]}
\newcommand{\vev}[1]{\langle #1 \rangle}
\newcommand{\be}{\begin{eqnarray}}

\newcommand{\ee}{\end{eqnarray}}

\newcommand{\tcm}{21$\,$cm\xspace}  

\newcommand{\vk}{\boldsymbol{k}}

\newcommand{\vq}{\boldsymbol{q}}

\newcommand{\vx}{\boldsymbol{x}}

\newcommand{\knl}{k_{\rm NL}}

\newcommand{\deltag}{\delta_{\rm g}}

\newcommand{\deltaHI}{\delta_{\rm HI}}
\newcommand{\Gtwo}{\mathcal{G}_2}
\newcommand{\Lag}{{\rm L}}
\newcommand{\Shi}{{\rm S}}
\newcommand{\Eul}{{\rm E}}

\newcommand{\PHI}{P_{\rm HI}}
\newcommand{\PN}{P_{\rm N}}
\newcommand{\Ptcm}{P_{21}}
\newcommand{\Ptcmi}{P_{21,i}}
\newcommand{\Ptcmj}{P_{21,j}}
\newcommand{\Pstoch}{P_{\rm stoch}^{(k\to 0)}}

\def\d{\partial}

\newcommand{\dirac}{\delta_{\rm D}}

\newcommand{\hinvMpc}{h\, {\rm Mpc}^{-1}}
\newcommand{\invMpc}{\, {\rm Mpc}^{-1}}

\newcommand{\Mpcinvh}{h^{-1}{\rm Mpc}}

\definecolor{purple}{rgb}{0.78,0.18,0.77}

\begin{document}

\title{Improving cosmological analyses of HI clustering by reducing stochastic noise}

\newcommand{\asu}{Department of Physics, Arizona State University,
550 E.\ Tyler Mall, Tempe, AZ 85287, USA}
\newcommand{\zurich}{Department of Astrophysics, University of Zurich, Winterthurerstrasse 190, 8057 Zurich, Switzerland}
\newcommand{\unifi}{Dipartimento di Fisica e Astronomia, Universit\'a di Firenze; Via G. Sansone 1; I-50019 Sesto Fiorentino, Italy}
\newcommand{\infn}{
INFN, Sezione di Firenze; Via G. Sansone 1; I-50019 Sesto Fiorentino, Italy}

\author{Simon Foreman}
\affiliation{\asu}
\author{Andrej Obuljen}
\affiliation{\zurich}
\author{Marko Simonovi\'{c}}
\affiliation{\unifi}
\affiliation{\infn}

\date{\today}

\begin{abstract}

High-number-density tracers of large-scale structure, such as the HI-rich galaxies measured by 21$\,$cm intensity mapping, have low sampling noise, making them particularly promising as cosmological probes. At large scales, this sampling noise can be subdominant to other scale-independent contributions to the power spectrum;
such contributions arise from nonlinear bias, and exceed the sampling noise if at least one of the associated bias coefficients is sufficiently large.
This has important consequences for cosmological constraints obtained from such tracers, since it indicates that using the power spectrum does not lead to optimal constraints even in the linear regime.
In this paper, we provide a conservative estimate of the possible improvement in constraining power of a \tcm survey if one were to use an optimal analysis strategy (such as field-level analysis), where only the true sampling noise enters the error budget.
We find that improvements in uncertainties on some cosmological parameters can be as large as 50$\%$, depending on redshift, foreground cleaning efficiency, scales used in the analysis, and instrumental noise.
One byproduct of our work is measurements of bias parameters and stochasticity for neutral hydrogen in the IllustrisTNG simulation over a wide range of redshifts; we provide simple fitting formulas for these measurements.
Our results motivate further exploration of new optimal analysis techniques and provide important insights into the constraining power of current and future \tcm surveys.
\end{abstract}

\maketitle

\section{Introduction}
\label{sec:intro}

Intensity mapping of neutral hydrogen (HI) via \tcm emission or absorption is a promising probe of cosmology, with the potential to take us beyond what has been possible with the cosmic microwave background (CMB), galaxy clustering, or weak lensing~\cite{Pritchard:2011xb,CosmicVisions21cm:2018rfq,Liu:2022iyy}.
If practical challenges (such as removal of radio foregrounds) can be overcome, the distribution of HI can be mapped at all times after recombination.
In this paper, we restrict our attention to the post-reionization ($z\lesssim 6$) universe, where the vast majority of HI mass is expected to be contained within galaxies, at such high densities that it is self-shielded from ionizing radiation (e.g.~\cite{Wolfe:2005jd,Villaescusa-Navarro:2018vsg}).

When using galaxy surveys as probes of cosmic large-scale structure (LSS), one must account for a sampling noise (or shot noise) contribution to measurements of clustering, which arises because galaxies are discrete tracers of a total-matter density field that can be treated as continuous on the scales of interest. Under the assumption that galaxies sample the continuous density field according to a Poisson process, this sampling noise appears in the galaxy power spectrum as a scale-independent term, equal to $1/\bar{n}_{\rm g}$ where~$\bar{n}_{\rm g}$ is the mean number density of galaxies. One of the features of HI that makes it such a powerful cosmological probe is that its sampling noise is expected to be extremely low compared to other galaxy samples targeted by optical or near-infrared surveys, due to the large number of lower-mass galaxies containing HI~\cite{Castorina:2016bfm,Villaescusa-Navarro:2018vsg,Spinelli:2019smg}.

However, high-number-density tracers such as HI-rich galaxies or low-mass dark matter halos can lead to some puzzles regarding the noise contribution to their power spectra. Measurements in N-body and hydrodynamical simulations have clearly shown that, in the low-$k$ limit, the power spectrum of a dense tracer has a constant contribution (often referred to as ``stochasticity") that is much larger than the $1/\bar{n}_{\rm g}$ estimate (e.g.~\cite{Villaescusa-Navarro:2018vsg,Hand:2017ilm,Hamaus:2010im,Schmittfull:2018yuk,Baldauf:2013hka,Obuljen:2022cjo}). 
If one defines a function $b_{\rm g}(k) = \sqrt{P_{\rm g}(k)/P_{\rm m}(k)}$ (sometimes referred to as galaxy bias, although we will use a different definition of galaxy bias in this work), the larger-than-expected constant contribution will manifest in strong scale-dependence of this function, and this has also been observed in several simulations of HI clustering~\cite{Villaescusa-Navarro:2018vsg,Modi:2019ewx,Wang:2019rub,Sarkar:2016lvb}.

The origin of the discrepancy between the Poisson sampling noise $1/\bar{n}_{\rm g}$ and the measured stochasticity is well understood. As argued in detail in Ref.~\cite{Schmittfull:2018yuk}, the main contribution to the stochasticity for dense tracers arises from second-order nonlinearities in the perturbative description of galaxies as biased tracers of the matter density. 
At quasi-linear scales, the galaxy overdensity $\delta_{\rm g}(\vk,z)$ can be written as
\beq
\delta_{\rm g}(\vx,z) = b_1(z) \delta_1(\vx,z) + b_2(z) \delta_1(\vx,z)^2 + N(z) + \cdots\ ,
\label{eq:bias-intro}
\eeq
where $\delta_1$ is the linear density field, $b_1$ and $b_2$ are bias coefficients, $N$ is a noise field whose power spectrum is a scale-independent constant (equal to $1/\bar{n}_{\rm g}$ in the Poisson case) and the ellipses contain other terms that will be described in the body of the paper (and are reviewed in Ref.~\cite{Desjacques:2016bnm}). The power spectrum of $\delta_{\rm g}$ contains the auto power spectrum of the $b_2 \delta_1^2$ term, and this term does not vanish in the low-$k$ limit, but instead asymptotes to a $k$-independent constant.
This may seem counterintuitive at first sight, since the nonlinearities are expected to be negligible at very large scales. However, one can check that due to the absence of mass and momentum conservation in counting discrete tracers, nonlinear terms such as~$b_2\delta_1^2$ can generate fluctuations with the constant power spectrum at arbitrarily low~$k$.

This fact has been known for a long time~\cite{Heavens:1998es}, and has been explored in various contexts, including the halo model~\cite{Baldauf:2013hka} and predictions for HI clustering~\cite{Umeh:2015gza,Umeh:2016thy,Penin:2017xyf}.
With the advent of renormalized bias and the effective field theory (EFT) of large-scale structure, it was realized that the low-$k$ contribution from terms like $b_2\delta_1^2$ should be absorbed into the leading stochastic term $N$ to ensure that the perturbative expansion is well-behaved at low~$k$~\cite{McDonald:2006mx,Assassi:2014fva,Angulo:2015eqa}. 
From the EFT perspective, this term has two contributions, which have not been carefully distinguished in some previous literature. The contribution from nonlinear modes ($k\gtrsim k_{\rm NL}$) should be completely removed by renormalization. The contribution from perturbative modes ($k\lesssim k_{\rm NL}$) is captured by the long-wavelength theory and should be retained as a finite contribution to the power spectrum. It is this latter contribution that is seen in simulations and gives rise to large observed stochasticity~\cite{Schmittfull:2018yuk,Obuljen:2022cjo,Cabass:2023nyo,Ivanov:2023yla}.

In a measured power spectrum, the sampling noise and bias-induced stochasticity are completely degenerate at low $k$, and only their sum can be constrained. This sum enters not only the power spectrum model, but also the covariance matrix, which determines the uncertainties on the inferred cosmological parameters. 
Note that the stochastic contribution to the noise can be up to ten times larger than the naive Poisson sampling noise for HI. In other words,~$b_2\delta_1^2$ interactions have the same effect on parameter uncertainties as reducing the number density of tracers by as much as a factor of ten.

On the other hand, it is intuitively clear that there should not be a fundamental obstacle to measuring the cosmological parameters with precision set only by the true sampling noise. Indeed, while bias-induced stochasticity may be large, it is still deterministic and calculable from the initial conditions. This has been shown in~\cite{Schmittfull:2018yuk} with full field-level modelling, using the same initial conditions for simulations and perturbative calculation to cancel the cosmic variance. Having this in mind, we can see that the degeneracy of the sampling noise and the stochasticity is just an artifact of using the simplest observable, the power spectrum, and that in principle, the contributions from the quadratic  bias can be used in a cosmological data analysis as signal rather than noise if all information in the data is utilized. For example, it has been shown recently that in an optimal field-level analysis, the uncertainties on cosmological parameters are indeed set by sampling noise alone~\cite{Cabass:2023nyo}.

This paper has several aims. The first aim is to briefly review the theoretical basis for bias-induced stochasticity in detail in the context of HI clustering. The second aim is to extend the results of Ref.~\cite{Obuljen:2022cjo}, which explored HI stochasticity with simulations and field-level modelling, to higher redshift, and provide fitting functions for bias coefficients and stochasticity that can be used in future forecasts or simulations. 
We also apply the methods of Ref.~\cite{Obuljen:2022cjo} to two different simulation suites, IllustrisTNG~\cite{TNGa,TNGb,TNGc,TNGd,TNGe} and Magneticum~\cite{Magneticum_Hirschmann2014,Magneticum_Teklu2015}, and demonstrate that our conclusions about HI stochasticity are robust to differences between these simulations.

The third aim is to estimate the difference in constraining power between a standard HI power spectrum analysis and a more optimal analysis that does not require bias-induced stochasticity to form part of the noise budget. A full comparison between a power spectrum analysis and a field-level analysis is still challenging, because the latter is theoretically and numerically challenging, although progress is being made on this front~\cite{Nguyen:2024yth}.
Nevertheless, a conservative estimate of this difference can still be made in a relatively straightforward manner, by performing two different Fisher forecasts: one using realistic effective noise in the covariance matrix to mimic the standard analysis, and the other using only Poisson noise, in order to mimic the more optimal estimate of cosmological parameters expected at the field level. 
We carry out these forecasts for the upcoming CHORD telescope~\cite{Vanderlinde:2019tjt} and the PUMA telescope concept~\cite{CosmicVisions21cm:2018rfq,PUMA:2019jwd,Castorina:2020zhz}.
The results depend on assumptions about the nonlinear scale, foreground cleaning, and so on, and we vary these assumptions over reasonable ranges. Our results generically suggest  that constraints on some important cosmological parameters, such as~$\sigma_8$, can differ by 10-50$\%$ between the two analyses. These results suggest that there is a relatively significant gain to be made using an optimal analysis and motivate further exploration in this direction.

In many cases, the results are determined by the level of thermal noise associated with a given \tcm survey. We also explore the implications of lowering the thermal noise in our forecasts, and find significant improvements in constraining power, both in a power spectrum analysis and in the difference between optimal and power spectrum analyses. For the eventual design of future \tcm surveys, this motivates targeting thermal noise levels that are as close as possible to the HI sampling noise, rather than the stochasticity.

This paper is structured as follows.
In Sec.~\ref{sec:theory}, we review the relevant aspects of the perturbative bias expansion in the effective field theory of large-scale structure (EFT of LSS), and discuss the distinction between model error, sampling noise, and stochasticity.
In Sec.~\ref{sec:simulations}, we apply the field-level modelling framework of Refs.~\cite{Schmittfull:2018yuk,Obuljen:2022cjo} to the IllustrisTNG and Magneticum simulations, measuring the different types of noise in the HI power spectrum and presenting fitting functions for linear, quadratic, and tidal biases (Sec.~\ref{sec:simulations:biascoefficients}) along with HI stochasticity (Sec.~\ref{sec:simulations:HIstochasticity}).
In Sec.~\ref{sec:forecasts}, we show forecasts for power spectrum or more-optimal analyses of \tcm intensity mapping by CHORD and PUMA, and in Sec.~\ref{sec:thermal_noise}, we highlight the role of thermal noise in these forecasts.
In Sec.~\ref{sec:discussion}, we briefly discuss the handling of bias-induced stochasticity in other \tcm forecasts in the literature.
Finally, we conclude in Sec.~\ref{sec:conclusion}.

The appendices contain details about different conventions for bias coefficients (Appendix~\ref{app:bias_params}), versions of our forecasts with alternative assumptions (Appendix~\ref{app:alternative_forecasts}), and thermal noise computations in our forecasts (Appendix~\ref{app:thermal_noise}).

\section{Theory}
\label{sec:theory}

In this section we briefly review the origin of the difference between sampling noise and stochasticity, and discuss the impact of this difference on the constraining power of a power-spectrum-based analysis. We begin by presenting the perturbative model for the HI density field (Sec.~\ref{sec:theory:theoretical_model}) and reviewing the definitions of sampling noise and model error (Sec.~\ref{sec:theory:sampling_noise}). We then discuss the relationship between these quantities and the stochasticity of the distribution of HI (Sec.~\ref{sec:theory:stochasticity}), and qualitatively discuss the implications of this distinction for cosmological constraints from \tcm intensity mapping surveys (Sec.~\ref{sec:theory:implications_estimators}).

\subsection{Theoretical model}
\label{sec:theory:theoretical_model}

We will use the EFT of LSS
(e.g.~\cite{Baumann:2010tm,Carrasco:2012cv,Perko:2016puo,Desjacques:2016bnm}) as our model for the large-scale (quasi-linear) distribution of HI, since this framework is sufficient to capture the key effects relevant to our discussion. Within this framework, there are several equivalent ways to describe the nonlinear density field of biased tracers in terms of the linear matter density field~$\delta_1$. 

For comparing theory to simulations or data at the map level, which we will need in this paper, it is convenient to express the deterministic part of the HI density field $\deltaHI(\vk, z)$ in terms of building blocks with the following form (for more details, see Ref.~\cite{Schmittfull:2018yuk}):
\beq
\label{eq:shifted_def}
\tilde{\mathcal{O}}(\bm{k},z) \equiv \int d^3\bm{q}\; \mathcal{O}(\bm{q},z) e^{-i\bm{k}\cdot(\bm{q}+\bm{\psi}_1(\bm{q},z)  ) } \ ,
\eeq
where $\bm q$ is the Lagrangian coordinate in the initial conditions and $\bm{\psi}_1$ is the Zel'dovich displacement field, given by
\beq
\bm{\psi}_1(\bm{k},z)=\frac{i\bm{k}}{k^2}\delta_1(\bm{k},z)\ .
\eeq
For simplicity, we will focus on the real-space density field in this section, but the same construction can be straightforwardly extended to redshift space~\cite{Schmittfull:2020trd}. The operators~$\mathcal O(\bm{q},z)$ are constructed from the Lagrangian-space linear density field~$\delta_1(\bm{q})$, and only a finite number of these operators contribute at each order in perturbation theory.\footnote{These operators appear as the main ingredients in evaluating the nonlinear density field in Fourier space using Lagrangian perturbation theory~\cite{Matsubara:2008wx,Carlson:2012bu,Vlah:2016bcl,Cuesta-Lazaro:2022dgr}.}
The function of Eq.~\eqref{eq:shifted_def} is to shift each operator from its initial (Lagrangian) position~$\bm{q}$ by the Zel'dovich displacement field $\bm{\psi}_1(\bm{q},z)$, non-perturbatively incorporating the large map-level effects of the displacements; for this reason, $\tilde{\mathcal{O}}(\bm{k},z)$ is referred to as a {\em shifted} operator.
As expected, the statistics of these operators can be shown to be the same as in the standard infrared-resummed Eulerian perturbation theory~\cite{Schmittfull:2018yuk,Chen:2020fxs}.\footnote{These equivalent approaches have been implemented in several publicly available codes, such as \texttt{CLASS-PT}~\cite{Chudaykin:2020aoj}, \texttt{PyBird}~\cite{DAmico:2020kxu}, \texttt{velocileptors}~\cite{Chen:2020fxs} and \texttt{CLASS-OneLoop}~\cite{Linde:2024uzr}.}
Therefore, this model properly describes all nonlinearities on large scales and effects of unknown short-scale physics (through counterterm operators), while faithfully reproducing nonlinear maps of biased tracers and capturing the broadening and shifting of the BAO peak in all correlation functions~\cite{Senatore:2014via,Baldauf:2015xfa,Vlah:2015zda,Blas:2016sfa}. 

For the purposes of this paper, we will focus only on the leading nonlinear contribution to the HI power spectrum, the so-called one-loop power spectrum. In this case, we have to keep all relevant operators~$\mathcal O$ up to cubic order in~$\delta_1$. Linear and quadratic contributions are given by~$\mathcal{O}=\{\delta_1, \delta_2\equiv(\delta_1^2-\sigma_1^2),\mathcal{G}_2\}$, where $\sigma_1^2\equiv\langle\delta_1^2 \rangle$ is the r.m.s.~fluctuation of the linear density field, while
\beq
\mathcal{G}_2\equiv\left[\frac{\partial_i \partial_j}{\nabla^2} \delta_1\right]^2 - \delta_1^2
\eeq
is the second-order bias operator induced by the tidal field. While constructing all cubic operators is also relatively straightforward, one can simplify their inclusion in the model by rewriting a generic cubic operator as~\cite{Schmittfull:2018yuk}
\beq
\tilde{\mathcal O}_3 = \frac{\langle \tilde{\mathcal O}_3  \tilde{\delta}_1 \rangle'}{\langle \tilde{\delta}_1 \tilde{\delta}_1 \rangle'} \tilde{\delta}_1 + \left( \tilde{\mathcal O}_3 - \frac{\langle \tilde{\mathcal O}_3  \tilde{\delta}_1 \rangle'}{\langle \tilde{\delta}_1 \tilde{\delta}_1 \rangle'} \tilde{\delta}_1 \right)\ ,
\eeq
where a prime in the correlation function~$\vev{\cdots}'$ indicates that the momentum-conserving delta function~$(2\pi)^3\dirac(0)$ is removed.
Let us define the second term in this sum to be
\beq
\tilde{\mathcal O}_3^\perp \equiv \tilde{\mathcal O}_3 - \frac{\langle \tilde{\mathcal O}_3  \tilde{\delta}_1 \rangle'}{\langle \tilde{\delta}_1 \tilde{\delta}_1 \rangle'} \tilde{\delta}_1\ .
\eeq
It is clear that~$\langle \tilde{\mathcal O}_3^\perp \tilde{\delta}_1 \rangle' = 0$, and in this sense, we can say that~$\tilde{\mathcal O}_3^\perp$ is orthogonal to~$\tilde{\delta}_1$. This implies that the only relevant contribution of cubic operators to the one-loop power spectrum is proportional to $\tilde{\delta}_1$. In other words, the combined effect of all cubic operators~$\tilde{\mathcal O}_3$ is a~$k$-dependent re-scaling of~$\tilde{\delta}_1$. Therefore, the {\em deterministic part} of the HI density field (i.e.\ the part that can be predicted from the linear density $\delta_1$) can be written as 
\beq
\label{eq:real_model}
\delta_\mathrm{HI}^{\rm model}(\bm{k}) = \bar\beta_1(k) \tilde{\delta}_1(\bm{k}) + b_2^\Shi \tilde{\delta}_2 (\bm{k}) + 
b_{\mathcal{G}_2}^\Shi \tilde{\mathcal{G}}_2 (\bm{k})\ ,
\eeq
where $\bar\beta_1(k)$ is a~$k$-dependent function given by
\be
\label{eq:beta1_model}
\bar \beta_1(k) =  b_1^\Shi +  c_s^2 k^2 +  b_{\Gamma_3}^\Shi \frac{\vev{\tilde {\Gamma}_3 \tilde\delta_1} '}{\vev{\tilde\delta_1 \tilde\delta_1 }'} - b_1^\Shi \frac{\vev{ \tilde {\mathcal S}_3 \tilde\delta_1}'}{\vev{\tilde \delta_1 \tilde \delta_1 }'} 
\ee 
and~$b_1^\Shi, c_s^2, b_2^\Shi, b_{\mathcal{G}_2}^\Shi$, and~$b_{\Gamma_3}^\Shi$ are unknown constant nuisance parameters.\footnote{The exact form of cubic operators $\Gamma_3$ and $\mathcal S_3$ can be found in Ref.~\cite{Obuljen:2022cjo}. Note that $\mathcal S_3$ comes from the second-order displacement acting on the tracer density field in Lagrangian coordinates. The Equivalence Principle therefore dictates that the coefficient of this operator is fixed to $-b_1^\Shi$~\cite{Schmittfull:2018yuk}. The other cubic operators only renormalize linear bias at this order in perturbation theory.} (We have dropped the redshift argument for brevity.) Note that $c_s^2$ should be understood as a sum of the usual dark matter one-loop counterterm~\cite{Carrasco:2012cv} and the nonlocal bias produced by the~$\nabla^2\delta$ term in the bias expansion. The superscript S indicates that these bias parameters multiply shifted fields. If one uses a more familiar Eulerian bias expansion
\be
\delta_\mathrm{HI}^{\rm model}(\bm{k}) = b_1^\Eul \delta(\bm{k}) + \frac{b_2^\Eul}2 \delta^2 (\bm{k}) + 
b_{\mathcal{G}_2}^\Eul \mathcal{G}_2 \, (\bm{k}) \ ,
\label{eq:deltaHI_Eul_maintext}
\ee
where~$\delta$ is the nonlinear dark matter field, then the relation to Eulerian bias parameters is~\cite{Schmittfull:2018yuk}
\be
b_1^\Eul = b_1^\Shi \, , \quad b_2^\Eul = 2b_2^\Shi \, , \quad b_{\mathcal{G}_2}^\Eul = b_{\mathcal{G}_2}^\Shi - \frac 27 b_1^\Shi \ .
\label{eq:Shi_to_Eul}
\ee
We derive relations to Lagrangian and Eulerian bias parameters in several other bases commonly used in the literature in Appendix~\ref{app:bias_params}.

For comparison to simulations, it is often more convenient to also orthogonalize quadratic and linear fields, in which case the transfer function multiplying~$\tilde\delta_1$ acquires additional contributions. We will describe this procedure in Sec.~\ref{sec:simulations:fieldlevel}. Let us finish by emphasising once more that this model for the map reproduces the usual one-loop power spectrum and tree-level bispectrum predictions for a biased tracer, with the same number of free parameters.

\subsection{Sampling noise and model error}
\label{sec:theory:sampling_noise}

The observed density field of HI can be written as a sum of two contributions:
\beq
\label{eq:deltaHI_det_stoch}
\deltaHI(\bm{k}) = \deltaHI^{\rm model}(\bm{k}) + \epsilon_\mathrm{HI}(\bm{k}) \ .
\eeq
The first term is the deterministic part of~$\delta_\mathrm{HI}(\bm{k})$, discussed in the previous section. The second term is the {\em stochastic part}, which we assume to be uncorrelated with~$\deltaHI^{\rm model}$. 

One can understand the origin of the stochastic part in the following way. In an ideal simulation, given the full realization of the initial conditions and a set of cosmological parameters, there is no stochastic contribution, because the time evolution on all scales is deterministic. However, in observations, we sample the continuous density field with a finite number of discrete tracers, with only partial information about their mass and velocities. Even if we had an exact forward model for the evolution of the density field, for fixed long-wavelength modes of~$\delta_1$, there would be many small-scale realizations of~$\delta_1$ which lead to the exact same distribution of tracers in space. Therefore, the contribution of the true small-scale realization to the continuous nonlinear density field is unknowable from observations, and we have no choice but to model this contribution, $\epsilon_\mathrm{HI}$, as the outcome of some random process (effectively marginalizing over all possible small-scale realizations). See Ref.~\cite{Feng:2018for} for a related discussion in the context of reconstruction of the initial conditions from the nonlinear matter density field. 

What are the expected properties of the stochastic term~$\epsilon_\mathrm{HI}$? Since it comes from averaging over the small-scale fluctuations, where ``small-scale" refers to modes of $\delta_1$ with wavelength smaller than the typical separation between the halos of a given tracer, its amplitude is expected to be inversely proportional to the number density of tracers,~$\bar{n}_{\rm g}$. Furthermore, in large-scale structure surveys we measure only positions of galaxies and have very little information about their mass. This introduces the effective violation of mass and momentum conservation in the observed galaxy samples and implies that the stochastic power is constant on large scales.\footnote{By locality, the contribution of short modes to the large-scale density field can be Taylor-expanded in $k$, with the first allowable term scaling like $k^0$. Mass and momentum conservation would force the $k^0$ and $k^1$ terms to vanish, but their absence in the observed galaxy samples implies that the $k^0$ term exists, and therefore dominates at low $k$. See Refs.~\cite{Zeldovich:1965gev,Peebles:1980yev,Mercolli:2013bsa} for further details.}

We also define the {\em model error power spectrum} as
\beq
\label{eq:model_error_pk}
P_{\rm{err}}(\bm{k}) \equiv \vev{|\delta_{\rm HI}(\bm{k})-\delta_{\rm 
HI}^{\rm model}(\bm{k})|^2}' 
=
\vev{|\epsilon_\mathrm{HI}(\bm{k})|^2}'\ .
\eeq
In the simplest model of Poisson sampling of the continuous density field, we get~\cite{Peebles:1980yev}
\beq
P_{\rm err} (k\to 0) = P_{\rm sampling} = \frac{1}{\bar{n}_{\rm g}}\ .
\eeq
However, various effects on small scales, such as nonlinear clustering, environment-dependence of galaxy formation, or halo exclusion effects can all modify this simple Poisson-sampling model. These modifications affect the amplitude only at the level of 
tens of percents, 
making the sampling noise sub- or super-Poissonian depending on the type of tracer~\cite{Baldauf:2013hka,Kokron:2021faa}. For this reason, the amplitude of the model error spectrum~$P_{\rm err}(k)$ on large scales is left as a free parameter in cosmological data analyses, but with a prior that it must be relatively close to the Poisson result, $P_{\rm sampling} = 1/\bar{n}_{\rm g}$. As we are now going to see, for very dense tracers, there are other contributions to the model error power spectrum at large scales that can be parametrically larger than the sampling noise.

\subsection{Large stochasticity for dense tracers}
\label{sec:theory:stochasticity}

Looking at the deterministic part of the tracer field given by~$\delta_{\rm HI}^{\rm model}$, one may expect that on very large scales, the only relevant fluctuations are given by a simple rescaling of the linear density field,~$b_1^\Eul\delta_1$. Naively, all other nonlinearities due to the shifts or nonlinear bias should be very small in the low-$k$ limit. This motivates the definition of the {\em stochasticity}~\cite{Hamaus:2010im}
of the HI density:
\beq
\label{eq:HI_stochasticity}
{\rm HI~stochasticity} \equiv \vev{|\delta_{\rm HI}(\bm{k})-b_1^\Eul \delta_1(\bm{k})|^2}'\ ,
\eeq
where~$b_1^\Eul$ is the measured linear bias of HI. We are interested in the constant low-$k$ limit of the stochasticitiy, which we define to be
\beq
\Pstoch \equiv \vev{|\delta_{\rm HI}(\bm{k})-b_1^\Eul \delta_1(\bm{k})|^2}' \Big|_{k\to 0}\ .
\eeq
Note that stochasticity differs from the model error in Eq.~\eqref{eq:model_error_pk} because the model error is the difference between the true density and the full deterministic model, while the stochasticity only uses the linear Eulerian contribution as the model.
If the only difference between the nonlinear HI density~$\delta_{\rm HI}$ and~$b_1^\Eul\delta_1$ was~$\epsilon_{\rm HI}$ on large scales,
then we would indeed have~$\Pstoch = P_{\rm sampling}$. 

This expectation is quite accurate for the massive dark matter halos that host galaxies observed in typical spectroscopic galaxy surveys. However, it is well-known that for lower-mass dark matter halos,~$\Pstoch$ and~$P_{\rm sampling}$ can be significantly different. This has been observed in N-body and hydrodynamical simulations~\cite{Hamaus:2010im,Baldauf:2013hka,Hand:2017ilm,Villaescusa-Navarro:2018vsg} and it has 
raised questions about the value of the expected shot noise in \tcm intensity mapping surveys~\cite{Villaescusa-Navarro:2018vsg}.

By now, the origin of the discrepancy between~$\Pstoch$ and~$P_{\rm sampling}$ is well understood, both in perturbation theory~\cite{Schmittfull:2018yuk} and within the framework of the halo model~\cite{Seljak:2009af}. The key insight is that nonlinearities in the bias expansion can produce long-wavelength fluctuations sourced by short modes, with  power that is not suppressed by~$k^2$ at low~$k$. This is a consequence of the fact that counting tracers without assigning correct mass to their host halos effectively leads to absence of mass and momentum conservation~\cite{Seljak:2009af}. For example, one of the contributions of the quadratic bias term~$b_2^\Eul\delta^2$ to the HI power spectrum is given by (for simplicity, we neglect shifts because they do not change the conclusions)
\beq
P_{\rm HI}^{\rm model}(k) \supset \frac{(b_2^{\Eul})^2}{2} \int \frac{d^3\bm{q}}{(2\pi)^3} P_{\rm lin}(q) P_{\rm lin}(|\bm{k}-\bm{q}|)\ ,
\eeq
where~$P_{\rm lin}(k)$ is the linear matter power spectrum. In the~$k\to 0$ limit, the integral on the right hand side does not vanish, but asymptotes to a constant. A part of this integral with~$q$ above the nonlinear scale~$k_{\rm NL}$ cannot be trusted and has to be absorbed by the appropriate counterterm in the language of the EFT of LSS. However, the dominant contribution to this integral in a~$\Lambda$CDM-like cosmology comes from perturbative scales, and this is the contribution we are interested in. One can easily check that with our choice of bias operators, this is the only term in the HI power spectrum with a constant low-$k$ limit. Therefore, on very large scales when~$k\to 0$, we can write 
\beq
\label{eq:Pstochseq}
\Pstoch = P_{\rm err}(k\to 0) + \frac{(b_2^{\Eul})^2}{2} \int_{q\lesssim k_{\rm NL}} \frac{d^3\bm{q}}{(2\pi)^3} P_{\rm lin}^2(q)\ .
\eeq

This is similar to the more familiar example of the one-loop matter power spectrum. In that case, the dominant contribution on large scales comes from the $P_{13}$ diagram, and in the limit~$k\to 0$, it is given by
\beq
\frac{P^{\rm matter}_\text{1-loop}(k)}{k^2P_{\rm lin}(k)}\Bigg|_{k\to 0} = -2c_s^2 - \frac{61}{315} \int_{q\lesssim k_{\rm NL}} dq\, P_{\rm lin}(q)\ ,
\eeq
where~$c_s^2$ is the one-loop counterterm. Even though the two constants on the right hand side are completely degenerate, they do not have the same meaning. The counterterm is supposed to absorb only the UV part of the integral, which comes from Fourier modes with~$k>k_{\rm NL}$ where the perturbative description is invalid. Indeed, the second term in the previous equation, which comes from large displacements, is numerically much bigger than~$c_s^2$ in a~$\Lambda$CDM-like cosmology, and we keep it explicitly every time we evaluate the one-loop matter power spectrum. 

Eq.~\eqref{eq:Pstochseq} clearly shows why the stochasticity is always different (and larger) than the model error. Furthermore, the second term on the right hand side is related to the variance of the square of the linear density field, and depends only weakly on the type of tracer, assuming that~$b_2^{\Eul}$ is generically of order unity. This is very different from~$P_{\rm sampling}$, which scales roughly as~$1/\bar{n}_{\rm g}$ and very sharply decreases as the number density of tracers increases. For this reason, for sparse tracers such as galaxies, the second term is a small correction, and we have~$\Pstoch \approx P_{\rm err}(k\to 0) \approx P_{\rm sampling}$. However, for very dense tracers such as HI
(and the Lyman-$\alpha$ forest~\cite{Ivanov:2023yla}), the second term can dominate and even surpass $P_{\rm sampling}$ by up to an order of magnitude. In Sec.~\ref{sec:simulations}, we will use hydrodynamical simulations to demonstrate that this indeed is the case for~HI.

One may wonder whether Eq.~\eqref{eq:Pstochseq} is significantly modified if one includes other operators (e.g.~$b_3^\Eul \delta^3$) at higher order in perturbation theory that generate constant contributions to the power spectrum in the low-$k$ limit. It is easy to show that these additional contributions are all loop-suppressed compared to the leading result from~$b_2^\Eul\delta^2$, and therefore much smaller. 
For this reason, we will neglect them in the rest of the paper.

\subsection{Implications for data analysis}
\label{sec:theory:implications_estimators}

Let us finally turn to consequences of previous discussion for cosmological data analysis. In the conventional approach based on the power spectrum, it is common to combine all terms with  constant power on large scales into a single contribution with an unknown amplitude that is fitted to the data. Schematically, 
\beq
\label{eq:Pg_split}
P_{\rm g} (k) = \bar P_{\rm model}(k) + \alpha \Pstoch ,
\eeq
where~$\alpha$ is a free parameter whose value is close to 1, and in~$\bar P_{\rm model}$, the constant contribution from the auto spectrum of the nonlinear bias term~$b_2^\Eul\delta^2$ (see Eq.~\ref{eq:Pstochseq}) has been subtracted.\footnote{For example, in  \texttt{CLASS-PT}~\cite{Chudaykin:2020aoj} and \texttt{PyBird}~\cite{DAmico:2020kxu}, this constant contribution is not included in the computed power spectrum by default, although it is included by default in  \texttt{velocileptors}~\cite{Chen:2020fxs}.}
The signal is assumed to only reside in~$\bar P_{\rm model}$, from which the cosmological and bias parameters are measured, while~$\alpha \Pstoch$ is treated as noise.\footnote{To be precise, for optical galaxy surveys, the noise term in Eq.~\eqref{eq:Pg_split} is usually taken to be $\alpha P_{\rm sampling}$ rather than $\alpha \Pstoch$. However, for these galaxy samples, $P_{\rm sampling}$ is typically much larger than the contribution to $\Pstoch$ from the $(b_2^{\Eul})^2$ term, such that $P_{\rm sampling}$ and $\Pstoch$ are interchangeable for practical purposes.} The same noise enters the covariance matrix, such that the cumulative signal to noise for the amplitude of the nonlinear power spectrum~$\bar P_{\rm model}$ in approximation of Gaussian covariance is given by the well-known formula (e.g.~\cite{1998ApJ...499..555T}) 
\begin{align}
\label{eq:SN-pk}
(S/N)^2 &= V \int \frac{d^3\bm{k}}{(2\pi)^3} \, \frac{\bar P_{\rm model}^2(k)}{2(\bar P_{\rm model}(k)+ \Pstoch)^2} \nonumber \\
& = \frac V2 \int_{k_{\rm min}}^{k_{\rm max}} \frac{k^2dk}{2\pi^2} \left( 1+ \frac{\Pstoch}{\bar P_{\rm model}(k)} \right)^{-2}\ .
\end{align}
Let us define 
\beq
N_{\rm pix} \equiv V \int^{k_{\rm max}} \frac{d^3\bm{k}}{(2\pi)^3}\ .
\eeq
These two equations show that, as expected, the signal to noise grows with the number of pixels as~$N_{\rm pix}^{1/2}$ as long as~$\Pstoch\ll \bar{P}_{\rm model}$. Given that~$\bar{P}_{\rm model}$ decays at high~$k$, this condition is violated at sufficiently high $k$. 

As we argued above, for dense tracers, the low-$k$ limit of the stochasticity in Eq.~\eqref{eq:Pstochseq} 
 can be much larger than the sampling noise. In the conventional power spectrum analysis, there is nothing that one can do to avoid having large noise given by~$\Pstoch$.\footnote{For tracers with~$\Pstoch\gg P_{\rm sampling}$, the best that one can do is to try to estimate the value of~$b_2^\Eul$ from the amplitude of~$\Pstoch$. Due to degeneracies with other bias parameters, this constraint may be a bit tighter than from the one-loop power spectrum, leading to more optimal estimate of cosmological parameters. However, note that this requires: ({\em a}) relatively good priors on~$P_{\rm err}$ and $k_{\rm NL}$; and ({\em b}) different renormalization conditions compared to the usual ones, where all bias parameters are defined in the~$k\to 0$ limit of various~$n$-point functions. 
 } This is a consequence of the fact that the estimator for the power spectrum in some sense ``averages''
over small-scale modes of $\delta_1$ that contribute to the stochasticity via their interactions through the~$b_2^\Eul\delta^2$ operator, such that the measured values of~$P_{\rm HI}$ on large scales are ``scattered'' more by the interacting small-scale modes than by the discrete sampling. 

On the other hand, an optimal method to extract cosmological parameters from data, such as field-level analysis, takes all relevant couplings and nonlinearities into account. Given that in these analyses, one uses the actual realization of large-scale structure without computing any particular summary statistic, contributions such as~$b_2^\Eul\delta^2$ become a part of the {\em signal} instead of the stochastic {\em noise}, and only the sampling noise limits the precision on inferred cosmological parameter values.
The full field-level analysis would lead to tighter error bars on cosmological parameters compared to the conventional approaches for several reasons, including more optimal BAO reconstruction and inclusion of higher-order correlation functions~\cite{Cabass:2023nyo}. If we want to focus only on the difference between~$\Pstoch$ and~$P_{\rm sampling}$, then the optimal analysis at the very least would lead to signal to noise given by
\beq
\label{eq:SN-opt}
(S/N)^2_{\rm opt.} = \frac V2 \int_{k_{\rm min}}^{k_{\rm max}} \frac{k^2dk}{2\pi^2} \left( 1+ \frac{P_{\rm sampling}}{\bar P_{\rm model}(k)} \right)^{-2}\ .
\eeq
In this paper, we focus only on this conservative\footnote{Here, we assume that no cosmological information can be obtained from the constant part of the power spectrum proportional to $(b_2^\Eul)^2$. In a truly optimal analysis, this would not be the case, but this assumption will be valid for an analysis based on an alternative estimator, as we will discuss below.}
estimate, and investigate what kind of improvements on cosmological parameters one would expect if the noise in the conventional analysis would be given by~$P_{\rm sampling}$ and not by~$\Pstoch$. This can be easily achieved by doing two Fisher forecasts for the power spectrum with two different noise amplitudes, as we explain below. 

One reason for focusing on this particular question is that, as argued in Ref.~\cite{Cabass:2023nyo}, one can hope to construct a simple estimator that removes the dominant contribution to the stochasticity, but that still only uses the power spectrum as a summary statistic. For instance, if we construct a new map\footnote{The nonlinear field used in this construction has to be appropriately smoothed to avoid large contributions from nonlinear scales.}
\beq
\hat\delta_{\rm HI} = \deltaHI - \frac{b_2^\Eul}{2(b_1^{\Eul})^2}\delta_{\rm HI}^2
\eeq
from the measured~$\deltaHI$ and compute the new power spectrum~$\hat P_{\rm HI}(k) = \vev{\hat{\delta}_{\rm HI}(\bm{k}) \hat{\delta}_{\rm HI}(-\bm{k}) }'$, the leading contribution to the stochasticity should be cancelled. In such a construction, the residual noise in~$\hat P_{\rm HI}$ should be closer to~$P_{\rm sampling}$ and the theoretical model for the power spectrum remains unchanged, since we have only performed a local transformation of the field. This is similar to BAO reconstruction, where one first ``removes'' large displacements using the map and then computes a new power spectrum that has cleaner BAO wiggles. The difference in the amount of information in~$P_{\rm HI}$ and~$\hat P_{\rm HI}$ is then dictated by the difference between~$\Pstoch$ and~$P_{\rm sampling}$, which is exactly what the two Fisher forecasts compute. 

Before describing our numerical tests, it is instructive to finish this section estimating the expected improvements in cosmological parameter estimation if one had~$P_{\rm sampling}$ instead of~$\Pstoch$ in the power spectrum analysis. For HI,~$\Pstoch$ can be up to ten times larger than~$P_{\rm sampling}$, and one may expect that this leads to dramatic change in the error bars. However, this is not the case, and the difference is bounded to be at most of~$\mathcal{O}(1)$. This is the consequence of the following arguments.

First, the effect is very significant only for tracers whose number density is already very high, such that the stochastic noise contribution to the covariance matrix is not large. Indeed, the dominant term in the stochasticity can be estimated at~$k>k_{\rm eq}$ as~\cite{Cabass:2023nyo}
\beq
\frac{(b_2^{\Eul})^2}{2} \int \frac{d^3\bm{q}}{(2\pi)^3} P_{\rm lin}(q) P_{\rm lin}(|\bm{k}-\bm{q}|) \approx \frac{(b_2^{\Eul})^2}{2} P_{\rm lin}(k) \Delta^2(k) \ ,
\eeq
where~$\Delta^2(k)$ is the contribution to the variance of the matter density field from modes with wavenumber $< k$, defined as
\beq
\Delta^2(k) \equiv \int^k \frac{d^3\bm{q}}{(2\pi)^3} P_{\rm lin}(q)\ .
\eeq
While this can be much larger than~$P_{\rm sampling}$, it is still smaller than the dominant part of the total error budget, given by the cosmic variance. The ratio of stochasticity and the HI power spectrum can be estimated as~$(b_2^\Eul/b_1^\Eul)^2 \Delta^2(k)$, which for realistic values of the biases can be at most~$\mathcal O(1)$ for $k$ close to the nonlinear scale. After all, the stochasticity is sourced by the loop corrections.  

The second reason why cosmological parameters may not be impacted very much is that both~$\Pstoch$ and~$P_{\rm sampling}$ can sometimes be smaller than the instrumental noise, which is an important part of the data covariance. If this is the case, neither $\Pstoch$ nor~$P_{\rm sampling}$ dominate the covariance, and there is nothing that one can do to make the analysis more optimal. As we are going to see, the assumptions about the instrumental noise are indeed important for our analysis, and we will argue that the optimal target for the thermal noise is to make it smaller than~$P_{\rm sampling}$. 

Finally, the impact of going from~$\Pstoch$ to~$P_{\rm sampling}$ is limited due to the fact that the total covariance matrix for the cosmological parameters depends not only on the data covariance, but also has additional contributions from the marginalization over biases and EFT counterterms. Schematically, 
\beq
C_{\rm cosmo.} = C_{\rm data} + C_{\rm marg}\ ,
\eeq
where~$C_{\rm marg}$ is of one-loop order, similar to the stochasticity~\cite{Chudaykin:2020hbf}. 
Therefore, marginalization over nuisance parameters can act to reduce the impact of changes in noise. Of course, this effect would be smaller if one had better priors on biases and EFT counterterms, which may be the situation in the future~\cite{2024arXiv240213310I}.

To conclude, performing the two Fisher forecasts with two different noise levels, given by~$\Pstoch$ and~$P_{\rm sampling}$, we expect at most a factor of two difference in error bars. The exact numbers will of course depend on the cosmological parameter, redshift, noise, scale cuts, and so on, but as we are going to see, the best and worst case scenarios for HI fall in this expected range. 

\section{Simulations}
\label{sec:simulations}

We have argued in the previous section that the difference between the low-$k$ limit of the HI stochasticity,~$\Pstoch$, and the model error power spectrum,~$P_{\rm sampling}$, can be very large. The exact values of these spectra depend significantly on the type of tracer and its properties, which is
why careful measurements in simulations are needed across a wide range of redshifts.\footnote{For neutral hydrogen at $z\lesssim 1$, the stochasticity and the model error have been measured from hydrodynamical simulations in Ref.~\cite{Obuljen:2022cjo}, while the model error was measured at higher redshifts from an N-body simulation augmented with an HI halo occupation distribution in Ref.~\cite{Modi:2019hnu}.} Even once~$\Pstoch$ and~$P_{\rm sampling}$ as a function of redshift are known, the way their difference impacts cosmological analysis still depends on the exact values of the bias parameters. For example, as we explained above, the expected improvement using optimal methods is quadratically sensitive to the ratio~$b_2^\Eul/b_1^\Eul$. For all these reasons, in order to have realistic results in Fisher forecasts, one has to have good estimates of all nuisance parameters that characterize the sample of interest. 

In this section, 
in order to avoid our results being sensitive to a single prescription for baryonic physics,
we use two different sets of hydrodynamical simulations in order to measure the noise and bias parameters
of the simulated HI sample across a wide range of redshifts. In what follows, we provide the details of the simulations we use (Sec.~\ref{sec:simulations:TNG300-1}-\ref{sec:simulations:Magneticum}) and describe our measurement procedure (Sec.~\ref{sec:simulations:fieldlevel}). We also find simple fitting formulas for the most important bias parameters (Sec.~\ref{sec:simulations:biascoefficients}), 
along with the HI stochasticity (Sec.~\ref{sec:simulations:HIstochasticity}),
as a function of redshift.
These formulas can be used in future forecasts or simulations that require fiducial values of HI bias or stochasticity.

\subsection{IllustrisTNG}
\label{sec:simulations:TNG300-1}

The first simulation set we use is IllustrisTNG~\cite{TNGa,TNGb,TNGc,TNGd,TNGe}, a suite of gravitational, 
magneto-hydrodynamical 
simulations run with the Arepo code~\cite{ArepoTNG}, with the subgrid physics 
model from Refs.~\cite{Pillepich:2017fcc,Weinberger2020TNG}. These simulations have been 
instrumental in understanding clustering properties of HI~\cite{Villaescusa-Navarro:2018vsg,Obuljen:2022cjo}. 
In Ref.~\cite{Obuljen:2022cjo} we have used the IllustrisTNG 
simulations to infer the HI clustering properties using the full field-level modeling, limiting our 
analysis to low redshifts $z=[0,1]$. Here we extend our analysis and consider snapshots at  $z=[0,0.5,1,1.5,2,3,5]$. We again consider only the largest simulation box with 
the highest mass resolution: TNG300-1, which has a box size of $L=205\Mpcinvh$ and 
$2\times2500^3$ dark matter and gas tracer particles (with masses $m_\mathrm{dm}=5.9\times10^7 M_\odot$ and $m_\mathrm{gas}=1.1\times10^7 
M_\odot$ respectively). To obtain the HI field from the simulation 
outputs, we follow the prescription described in Ref.~\cite{Villaescusa-Navarro:2018vsg} to 
compute the HI masses from the gas and star-forming particles. Finally, we assign HI particles to 
a regular $256^3$ grid using Cloud-In-Cell (CIC) mesh scheme to obtain the simulated HI field at 
different redshifts. 

\subsection{Magneticum}
\label{sec:simulations:Magneticum}

The second set of simulations we use is Magneticum~\cite{Magneticum_Hirschmann2014,Magneticum_Teklu2015}, a suite 
of cosmological hydrodynamical simulations run using N-body/Smoothed Particle 
Hydrodynamics (SPH) code Gadget3 \cite{Gadget, Gadget_SPH}, with improved SPH 
prescription \cite{Dolag2004,Dolag2005,Dolag2006}. These simulations use a subgrid physics 
model described in Refs.\ 
\cite{Magneticum_Hirschmann2014,Magneticum_Teklu2015,Magenticum_Steinborn2016}. For 
the purposes of this work, we chose a large simulation box with a mass resolution sufficient to 
resolve the halos in the halo mass range $10^{10-12} h^{-1} M_\odot$ which host the majority of 
the HI content in the late universe. In particular, we use Box2b/hrm with the box size of $L=640\hinvMpc$ 
and $2\times 2880^2$ particles, providing the dark matter and gas particle mass resolution of 
$m_\mathrm{dm}=6.9\times10^8 h^{-1} M_\odot$ and $m_\mathrm{gas}=1.4\times10^8 h^{-1} 
M_\odot$, respectively. 

Due to the lower mass resolution, we focus on output redshifts $z<2$, in particular $z=0.25$ and 
$z=1.18$. We apply the same post-processing procedure to the simulation outputs as we did in 
the case of IllustrisTNG, following Ref.~\cite{Villaescusa-Navarro:2018vsg}, to obtain the
HI masses for each gas and star-forming particle. We then assign the particles to a regular 
$512^3$ grid weighted by their HI mass to obtain the final simulated HI density field.

\subsection{HI field-level modelling}
\label{sec:simulations:fieldlevel}

We have already presented the theoretical model for the maps of biased tracers in real space in Sec.~\ref{sec:theory}. In order to simplify comparison to simulations in practice, it is convenient to slightly modify this model and rewrite it in terms of fields that do not correlate to each other. This can be simply achieved by introducing a basis of orthogonal operators~$\tilde{\mathcal O}^\perp$, such that~$\langle \tilde{\mathcal O}_a^\perp \tilde{\mathcal O}_b^\perp \rangle = 0$ for any $a\neq b$. This orthogonalization is just a linear transformation of the original operator basis~$\tilde{\mathcal O}$, which does not change the statistical properties of the map. Using this new basis, we can write
\beq
\label{eq:real_model_orht}
\begin{split}
\delta_\mathrm{HI}^{\rm{model}}(\bm{k}) =\, & \beta_1(k)\tilde{\delta}_1(\bm{k}) + 
\beta_2(k)\tilde{\delta}_2^\perp(\bm{k}) + \beta_{\mathcal{G}_2}(k)\tilde{\mathcal{G}}_2^\perp(\bm{k}) \ ,
\end{split}
\eeq
where we have promoted all constant bias parameters to scale-dependent functions~$\beta_i(k)$. Note that after orthogonalization, the transfer function multiplying~$\tilde{\delta}_1$ acquires two additional contributions from quadratic fields compared to~$\bar\beta_1(k)$ in Eq.~\eqref{eq:beta1_model}. More explicitly,
\beq
\beta_1(k) = \bar{\beta}_1(k) + b_{\mathcal G_2} \frac{\vev{\tilde {\mathcal G}_2 \tilde\delta_1} '}{\vev{\tilde\delta_1 \tilde\delta_1 }'} + b_2 \frac{\vev{\tilde \delta_2 \tilde\delta_1} '}{\vev{\tilde\delta_1 \tilde\delta_1 }'}   \ .
\eeq
The transfer functions can be measured from simulations by performing least-square minimization of the error power spectrum
\beq
P_{\rm{err}}(\bm{k}) = \vev{|\delta_{\rm HI}^{\rm sim}(\bm{k})-\delta_{\rm 
HI}^{\rm model}(\bm{k})|^2}' \ .
\eeq
The advantage of orthogonal basis is that the best-fit transfer functions are then simply given by the cross-spectra of~$\tilde{\mathcal O}^\perp$ and~$\delta_{\rm HI}^{\rm sim}$:
\beq
\beta_i^\text{best-fit}(k) = \frac{\vev{\tilde{\mathcal O}_i^\perp(\bm{k}) \delta_{\rm HI}^{\rm sim} (-\bm{k}) }'}{\vev{\tilde{\mathcal O}_i^\perp(\bm{k}) \tilde{\mathcal O}_i^\perp (-\bm{k}) }' }\ .
\eeq
These best-fit transfer functions can then be fit using the theoretical model in order to obtain the best-fit bias parameters inferred from the map. In practice, we use the publicly available code \texttt{Hi-Fi mocks}\footnote{\url{https://github.com/andrejobuljen/Hi-Fi_mocks}}\cite{Obuljen:2022cjo} to compute the theoretical model with the same initial conditions used to simulate the HI field, and to obtain the best-fit transfer functions.

\begin{figure*}[t]
\includegraphics[width=\textwidth, trim=0 20 0 0]{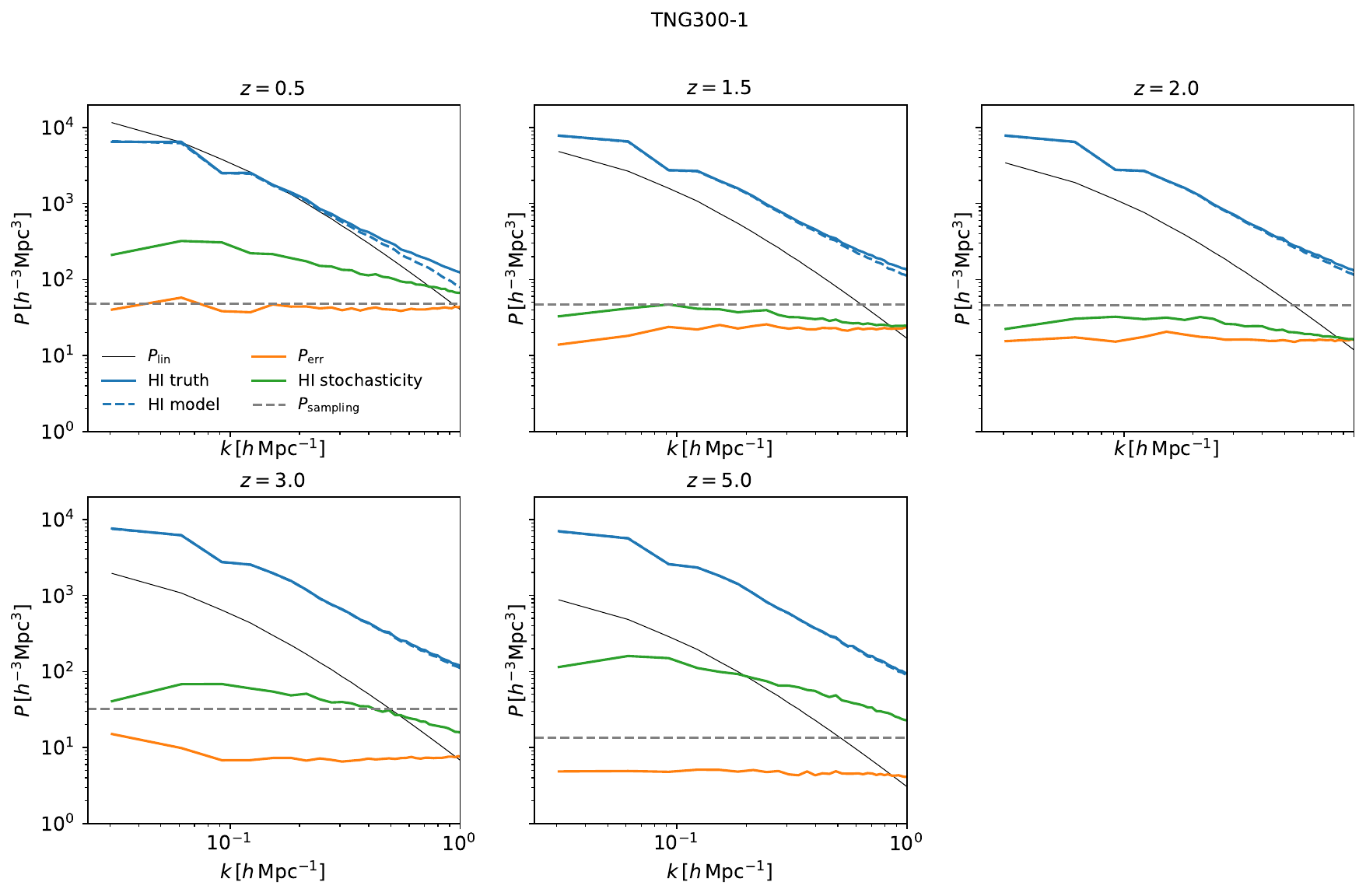}
\caption{IllustrisTNG HI power spectrum: simulated (``truth", {\em blue solid line}), best-fit cubic model ({\em blue dashed line}), error power spectra $P_{\rm err}$ ({\em orange solid line}), HI stochasticity ({\em green solid line}), and $M_\mathrm{HI}$-weighted sampling noise estimate ({\em dashed grey line}) for comparison. Similar to previous results from Ref.~\cite{Obuljen:2022cjo} at $z=0$ and $z=1$, we find that $P_{\rm err}$ is roughly independent of $k$, and HI stochasticity is larger than $P_{\rm err}$ at all redshifts. Most of the difference between $P_{\rm err}$ and the stochasticity can be explained by the quadratic bias operator contribution, and the smallest difference is at redshift $z\approx2$ where $b_2^\Shi$ is close to zero (see text for details). The $M_\mathrm{HI}$-weighted sampling noise shows behavior similar to that at redshifts $z=0$ and $z=1$ from Ref.~\cite{Obuljen:2022cjo}, being comparable or larger than $P_{\rm err}$, and generally lower than HI stochasticity at all redshifts.}
\label{fig:TNG_summary}
\end{figure*}

\begin{figure*}
\includegraphics[width=0.7\textwidth, trim=0 20 0 0]{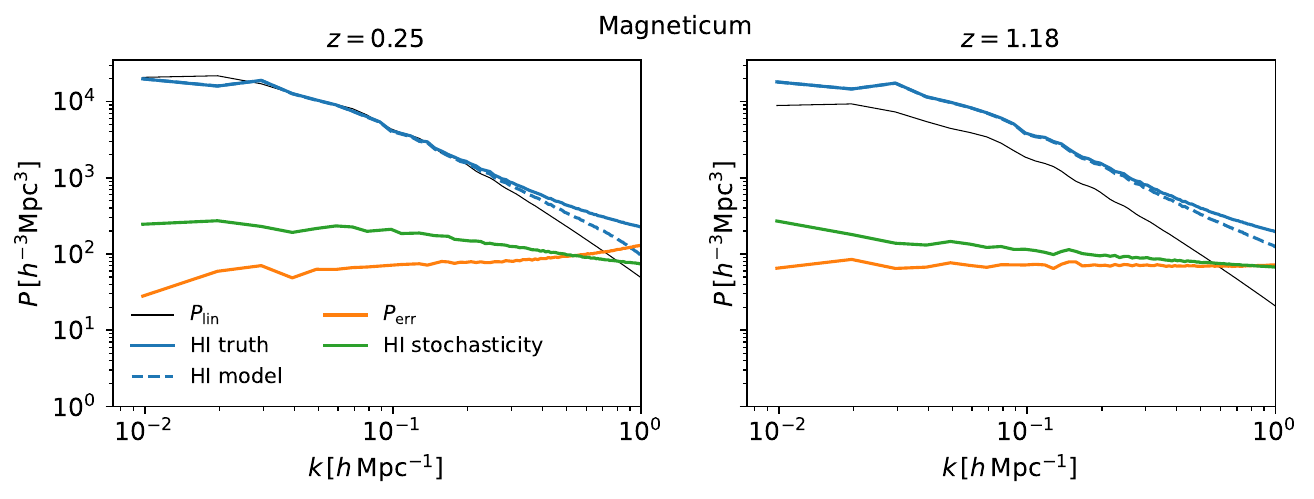}
\caption{Magneticum HI power spectrum: simulated (``truth", {\em blue solid line}), best-fit cubic model ({\em blue dashed line}), error power spectra $P_{\rm err}$ ({\em orange solid line}), and HI stochasticity ({\em green solid line}). We find similar quantitative behavior of $P_{\rm err}$ and HI stochasticity compared to results from IllustrisTNG.}
\label{fig:magneticum_summary}
\end{figure*}

Figure~\ref{fig:TNG_summary} compares the HI power spectrum measured from IllustrisTNG with the best-fit power spectrum of the model in Eq.~\eqref{eq:real_model_orht} at five different redshifts. The figure also shows the model error power spectrum and HI stochasticity computed using the best-fit model. At all redshifts we consider, we find that the HI stochasticity exceeds the model error, sometimes by an order of magnitude.\footnote{Note that this finding, and those of Ref.~\cite{Obuljen:2022cjo}, effectively disprove the claim of Ref.~\cite{Umeh:2021xqm} that bias-induced stochasticity is unphysical and only appears in theoretical predictions that ignore the finite sizes of halos.} The smallest difference between stochasticity and model error is observed in the $z=2$ snapshot, and we will see in Sec.~\ref{sec:simulations:biascoefficients} that $|b_2^\Shi|$ reaches a minimum at this redshift, in agreement with our expectation that the difference is caused by a term in the model involving~$b_2^\Shi$.

The dashed grey lines in Figure~\ref{fig:TNG_summary} show estimates for the $M_\mathrm{HI}$-weighted sampling noise of HI \cite{Villaescusa-Navarro:2018vsg,Obuljen:2022cjo}. We compute it as the low-$k$ limit of the 1-halo term in the HI power spectrum, using the HI mass-halo mass relation measured from IllustrisTNG (see Sec.~15 of Ref.~\cite{Villaescusa-Navarro:2018vsg}). An alternative way to estimate the HI sampling noise is to place the total amount of HI inside each halo at the center of that halo and take the amplitude of the high-$k$ plateau of the $M_\mathrm{HI}$-weighted halo power spectrum. We previously found good agreement between these two estimates at low redshifts \cite{Obuljen:2022cjo}, we have also found good agreement to hold at higher redshifts for TNG300-1.

At $z=0.5$, we find that $P_{\rm err}(k)$ is comparable to $P_{\rm sampling}$, while at higher redshifts, $P_{\rm err}(k)$ is consistently lower than $P_{\rm sampling}$. Sub-Poissonian model error has previously been seen in simulations for higher-mass halos (e.g.~\cite{Schmittfull:2018yuk}), and is usually attributed to halo exclusion effects.
This could potentially be related to the low fraction of HI contained in satellite galaxies at the relevant halo masses~\cite{Villaescusa-Navarro:2018vsg,Kokron:2021faa}, but we leave it to future work to investigate the physical origin of our observed sub-Poissonian model error for HI.

In Figure~\ref{fig:magneticum_summary}, we show analogous measurements from Magneticum at two redshifts. We observe behavior that is consistent with IllustrisTNG, implying that our conclusions about stochasticity and model error are not unique to the specific properties (e.g.\ subgrid models) of either simulation suite.

\subsection{Bias coefficients}
\label{sec:simulations:biascoefficients}

Having obtained the best-fit transfer functions from the simulated HI fields at different redshifts, we follow the approach from Ref.~\cite{Obuljen:2022cjo} and use polynomials to describe the transfer functions as follows:
\begin{align}
\begin{split}
\beta_1(k) &= a_0 + a_1k + a_2k^2 + a_4k^4,  \\
\beta_{i\ne1}(k) &= a_0 + a_2k^2 + a_4k^4.
\end{split}
\end{align}
We obtain the free coefficients $a_i$ from the best-fit of the polynomials to the measured transfer functions, weighting each $k$ bin by $k$ to account for the different number of modes and only using scales below $k_\mathrm{max}=1~\Mpcinvh$. 

\begin{figure}[t]
\includegraphics[width=0.48\textwidth, trim = 0 10 0 0]{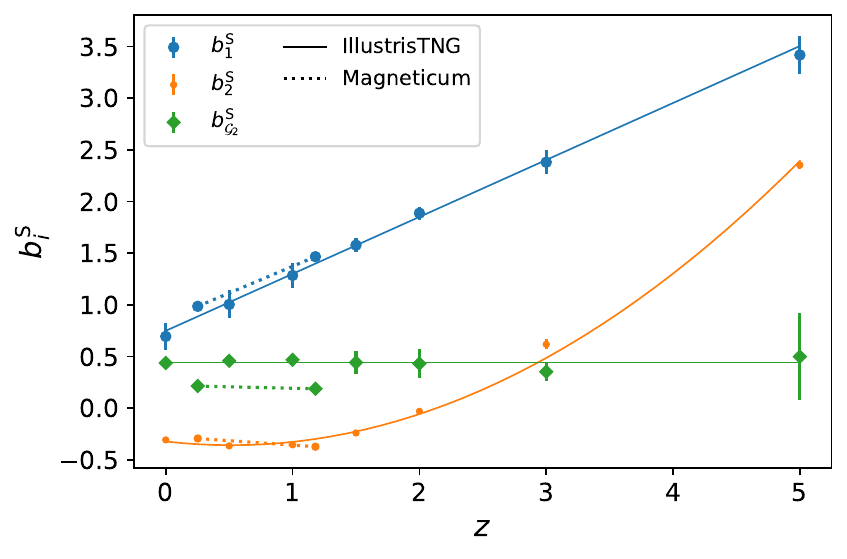}
\caption{HI linear, quadratic and tidal bias measurements from two different hydrodynamical 
simulations: IllustrisTNG ({\em solid lines}) and Magneticum ({\em dotted lines}). Each line is a best-fit
function for linear, quadratic, or tidal bias assuming linear, quadratic and constant dependence 
on redshift, respectively. We obtain the bias values as the the low-$k$ limit of the polynomial 
transfer functions using the best-fit field-level model. The uncertainties are estimated from the 
residual scatter in the transfer functions around the polynomial best-fit, considering scales below 
$k_\mathrm{max}=0.3\hinvMpc$. The uncertainties in the bias measurements are smaller in the 
case of Magneticum due to its larger box size.}
\label{fig:z_bias}
\end{figure}

We identify the constant pieces $a_0$ of different best-fit polynomials as values of the corresponding HI bias parameters, i.e.\ $a_0$ of $\beta_1$, $\beta_2$ and $\beta_{\mathcal{G}_2}$ as $b_1^\Shi$, $b_2^\Shi$ and $b_{\mathcal{G}_2}^\Shi$, respectively. We show the measured HI bias parameters as a function of redshift in Fig.~\ref{fig:z_bias} from both IllustrisTNG and Magneticum simulations. We estimate the uncertainties on the bias parameters from the residual scatter of transfer functions around the polynomial best-fits, considering only scales below $k_{\rm max}=0.3~\Mpcinvh$.\footnote{The best-fit $a_i$ coefficients were fit up to $k_\mathrm{max}=1\,\Mpcinvh$ to allow for the \texttt{Hi-Fi mocks} code to generate HI mocks that match the HI clustering in IllustrisTNG up to that $k_\mathrm{max}$, while we estimate uncertainties using measurements up to $k_{\rm max}=0.3~\Mpcinvh$ to ensure that these uncertainties only capture information on quasi-linear scales where the transfer functions can be reliably related to perturbative predictions.}
We find smaller uncertainties in bias measurements from Magneticum due to its larger box size compared to IllustrisTNG. 
For the linear and quadratic biases $b_1^\Shi$ and $b_2^\Shi$, we find excellent agreement between the two simulations, with $b_1^\Shi$ scaling linearly with $z$ and $b_2^\Shi$ scaling quadratically with $z$. The tidal bias $b_{\mathcal{G}_2}^\Shi$ does not show evidence for redshift evolution in either simulation, and the size of this bias differs somewhat between the two simulations in the overlapping redshift range. We leave to future work an investigation of the specific properties of the simulations that drive the agreement in $b_1^\Shi$ and $b_2^\Shi$ and the discrepancy in $b_{\mathcal{G}_2}^\Shi$.

Motivated by these results,
 we use linear, quadratic and constant functions of redshift to model the linear, quadratic and tidal HI biases. We then obtain the weighted best-fit parameters for these functions using the TNG300-1 measurements and HI bias uncertainties from above, and we find the following:
\begin{align*}
\label{eq:fitting_functions}
b_1^\Shi(z) & = 0.75(0.03) + 0.55(0.02) z, \\
b_2^\Shi(z) & = -0.32(0.02) - 0.14(0.03) z + 0.14(0.01) z^2, \\
b_{\mathcal{G}_2}^\Shi(z) & = 0.44(0.01),
\numberthis
\end{align*}
where the values in parentheses are the uncertainties on parameters obtained from the weighted best-fit. 
Using Eq.~\eqref{eq:Shi_to_Eul}, these can be converted to Eulerian bias parameters in the conventions of Eq.~\eqref{eq:deltaHI_Eul_maintext}:
\begin{align*}
\label{eq:fitting_functions_Eul}
b_1^\Eul(z) & = 0.75 + 0.55 z, \\
b_2^\Eul(z) & = -0.64 - 0.28 z + 0.28 z^2, \\
b_{\mathcal{G}_2}^\Eul(z) & = 0.23 - 0.16 z.
\numberthis
\end{align*}
(See Appendix~\ref{app:bias_params} for translations between the shifted-operator biases and other conventions for bias coefficients.)

\subsection{HI stochasticity}
\label{sec:simulations:HIstochasticity}

As argued above, in a forecast involving the HI power spectrum, one should include the HI stochasticity in the power spectrum covariance. To facilitate this in future forecasts, we provide the following fitting function for our measurements of $\Pstoch$ from IllustrisTNG:
\beq
\label{eq:stoch_fitting_function}
\Pstoch(z) = 
\begin{cases} 
471 - 485 z + 132 z^2, & \text{if } z \leq 2 \\
-33 + 31 z, & \text{if } 2 < z \leq 5\ .
\end{cases}
\eeq
Specifically, we measure $\Pstoch$ at each redshift from the full $k$-dependent stochasticity by taking an average value on large scales ($k<0.16\,\hinvMpc$), using the scatter on those scales as an estimate of the uncertainty. We then perform a weighted least-squares fit using quadratic and linear relations at lower ($z \leq 2$) and higher redshifts ($z \geq 2$), respectively.

\begin{figure}[t]
\includegraphics[width=\columnwidth, trim=0 15 0 0]{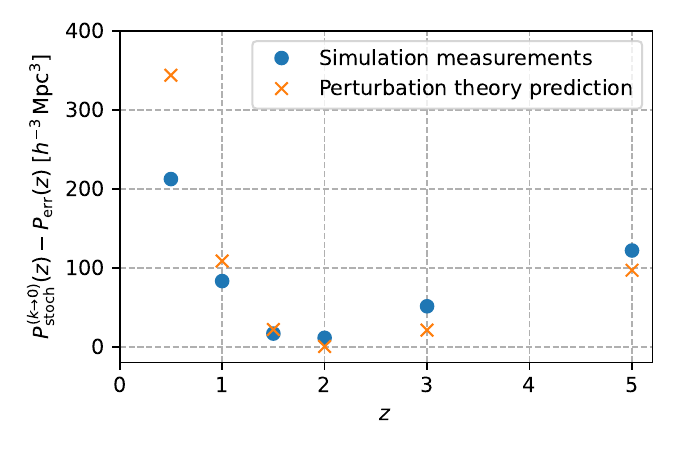}
\caption{%
Comparison between simulation measurements of the difference between stochasticity and sampling noise \textit{(blue points)} and perturbation theory predictions for bias-induced stochasticity at leading order \textit{(orange crosses)}. The two sets of points agree within a factor of 2 except at $z=2$. We do not attempt a statistically rigorous comparison in this work, but note that the similar redshift evolution of the prediction and simulation measurements supports our claim that nonlinear bias is the leading source of excess stochasticity.
}
\label{fig:stoch_comparison}
\end{figure}

In Figure~\ref{fig:stoch_comparison}, we show the difference between these $\Pstoch$ estimates and $P_{\rm err}$ values from simulations, where the latter is determined from the average of the measured $P_{\rm err}(k,z)$ over $0.1\hinvMpc < k < 0.3\hinvMpc$ (where we consistently observe roughly flat behavior in~$k$). From the arguments in Sec.~\ref{sec:theory}, this difference should be equal to the low-$k$ limit of the $b_2^2$ term in the power spectrum model (see Eq.~\ref{eq:Pstochseq}). We also plot a prediction for this term, using our measured value for $b_2^\Shi$ and the forecasting code described in Sec.~\ref{sec:forecasts:setup:predictions} below.

We find that the perturbation theory prediction agrees with the simulation measurements within a factor of~2 at all redshifts except $z=2$, where prediction is very small due to the small measured value of $b_2^\Shi$ ($\approx -0.04$).\footnote{Note that the prediction extracted from the forecasting code is computed integrating up to a large $k_{\rm max}$ that is unrelated to $k_{\rm NL}$, while in detail one would want to set $k_{\rm max}=k_{\rm NL}$ in this computation. As a result, the discrepancy between prediction and simulation measurements is larger at lower redshifts, where $k_{\rm NL}$ is smaller.}
We have not attempted to associated errorbars to the simulation measurements in this comparison, since those errorbars would need to combine uncertainty from the method used to measure the points, sample variance from the finite number of modes in the simulation, and other potential sources of error. We leave such a rigorous comparison for future work, but note that the agreeing trends between perturbation theory and simulation measurements supports our claim that nonlinear bias is a strong contributor to the measured excess stochasticity.

\section{Forecasts}
\label{sec:forecasts}

To quantify the impact of the HI model error on future analyses of \tcm measurements, in this 
section we perform a series of forecasts for different surveys and science targets. We describe 
the setup for these forecasts in Sec.~\ref{sec:forecasts:setup}, and compare fiducial power 
spectra for signal, thermal noise, model error, and stochasticity in 
Sec.~\ref{sec:forecasts:powerspectra}. We then present the results for cosmological distance 
measures in Sec.~\ref{sec:forecasts:distancemeasures}, structure growth in 
Sec.~\ref{sec:forecasts:growth}, and cosmological parameters in 
Sec.~\ref{sec:forecasts:parameters}.

In this section, we are concerned with comparing an analysis of the HI power spectrum with an analysis that uses beyond-2-point information, either via forward modelling of the full density field or a simpler estimator that nulls the dominant stochastic noise~\cite{Cabass:2023nyo}. We will not distinguish between these options in what follows, but will simply refer to a ``more optimal" analysis as one with reduced stochastic noise.

We also note that the main goal of these forecasts is to estimate the {\em improvement} in cosmological constraints obtained from a more optimal analysis compared to a power spectrum analysis. To illustrate the context of this improvement, we also show forecasts for the absolute constraining power of a power spectrum analysis. We have attempted to make reasonable assumptions about the \tcm surveys we consider, but we caution the reader that the actual achievable constraining power of these surveys will strongly depend on several factors that are difficult to predict, including foreground cleaning efficiency, data loss due to radio-frequency interference (ignored in our forecasts), and so on. However, we expect the {\em relative} constraining power between power spectrum and more optimal analyses to be more robust to these factors than the absolute constraining power of either analysis.

\subsection{Setup}
\label{sec:forecasts:setup}

Our forecasts mostly follow the methods of Ref.~\cite{Sailer:2021yzm}, with a few distinctions 
that we will describe below. All forecasts use a modified version of the public \texttt{FishLSS} 
code\footnote{Our version of the code is located at 
\url{https://github.com/sjforeman/FishLSS/tree/HI}, 
and files that can be used to reproduce our forecasts are located at 
\url{https://github.com/sjforeman/HI_forecasts}.}.

\subsubsection{Predictions}
\label{sec:forecasts:setup:predictions}

We use the one-loop Lagrangian Perturbation Theory (LPT) model for biased tracers in redshift 
space presented in Refs.~\cite{Chen:2020fxs,Chen:2020zjt}, as implemented in the public 
\texttt{velocileptors} code\footnote{\url{https://github.com/sfschen/velocileptors}}. Linear matter 
power spectra are computed using 
\texttt{CLASS}\footnote{\url{http://class-code.net}}~\cite{Blas:2011rf}. 
At one-loop order, the LPT prediction is equivalent to that of the 
shifted-operator 
formalism from Sec.~\ref{sec:simulations:fieldlevel} (up to higher-order differences and implementation 
details), since both formalisms non-perturbatively incorporate the effects of linear displacements 
(e.g.~\cite{Schmittfull:2018yuk}). At this order, the power spectrum prediction involves the linear 
bias $\tilde{b}_1^{\rm E}$, quadratic bias $\tilde{b}_2^{\rm E}$, and tidal bias $\tilde{b}_s^{\rm E}$, which we relate to the bias parameters 
measured in Sec.~\ref{sec:simulations} using the expressions in 
Appendix~\ref{app:bias_params}. As in Ref.~\cite{Sailer:2021yzm}, cubic operators are also 
included in the bias expansion, but their coefficients are assumed to be fixed by time evolution in 
LPT.

We also include the counterterms and stochastic terms that are relevant at this order:
\begin{align*}
\PHI(k, \mu) &= \PHI^\text{1-loop}(k, \mu) \\
&\quad
	+ \lp \alpha_0 + \alpha_2 \mu^2 + \alpha_4 \mu^4 \rp 
	\frac{k^2}{k_*^2} P_{\rm cb}^{\rm Zel}(k) \\
&\quad
	+ N_0 + N_2 (\mu k)^2 + N_4 (\mu k)^4\ ,
	\numberthis
	\label{eq:PHI_prediction}
\end{align*}
where $k_* \equiv 1\hinvMpc$, $P_{\rm cb}^{\rm Zel}$ is the CDM+baryon power spectrum in the 
Zel'dovich approximation, and we have omitted the redshift-dependence for brevity. The isotropic 
stochastic parameter $N_0$ is typically set to the sampling noise for the tracer under 
consideration.
In our forecasts, 
we instead set $N_0$ to the estimate of the constant model error power spectra $P_{\rm err}$ 
measured in Sec.~\ref{sec:simulations:HIstochasticity}, since this more accurately reflects the noise not captured by the model, and in some cases differs significantly from the Poisson expectation for the sampling noise (see Figure~\ref{fig:TNG_summary}).

For the other parameters in Eq.~\eqref{eq:PHI_prediction}, we follow the choices of  
Ref.~\cite{Sailer:2021yzm}. For the counterterms, we use $\alpha_0(z) = 1.22 + 
0.24b_1(z)^2(z-5.96)$ and $\alpha_2=\alpha_4=0$. For the $(k\mu)^2$ stochastic parameter, we 
use $N_2=-N_{\rm sampling} \sigma_{v}^2$, where $N_{\rm sampling}$ is the HI sampling noise 
measured in Sec.~\ref{sec:simulations:fieldlevel} and
\beq
\label{eq:sigv}
\sigma_{v}(z) = (10\,{\rm km}\, {\rm s}^{-1})(1+z) H(z)^{-1}\ .
\eeq
We set $N_4=0$.

The normalization of $\sigma_{v}(z)$ in Eq.~\eqref{eq:sigv} is reflective of the physical velocity dispersions 
that generate Finger-of-God effects in the distribution of 
HI~\cite{Sailer:2021yzm,Villaescusa-Navarro:2018vsg}. 
However, $\sigma_{v}(z)$ can also be fit to the measured error power spectrum $P_{\rm err}$ associated with field-level modelling of HI, and Ref.~\cite{Obuljen:2022cjo} found values of $\sigma_v(z=0) = 2.87\Mpcinvh$ and $\sigma_v(z=1) = 2.54\Mpcinvh$ when carrying out these fits using IllustrisTNG. The $z=0$ value is equivalent to using $287\,{\rm km}\, {\rm s}^{-1}$ in Eq.~\eqref{eq:sigv}. We leave an investigation of this discrepancy to future work, but we have carried out a version of our forecasts that uses this higher normalization of $\sigma_v(z)$, and we found that it does not significantly affect our results (see Appendix~\ref{app:alternative_forecasts:sigmav} for details).

To construct the observed \tcm power spectrum $\Ptcm$, $\PHI$ is multiplied by the square of 
the mean \tcm brightness temperature $T_{\rm b}$, and gets added to an instrumental noise 
power spectrum $\PN$:
\beq
\Ptcm(k, \mu) = T_{\rm b}^2 \PHI(k, \mu) + \PN(k, \mu)\ .
\eeq
The mean brightness temperature is given by (e.g.~\cite{CosmicVisions21cm:2018rfq})
\beq
T_{\rm b}(z) = 188 h \frac{H_0}{H(z)} \Omega_{\rm HI}(z) (1+z)^2\; {\rm mK} \ ,
\label{eq:Tbz}
\eeq
where we use the fitting formula from Ref.~\cite{Crighton:2015pza} for the HI density parameter 
$\Omega_{\rm 
HI}(z) \equiv \bar{\rho}_{\rm HI}(z) / \rho_{\rm c,0}$. In our forecasts, we also 
include the Alcock-Paczynski effect as described in Ref.~\cite{Sailer:2021yzm}.

We use fiducial cosmological parameters from Planck 2018~\cite{Planck:2018vyg} 
($\omega_{\rm 
b} = 0.0224$, $\omega_{\rm c} = 0.120$, $h = 0.673$, $A_{\rm s} = 2.10\times 
10^{-9}$, $n_{\rm s} = 0.966$, and $\tau=0.543$). Our fiducial model also takes $N_{\rm 
eff}=3.046$ 
and assumes 3 neutrino species with masses $\{0, 0.01, 0.05\}\,{\rm eV}$.

When considering the PUMA survey (see Sec.~\ref{sec:forecasts:setup:surveys}), we perform forecasts up to $z=6$, but we note that there is growing evidence from observations (e.g.~\cite{Kulkarni:2018erh,Bosman:2021oom}) and simulations (e.g.~\cite{Keating:2019qda,Giri:2024xet}) that reionization is likely still ongoing at $z=6$. A perturbative bias treatment may still apply~\cite{McQuinn:2018zwa,Qin:2022xho}, but the values and interpretation of the bias coefficients will be different from the post-reionization case we assume in this work. The reader should keep this caveat in mind when examining our forecasts for $5\lesssim z \lesssim 6$. Our forecasts for cosmological parameters in Sec.~\ref{sec:forecasts:parameters} are dominated by contributions from lower redshifts, so our conclusions would not change significantly if $5\lesssim z \lesssim 6$ was omitted completely.

\subsubsection{Fisher matrices}

We use Fisher matrices (e.g.~\cite{Tegmark:1996bz}) to forecast the measurement uncertainty 
on model parameters $\theta_a$.  We divide the redshift range of a given survey into equal-width 
redshift bins with central redshifts $z_m$ and comoving volumes $V_m$, and compute a 
separate Fisher matrix for each bin. At each redshift, we further divide the accessible ranges of 
$k$ and $\mu$ into logarithmically-spaced and linearly-spaced bins with widths $\Delta \log k$ 
and $\Delta\mu$ respectively, and index each pair of $(k, \mu)$ bins by $i$ or $j$. The Fisher 
matrix for redshift bin $m$ is then given by
\beq
F_{ab}(z_m) = \sum_{ij} \frac{\d\Ptcmi(z_m)}{\d\theta_a} 
	\frac{1}{\sigma_i(z_m)^2}
	\frac{\d\Ptcmj(z_m)}{\d\theta_b} \ ,
	\label{eq:Fab}
\eeq
where $\Ptcmi$ denotes the value of $\Ptcm$ within $(k, \mu)$ bin $i$, and the variance of a 
measurement of $\Ptcmi$ is
\beq
\sigma_i(z_m)^2 = \frac{4\pi^2}{k_i^3 V_m\, \Delta \mu\, \Delta \log k} \Ptcmi(z_m)^2\ .
\label{eq:sigma_i}
\eeq
Since $\Ptcmi$ includes both the model and noise contributions, $\sigma_i(z_m)^2$ includes 
cosmic variance as well as observational uncertainty. We have assumed a diagonal covariance in 
$i$ and $j$, which allows us to only specify the variance in Eq.~\eqref{eq:Fab}.

For forecasts that combine information from multiple redshift bins, we simply sum the Fisher 
matrices from each bin, assuming zero correlation between bins:
\beq
F_{ab} = \sum_m F_{ab}(z_m)\ .
\eeq
The best-case measurement uncertainty on parameter $\theta_a$, when all other free 
parameters are marginalized over, is then given by the Cram\'{e}r-Rao inequality:
\beq
\sigma(\theta_a) \approx \lp  \lb F^{-1} \rb_{aa} \rp^{1/2}\ .
\eeq

\subsubsection{HI stochasticity}
\label{sec:forecasts:setup:stoch}

As discussed in Sec.~\ref{sec:theory} and verified with simulations in Sec.~\ref{sec:simulations}, the 
large-scale distribution of HI possesses a  stochastic noise contribution arising from nonlinear 
couplings between smaller-scale modes. Specifically, this contribution arises from the nonzero 
low-$k$ limit of the $\langle \tilde{\delta}_2 \tilde{\delta}_2 \rangle$ term in the shifted-operator 
prediction for the HI power spectrum. If one is using an operator basis where the tidal term is 
represented as $s^2$ instead of $\Gtwo$, as in the LPT formalism we use for our forecasts, the 
$\langle 
\tilde{s}^2 \tilde{s}^2 \rangle$ and $\langle \tilde{\delta}_2 \tilde{s}^2 \rangle$ terms will 
also have nonzero low-$k$ limits that will contribute in the same way.
(This is because the auto power spectrum of $\Gtwo$ and its cross power spectrum with $\delta_1^2$ vanish in the low-$k$ limit, unlike $s$.)

In the standard scheme of renormalizing bias parameters in the $k\to 0$ limit, this noise is an 
unavoidable contribution to the model error in a power spectrum analysis, but it can become part of the model {\em prediction} in a 
more optimal
analysis, and therefore does not contribute to the model error. To assess how this affects 
cosmological constraints, we perform two sets of forecasts:
\begin{enumerate} 
\item In the first set, when we evaluate the power spectra appearing in Eqs.~\eqref{eq:Fab} 
and~\eqref{eq:sigma_i},
we subtract the low-$k$ limits from the $\langle \delta_2 \delta_2 \rangle$, $\langle s^2 s^2 
\rangle$, and $\langle \delta_2 s^2 \rangle$ terms.
Subtracting these contributions from Eq.~\eqref{eq:sigma_i} implies that they do not contribute to 
the noise variance, and this lower noise level is meant
to mimic the constraining power of a more optimal analysis. 
\item In the second set, we do not subtract these contributions from the variance factor 
[Eq.~\eqref{eq:sigma_i}] in the Fisher matrix.
This case mimics 
a power spectrum analysis, in which these low-$k$ contributions cannot be distinguished 
from the model error, and therefore contribute to Eq.~\eqref{eq:sigma_i}.
\end{enumerate}

Thus, the difference between the first and second sets of forecasts indicates the improvement in 
constraining power available in a more optimal analysis over a power spectrum analysis. In 
fact, we expect this difference to underestimate the improvement, because in a more optimal 
analysis one will likely also have access to information that is contained in higher-point functions, 
which we do not attempt to incorporate here.

Note that in both sets of forecasts, we subtract the low-$k$ limits from the $\langle \delta_2 
\delta_2 \rangle$, $\langle s^2 s^2 \rangle$, and $\langle \delta_2 s^2 \rangle$ terms in the power 
spectrum derivatives in Eq.~\eqref{eq:Fab}. This is the correct procedure when the bias 
coefficients are renormalized in the $k\to 0$ limit, and implies that we do not attempt to extract 
information about the values of quadratic or tidal bias coefficients from $\mathcal{O}(k^0)$ 
contributions to the corresponding terms. See Refs.~\cite{Cabass:2023nyo,Rubira:2023vzw} for discussions of how to extract this information 
via alternative bias renormalization schemes.

\subsubsection{Nonlinear scale}
\label{sec:forecasts:setup:knl}

Our results depend on the nonlinear scale $\knl$ beyond which we expect the perturbative 
approach to bias and/or gravitational nonlinearity to break down. If gravitational nonlinearity were the dominant effect in determining the redshift-dependence of $\knl$ for HI, we would determine $\knl(z)$ as the scale where the variance of the matter density becomes order unity (commonly written as $\Delta^2(\knl,z) \sim 1$). For a power-law universe $P_{\rm lin}(k)\sim k^n$, one finds that $k_{\rm NL} \sim D(z)^{-\frac 2{3+n}}$. For a $\Lambda$CDM-like cosmology, $k_{\rm NL}$ approximately scales as $D(z)^{-1}$ close to $z=0$ and increases much faster at high redshifts.\footnote{One sometimes obtains $\knl(z) \propto D(z)^{-1}$ by assuming that the nonlinear scale is set by the typical displacements in the Zel'dovich approximation, as done in some previous forecasts in the literature (e.g.~\cite{CosmicVisions21cm:2018rfq,Sailer:2021yzm}). However, such nonlinearities can be treated exactly by the infrared resummation and this estimate is very conservative, particularly at high redshifts.} However, for biased tracers, it is common for nonlinear bias to be more important than gravitational nonlinearity in determining $\knl(z)$ (e.g.~\cite{Schmittfull:2018yuk}). 
In an extreme case, for a passively evolving population of tracers, we expect the $n$th-order Eulerian bias parameters to be $n$th-order polynomials in $D(z)^{-1}$ (e.g.~\cite{Desjacques:2016bnm}), while the corresponding Eulerian operators scale 
like $D(z)^{n}$. 
In this scenario, 
the relative strength of bias-induced nonlinearities in the tracer overdensity 
would not change significantly with redshift.
and we would expect $\knl$ to be roughly constant in time.
While we do not expect this to be exactly true for HI (both on theoretical grounds and based on simulation measurements), nonlinear bias might still be expected to cause $\knl(z)$ to scale more weakly with redshift than the pure-gravity case.

\begin{figure}[t]
\includegraphics[width=\columnwidth, trim=10 20 10 0]{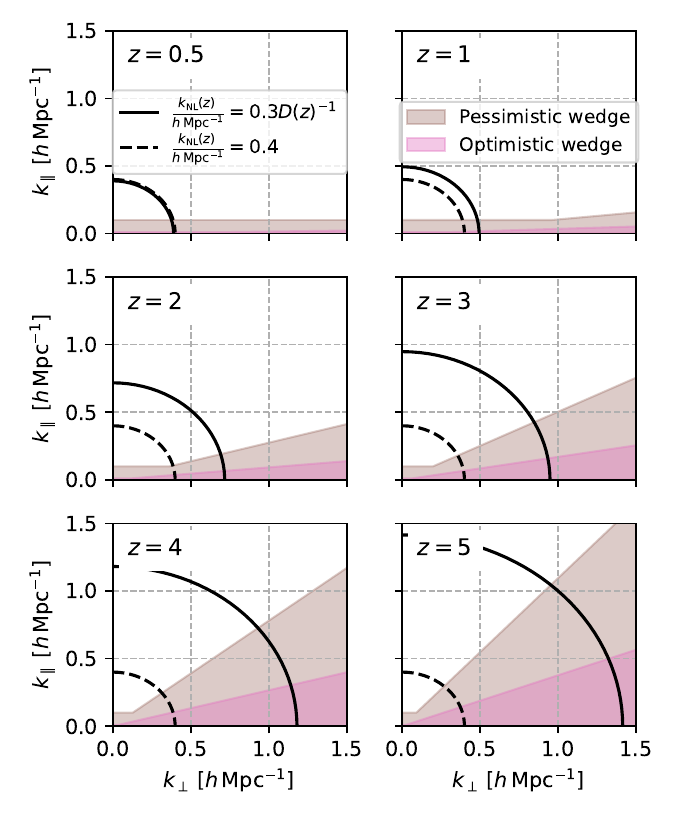}
\caption{%
Assumptions about the nonlinear scale $k_{\rm NL}(z)$ and the \tcm foreground wedge used in 
our forecasts. The {\em black lines} in each panel denote the two options for $k_{\rm NL}(z)$ we 
consider, while the {\em brown region} and {\em pink region} denote pessimistic and optimistic 
assumptions about the Fourier modes that are obscured by \tcm foregrounds. See 
Sec.~\ref{sec:forecasts:setup:knl} and~\ref{sec:forecasts:setup:surveys} for the details of these 
assumptions.
}
\label{fig:wedge}
\end{figure}

To account for uncertainty in the true nonlinear scale for HI, we present forecasts for two choices: $\knl(z)=0.4\hinvMpc$ (independent of redshift), and $\knl(z)=0.3D(z)^{-1}\hinvMpc$. 
We show these choices at a set of representative redshifts in Fig.~\ref{fig:wedge}. These 
assumptions are both pinned to $\knl(z=0.5)\approx 0.4\hinvMpc$, motivated by the promising 
performance of field-level modelling of HI in simulations (see e.g.\ Sec.~\ref{sec:simulations:fieldlevel} and Ref.~\cite{Modi:2019hnu}), and also the 
expectation that the Finger of God effect (which is important in determining $\knl$ for 
spectroscopic surveys~\cite{Ivanov:2019pdj}) will be smaller for HI than for currently measured galaxy 
samples such as luminous red galaxies~\cite{Villaescusa-Navarro:2018vsg,Obuljen:2022cjo}. 
In Appendix~\ref{app:alternative_forecasts:kNL}, we explore how other assumptions for $\knl(z)$ 
affect our results.

In Ref.~\cite{Sailer:2021yzm}, the authors also omitted scales where the leading Finger-of-God 
term $N_2 (\mu k)^2$ is greater than 20\% of the total power spectrum. Due to the expected low 
velocity dispersion of HI, we find that there are no wavenumbers that satisfy this condition within 
any of our $k_{\rm NL}(z)$ cuts, so Fingers of God are never a limiting factor in our forecasts.

\subsubsection{Surveys}
\label{sec:forecasts:setup:surveys}

We consider three surveys in our forecasts. The first two are different versions of the Packed 
Ultra-wideband Mapping Array (PUMA) concept, as described in 
Refs.~\cite{CosmicVisions21cm:2018rfq,PUMA:2019jwd,Castorina:2020zhz}. This concept 
consists of a hex-packed, half-filled dense array of 32000 (``PUMA-32k") or 5000 (``PUMA-5k") 
$6\,{\rm m}$ dishes, observing as a drift-scan interferometer over $200$-$1100\,{\rm MHz}$ 
($0.3<z<6$ in \tcm). This design is motivated by its potential for constraining the physics of dark 
energy and cosmic inflation, along with its capabilities for observing fast radio bursts, pulsars, 
and other radio transients~\cite{CosmicVisions21cm:2018rfq}. We compute the instrumental 
noise power spectrum for the two PUMA versions following the formalism of 
Ref.~\cite{CosmicVisions21cm:2018rfq}, as implemented in \texttt{FishLSS} based on the 
\texttt{PUMANoise} code\footnote{\url{https://github.com/slosar/PUMANoise}}.

As our third survey, we consider an approximation of the Canadian Hydrogen Observatory and 
Radio-transient Detector (CHORD), a successor to CHIME that is currently under 
development~\cite{Vanderlinde:2019tjt}. The central array of CHORD is planned to consist of 512 
$6\,{\rm m}$ dishes in a close-packed rectangular array, observing over $300$-$1500\,{\rm 
MHz}$ 
(covering $0<z<3.7$ in \tcm). In order to use the instrumental noise model from 
Ref.~\cite{CosmicVisions21cm:2018rfq}, we approximate this as a square array of $23^2=529$ 
dishes. 

For each survey, we assume 5 years of usable observing time and observations over half the sky 
(i.e.\ $f_{\rm sky}=0.5$).

\subsubsection{System temperature}
\label{sec:forecasts:setup:system_temperature}

The thermal noise power spectrum is proportional to the square of the system temperature $T_{\rm sys}(z)$. 
Ref.~\cite{CosmicVisions21cm:2018rfq} models the system temperature as
\beq
T_{\rm sys}(\nu) 
	= \frac{1}{\eta_{\rm c} \eta_{\rm s}} T_{\rm ampl} 
	+ \frac{1-\eta_{\rm s}}{\eta_{\rm s}} T_{\rm ground}
	+ T_{\rm sky}(\nu)\ ,
	\label{eq:Tsys}
\eeq
containing three components: an amplifier noise temperature $T_{\rm ampl}=50\,{\rm K}$ divided by an 
optical efficiency $\eta_{\rm c}=0.9$ and a sky coupling $\eta_{\rm s}=0.9$; a ground temperature $T_{\rm 
ground}=300\,{\rm K}$, of which a fraction $(1-\eta_{\rm s})/\eta_{\rm s}$ is picked up by the antenna; and a 
sky contribution $T_{\rm sys}(\nu)$, which is dominated by Galactic synchrotron and the CMB. Under these 
assumptions, the non-sky components sum to $95\,{\rm K}$.

On the other hand, measurements of system components for CHORD~\cite{CHORD-feed} (which will use 
$6\,{\rm m}$ dishes and wide-band feeds, as envisioned for PUMA), combined with estimates for 
ground-spill, 
have shown that the sum of non-sky components is below $30\,{\rm K}$ at $\nu\gtrsim 400\,{\rm MHz}$ 
($z\lesssim 2.5$).\footnote{CHIME has also measured total system temperatures in the range of 
$50$-$70\,{\rm 
K}$, depending on frequency and antenna polarization~\cite{CHIME:2022dwe}. These are well below 
the assumption of Ref.~\cite{CosmicVisions21cm:2018rfq} for PUMA, and provide further evidence that this 
assumption may be overly pessimistic.} 
For our baseline forecasts, we will
suppose that versions of CHORD feeds and amplifiers could be engineered to stay below this $30\,{\rm K}$ 
threshold over the envisioned PUMA band ($200\,{\rm MHz} < \nu < 1100\,{\rm MHz}$, or $6 \gtrsim z 
\gtrsim 0.3$), and therefore we will use $T_{\rm sys}(\nu) = T_{\rm sky}(\nu) + 30\,{\rm K}$ for both PUMA 
and CHORD. In Appendix~\ref{app:alternative_forecasts:Tsys}, we present alternative forecasts that follow 
the assumptions of Ref.~\cite{CosmicVisions21cm:2018rfq} for the system temperature, with $T_{\rm 
ampl}=50\,{\rm 
K}$ for PUMA and $30\,{\rm K}$ for CHORD.

\subsubsection{Accessible wavenumbers}
\label{sec:forecasts:setup:accessible_wavenumbers}

In an interferometric telescope, the autocorrelations of each receiver are usually excluded from 
mapmaking and other analyses of sky signals, since they are contaminated by a noise bias that 
is difficult to subtract~\cite{Liu:2019awk}. Thus, there is a minimum baseline (i.e.\ inter-antenna 
distance) that is used to measure large-scale structure, and this corresponds to a 
(redshift-dependent) 
minimum $k_\perp$ value that is accessible. In our forecasts, we assume that this 
minimum baseline is equal to the dish diameter ($6\,{\rm m}$).\footnote{In a real array, the 
minimum baseline will be larger than the dish diameter, in order to leave space between the 
edges of neighboring dishes. This will slightly increase the minimum accessible $k_\perp$ 
compared to what we assume here.} We will show how this translates into $k_{\perp,{\rm 
min}}(z)$ 
in Sec.~\ref{sec:forecasts:powerspectra}.

The effectiveness of \tcm surveys in obtaining cosmological constraints will strongly depend on 
the efficiency with which radio foregrounds (dominantly, radio point sources and diffuse emission 
from the Milky Way) can be cleaned from the observations. These foregrounds are expected to 
be intrinsically spectrally smooth, and therefore obscure Fourier modes of large-scale structure 
with low line-of-sight wavenumbers ($k_\parallel$). However, the intrinsic chromaticity of 
observations by a given baseline will generically spread foreground power to higher-$k_\parallel$ 
modes, with the severity of the spread increasing linearly with baseline length (and therefore 
$k_\perp$). This latter effect, the so-called foreground wedge 
(e.g.~\cite{Morales:2012kf,Parsons:2012qh,Liu:2014bba}), can be written (under certain 
assumptions about the primary beam shape~\cite{CosmicVisions21cm:2018rfq}) as foregrounds 
contaminating modes with
\beq
k_\parallel < \beta(z) k_\perp\ ,
\label{eq:wedge-general}
\eeq
where
\begin{align}
\label{eq:wedge-beta}
\beta(z) 
	&\equiv \frac{\chi(z) H(z)}{c(1+z)} \sin[\theta_{\rm w}(z)]\ , \\
\label{eq:wedge-thetaw}
\theta_{\rm w}(z) 
	&= N_{\rm w} \frac{1.22}{2\sqrt{0.7}} \frac{\lambda_{\rm obs}(z)}{D_{\rm phys}}\ ,
\end{align}
with $\lambda_{\rm obs}(z) \equiv 21(1+z)\,{\rm cm}$, $D_{\rm phys}$ as the dish diameter, and 
$N_{\rm w}$ parameterizing the severity of the wedge.\footnote{In detail, 
Eq.~\eqref{eq:wedge-thetaw} 
assumes that foreground contamination is adequately suppressed for sources at a 
distance from the beam center that is greater than $N_{\rm w}$ times the primary beam width. 
The primary beam is here idealized as an Airy disk. See Ref.~\cite{CosmicVisions21cm:2018rfq} 
for further discussion.}

The true shape of the wedge in future surveys (along with the minimum accessible $k_\parallel$ 
value) will depend on the level of systematics control that is achieved.
To account for this uncertainty, we show results for two different assumptions, following 
Ref.~\cite{Sailer:2021yzm}: an ``optimistic" assumption, where $k_{\parallel,{\rm 
min}}=0.01\hinvMpc$ 
and $N_{\rm w}=1$, and a ``pessimistic" case where $k_{\parallel,{\rm 
min}}=0.1\hinvMpc$ 
and $N_{\rm w}=3$. We visualize each case at several redshifts in 
Fig.~\ref{fig:wedge}.

\subsection{Power spectra}
\label{sec:forecasts:powerspectra}

\begin{figure*}[t]
\includegraphics[width=\textwidth, trim=0 20 0 0]{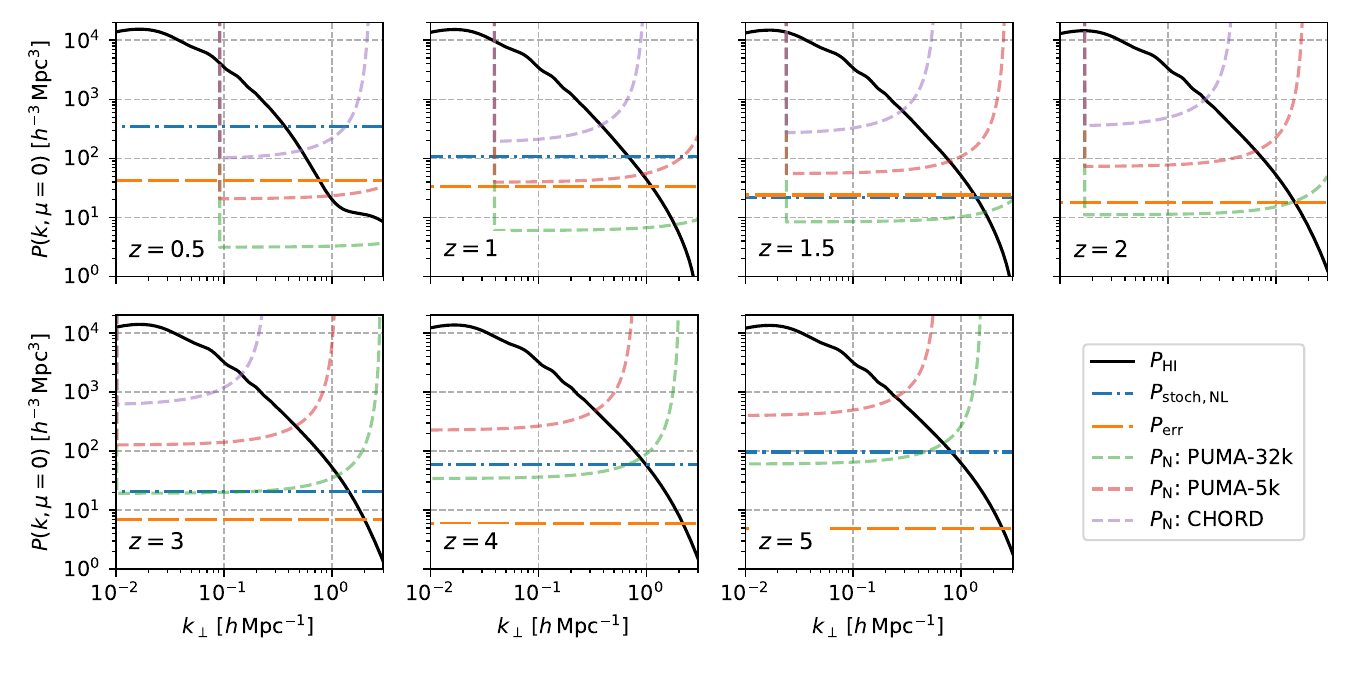}
\caption{%
Signal and noise contributions to the power spectrum of HI, evaluated at $\mu=0$.
{\em Black solid curves}: HI signal power spectrum, without subtracting the nonlinearity-induced 
stochasticity from the perturbative prediction (see Sec.~\ref{sec:forecasts:setup:stoch}).
The plateau at $z=0.5$ and $k\gtrsim 1\hinvMpc$ is a coincidence caused by the specific bias parameter values we use, but is unimportant because it is outside of the range of validity of the perturbative prediction.
{\em Blue dot-dashed curves}: Nonlinearity-induced stochasticity $P_{\rm stoch,NL}$, given by 
the low-$k$ limits of the quadratic terms in $P_{\rm HI}$. This contribution is present in a 
power spectrum
analysis, but absent from a more optimal analysis.
{\em Orange long-dashed curves}: Model error power spectrum $P_{\rm err}$, as measured in 
simulations in Sec.~\ref{sec:simulations}.
{\em Short-dashed curves}: Thermal noise power spectra $P_{\rm N}$ for PUMA-32k ({\em 
green}), PUMA-5k ({\em pink}), and CHORD ({\em purple}). 
The vertical features in these curves at $z\leq 2$ denote the minimum $k_\perp$ value accessible to baselines with nonzero length in each array (assuming antenna auto-correlations will not be used for cosmology).
For redshifts and surveys where 
$P_{\rm stoch,NL}\gg P_{\rm err}$ and $P_{\rm stoch,NL}\gtrsim P_{\rm N}$, we expect a 
more optimal
\tcm analysis to have better constraining power than a power spectrum analysis.
}
\label{fig:power_spectra}
\end{figure*}

In Fig.~\ref{fig:power_spectra}, we show the HI power spectrum from 
Eq.~\eqref{eq:PHI_prediction} at $\mu=0$ and a set of representative redshifts. Since the linear 
HI bias scales roughly like $D(z)^{-1}$ with redshift (see Fig.~\ref{fig:z_bias}), the overall power 
spectrum amplitude is roughly constant in redshift.

We also show the 
contribution to the stochasticity
 from the sum of the low-$k$ limits of the quadratic 
terms in the power spectrum, analogous to the term depending on $b_2^\Eul$ in Eq.~\eqref{eq:Pstochseq}. We denote this term by $P_{\rm stoch,NL}$, such that, in the notation of Sec.~\ref{sec:theory:stochasticity},
\beq
\Pstoch = P_{\rm err}(k\to 0) + P_{\rm stoch,NL}\ .
\eeq
In addition, we an approximate $k$-independent model error power spectrum, as described in  Sec.~\ref{sec:simulations:HIstochasticity}. Note that we have not subtracted 
$P_{\rm stoch,NL}$ from the black curves.

The redshift-dependence of $P_{\rm stoch,NL}$ depends on
 the quadratic bias $b_2^\Shi$ measured in Sec.~\ref{sec:simulations:biascoefficients} (recall Fig.~\ref{fig:stoch_comparison}), reaching a 
minimum at $z=2$ and increasing at higher and lower $z$.  The relative hierarchy between 
$P_{\rm stoch,NL}$ and $P_{\rm err}$ indicates the advantage of a more optimal analysis over a 
power spectrum analysis: when $P_{\rm stoch,NL} > P_{\rm err}$, a more optimal analysis will 
achieve a lower level of model error because the $P_{\rm stoch,NL}$ contribution will be absent 
from the covariance in that case.

However, the usefulness of more-optimal constraints will also be set by the thermal noise level of the 
experiment, shown (again at $\mu=0$) in the dashed curves for PUMA-32k, PUMA-5k, and 
CHORD. The low-$k_\perp$ cutoff in these spectra is set by the minimum baseline in each 
instrument; this corresponds to $k_\perp \approx 0.09\hinvMpc$ at $z=0.5$, $0.04\hinvMpc$ at 
$z=1$, and less than $0.01\hinvMpc$ at $z\gtrsim 2$. The upturn in these spectra at high 
$k_\perp$ is set by the size of the synthesized beam, or equivalently, the longest baseline in 
each instrument.

At $z=0.5$, $P_{\rm stoch,NL}$ is significant compared to $P_{\rm HI}$ in the quasi-linear 
regime, and is also above the thermal noise for both versions of PUMA and 
CHORD. Thus, we'd expect an analysis without $P_{\rm stoch,NL}$ to incur a significant increase 
in constraining power at this redshift. On the other hand, in cases where $P_{\rm stoch,NL}$ is 
well below the thermal noise, and/or small compared to $P_{\rm HI}$ on quasi-linear scales (e.g.\ 
$1.5<z<3$ in Fig.~\ref{fig:power_spectra}), we expect it to have much less of an impact on our 
forecasts.

At $z\gtrsim 4$, $P_{\rm stoch,NL}$ is more than an order of magnitude above $P_{\rm err}$. For 
PUMA-32k, the thermal noise is of the same order as $P_{\rm stoch,NL}$, so the associated 
cosmological constraints will be limited by thermal noise. However, if one had access to a survey 
with much lower thermal noise, $P_{\rm stoch,NL}$ would be the limiting factor, and therefore a 
more optimal analysis would become much more advantageous. We will discuss this point further in 
Sec.~\ref{sec:thermal_noise}.

\subsection{Distance measures}
\label{sec:forecasts:distancemeasures}

Our first set of forecasts is for the distance measures that can be obtained from the transverse 
and radial BAO scales. Following Ref.~\cite{Sailer:2021yzm}, we assume that BAO 
reconstruction\footnote{See Ref.~\cite{Obuljen:2016urm} for a proposal for implementing 
reconstruction with maps obtained from single-dish \tcm observations.} has been performed, and 
use the LPT-based formalism from Ref.~\cite{Chen:2019lpf} for predicting the post-reconstruction 
power spectrum. We fix the shape of the linear power spectrum and the values of $\tilde{b}_2^\Eul$ and 
$\tilde{b}_s^\Eul$, and marginalize over $\tilde{b}_1^\Eul$ along with 15 power-law terms in $k$ and $\mu$ that 
parameterize the broad-band power spectrum shape. The resulting constraints on the 
Alcock-Paczynski 
parameters $\alpha_\perp$ and $\alpha_\parallel$ can be interpreted as constraints 
on $D_{\rm A}/r_{\rm d}$ and $Hr_{\rm d}$ in each redshift bin, where $D_{\rm A}$ is the angular 
diameter distance and $r_{\rm d}$ is the sound horizon during the baryon drag epoch.

\begin{figure*}[t]
\includegraphics[width=\textwidth, trim=0 20 0 0]{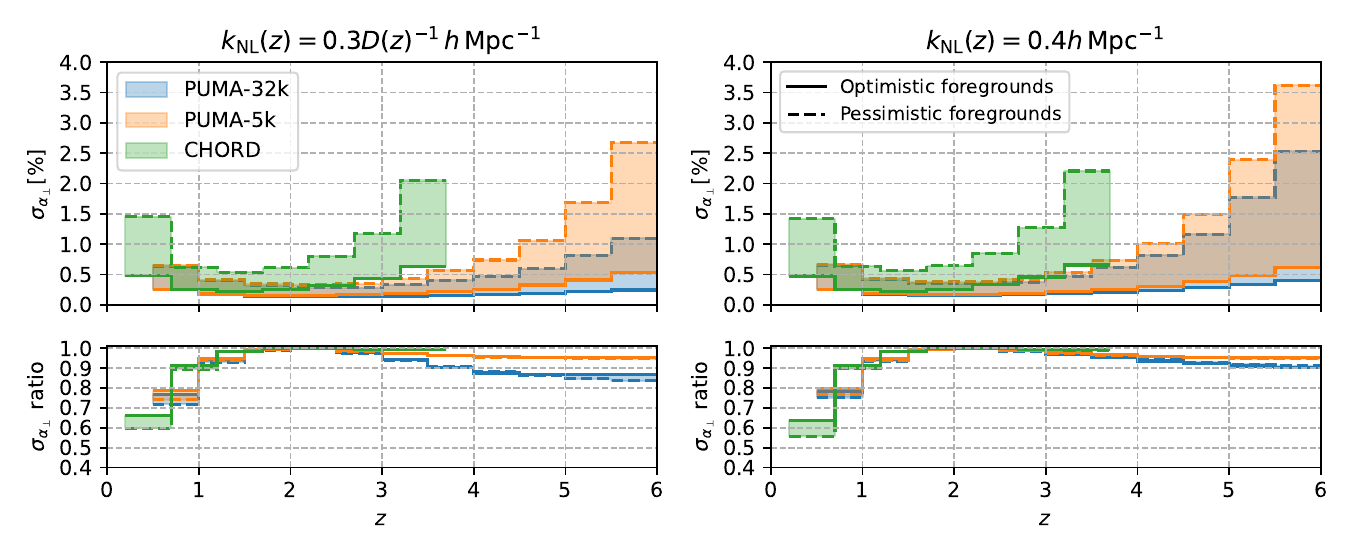}
\caption{%
Forecasted constraints on the Alcock-Paczynski parameter $\alpha_\perp$, which can be 
interpreted as constraints on $D_{\rm A}/r_{\rm d}$. We assume that BAO reconstruction has 
been performed, and marginalize over the linear bias and the broadband shape of the power spectrum. 
{\em Upper panels}: Constraints relative to fiducial parameter values. {\em Lower panels}: Relative improvement in 
constraining power from performing a more optimal analysis instead of a power spectrum 
analysis.
The linestyles of the upper and lower edges of each shaded region indicate whether each edge
corresponds to optimistic or pessimistic assumptions about foreground removal, according to the
legend in the upper right panel.
 A more optimal analysis matters the most at $z\lesssim 1$, where 
nonlinearity-induced stochasticity has the largest relative impact on the power spectrum.
}
\label{fig:bao_aperp}
\end{figure*}

\begin{figure*}[t]
\includegraphics[width=\textwidth, trim=0 20 0 0]{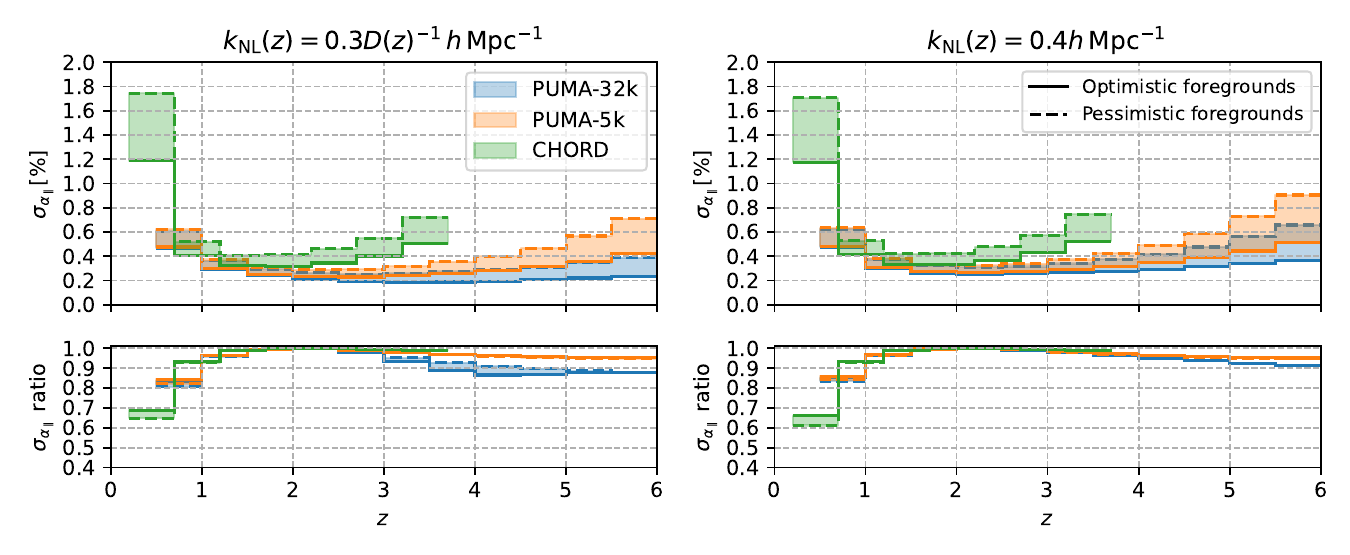}
\caption{%
As Fig.~\ref{fig:bao_aperp}, but for the line-of-sight Alcock-Paczynski parameter 
$\alpha_\parallel$, 
whose constraints can be interpreted as $Hr_{\rm d}$. Similarly to
$\alpha_\perp$, 
the constraints improve the most at lower redshift when a more optimal analysis is 
carried out.
}
\label{fig:bao_apar}
\end{figure*}

Figures~\ref{fig:bao_aperp} and~\ref{fig:bao_apar} show our forecasts for $1\sigma$ 
uncertainties for $\alpha_\perp$ and $\alpha_\parallel$. The upper panels only show relative 
uncertainties on each parameter in the case of a power spectrum analysis, while the lower 
panels show the relative {\em improvement in the uncertainties} when a more optimal analysis is 
done instead. 

As expected from the arguments in Sec.~\ref{sec:forecasts:powerspectra}, the lowest-redshift 
bins of each survey exhibit the largest difference: for $\alpha_\perp$, a 40-45\% 
improvement for the $0.2<z<0.7$ bin of CHORD, and a 20-30\% improvement for the $0.5<z<1$ bin of 
PUMA-5k and PUMA-32k.
At higher redshifts, the CHORD results are insensitive to the assumed form of $k_{\rm NL}(z)$, since the 
maximum $k$ is limited by the longest baseline. The improvement at higher redshifts is less than 10\% for 
PUMA-5k and less than 20\% for PUMA-32k.

\subsection{Structure growth}
\label{sec:forecasts:growth}

\begin{figure*}[t]
\includegraphics[width=\textwidth, trim=0 20 0 0]{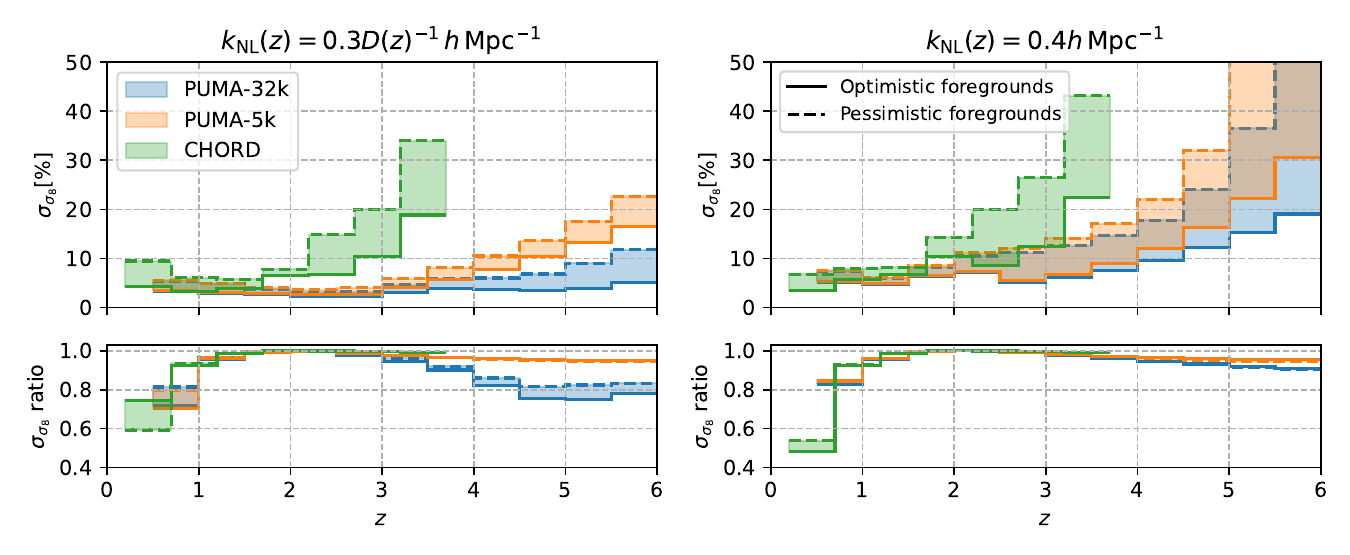}
\caption{%
As Figs.~\ref{fig:bao_aperp}-\ref{fig:bao_apar}, but for the fluctuation amplitude $\sigma_8(z)$, 
assuming the shape of the linear power spectrum is fixed (which reduces the 
$\Lambda$CDM-dependence 
of the constraints). Note that some shaded bands extend beyond the upper limit of 
the plot. The constraints for $k_{\rm NL}(z)=0.3D(z)^{-1}\hinvMpc$ are tighter than those for 
$k_{\rm NL}(z)=0.4\hinvMpc$ because higher-$k$ modes are important for breaking the linear 
degeneracy between $T_{\rm b}(z)$ and $\sigma_8(z)$. As for the BAO constraints, a more optimal 
analysis has the largest impact at $z\lesssim 1$.
}
\label{fig:sigma8_fixedshape_freeTb}
\end{figure*}

Next, we perform forecasts for measurements of $\sigma_8(z)$, the amplitude of linear density 
fluctuations within $8h^{-1}\,{\rm Mpc}$ spheres, which is valuable for constraining neutrino 
mass and deviations from general relativity (among other things). In these forecasts, we fix the 
shape of the linear matter power spectrum by fixing the $\Lambda$CDM parameters 
($\omega_{\rm 
b}$, $\omega_{\rm c}$, $h$, $n_{\rm s}$, $\tau$), while varying $\{ T_{\rm b}, b_1, 
b_2, b_s, N_0, N_2, N_4, \alpha_0, \alpha_2, \alpha_4 \}$ separately in each redshift bin; this 
removes much of the dependence of the constraints on the $\Lambda$CDM model, and is 
suitable for extracting possible deviations from $\Lambda$CDM (e.g.\ due to differences in linear 
growth)~\cite{Sailer:2021yzm,Yu:2018tem}. We have also considered forecasts where all 
cosmological parameters are varied, and the trends are qualitatively similar to the fixed-shape 
case.

We show the results in Fig.~\ref{fig:sigma8_fixedshape_freeTb}, with the same conventions as 
Figs.~\ref{fig:bao_aperp}-\ref{fig:bao_apar}. Qualitatively, the conclusions are the same as those 
for the distance measures in Sec.~\ref{sec:forecasts:distancemeasures}: the improvement from a 
more optimal analysis is most noticeable in the lowest redshift bin of each survey, reaches a 
minimum at $z\sim 2$, and is relatively mild at higher~$z$. Quantitatively, a more optimal analysis 
makes more of a difference for $\sigma_8(z)$ than for the distance measures. For example, for 
$\sigma_8(z)$, 
we observe roughly a 25-50\% improvement in the forecasted uncertainties in the 
$0.2<z<0.7$ bin for CHORD, while the improvement for PUMA in the $0.5<z<1$ bin ranges from 
15-30\%, depending on the survey configuration and assumption about $k_{\rm NL}(z)$. 
However, it's worth noting that the forecasted low-$z$ constraints from CHORD 
($\sigma_{\sigma_8} 
\gtrsim 3\%$) are unlikely to be competitive with combined LSS-CMB 
analyses that will be possible on similar timescales (e.g.~\cite{Sailer:2021yzm,Yu:2018tem})

Due to the perfect degeneracy between $\sigma_8(z)$ and $T_{\rm b}(z)$ in the linear regime, 
our ability to constrain $\sigma_8(z)$ strongly depends on our sensitivity to nonlinear scales 
where this degeneracy is broken~\cite{Castorina:2019zho}. The relative importance of the 
stochastic noise increases at higher $k$, where nonlinearities are stronger, and therefore the 
presence or absence of this noise has more of an impact on our ability to constrain 
$\sigma_8(z)$. 
This also explains why the constraints are tighter if $k_{\rm NL}(z) = 0.3D(z)^{-1}\hinvMpc$ 
rather than $k_{\rm NL}(z)=0.4\hinvMpc$, since the former assumption results in a higher $k_{\rm NL}$ at 
high $z$.

In Fig.~\ref{fig:sigma8_fixedshape_freeTb}, we have made the conservative choice of not 
assuming any prior on $T_{\rm b}(z)$. In reality, such a prior may be available from 
measurements of damped Lyman-alpha absorbers~\cite{Crighton:2015pza}, stacking of 
high-resolution 
 observations of HI-rich galaxies~\cite{Chowdhury:2020uqa}, or direct \tcm galaxy 
surveys~\cite{Jones:2018mnras}, depending on the redshift range (see 
Ref.~\cite{Chen:2021mnras} for a recent summary of estimates of the mean HI density 
$\Omega_{\rm 
HI}(z)$, which is directly related to $T_{\rm b}(z)$ by Eq.~\eqref{eq:Tbz}).\footnote{See 
Ref.~\cite{Obuljen:2017jiy} for other proposals to break the $T_{\rm b}-b_{\rm HI}$ degeneracy.} The 
achievable precision of these measurements is an open question, but we can get a sense of how 
a $T_{\rm b}(z)$ prior would influence our results by repeating our $\sigma_8(z)$ forecasts with 
$T_{\rm b}(z)$ held fixed. For the optimistic-foreground case, constraints on $\sigma_8(z)$ 
improve by around a factor of~5 in many redshift bins compared to the case where $T_{\rm b}(z)$ 
is left free (see Appendix~\ref{app:alternative_forecasts:fixed_Tb}), while the relative 
improvement obtained from a more optimal vs.\ power spectrum analysis does not change by 
more than 10\%.

\subsection{Cosmological parameters}
\label{sec:forecasts:parameters}

\begin{figure}[h!]
\includegraphics[width=\columnwidth, trim=0 10 0 0]{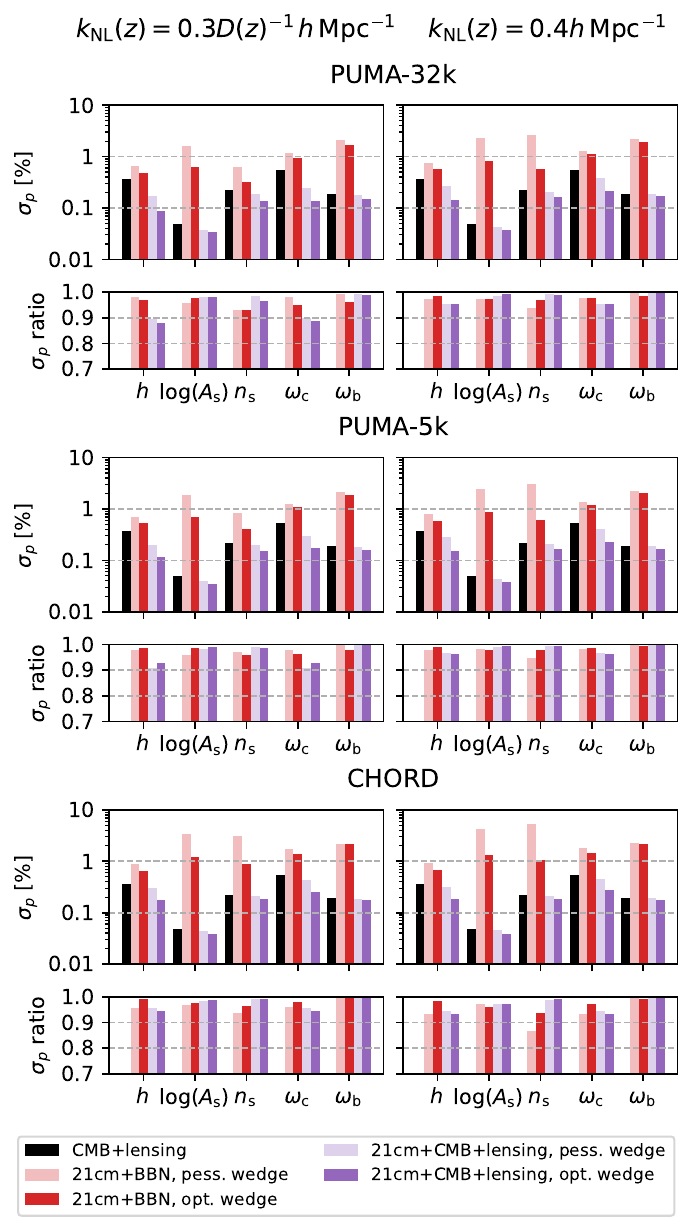}
\caption{%
Forecasts for standard $\Lambda$CDM parameters, where the upper panel for each survey 
shows the constraints from a power spectrum analysis relative to the fiducial parameter values, and the 
lower panel shows the improvement from a more optimal \tcm analysis. The {\em black bars} show 
reference constraints using low-$\ell$ CMB from {\em Planck} and higher-$\ell$ CMB from 
Simons Observatory, combined with CMB lensing from Simons Observatory. The {\em colored 
bars} are as indicated in the legend, where ``BBN" denotes a big bang nucleosynthesis prior on 
$\omega_{\rm 
b}$ and ``CMB+lensing" denotes the same information used for the black bars. In 
most cases, a more optimal \tcm analysis improves the constraining power by less than 10\%,
while for PUMA-32k with optimistic foregrounds and $\knl(z)$, the improvement 
is slightly better.}
\label{fig:lcdm}
\end{figure}

\begin{figure}[h!]
\includegraphics[width=\columnwidth, trim=0 10 0 0]{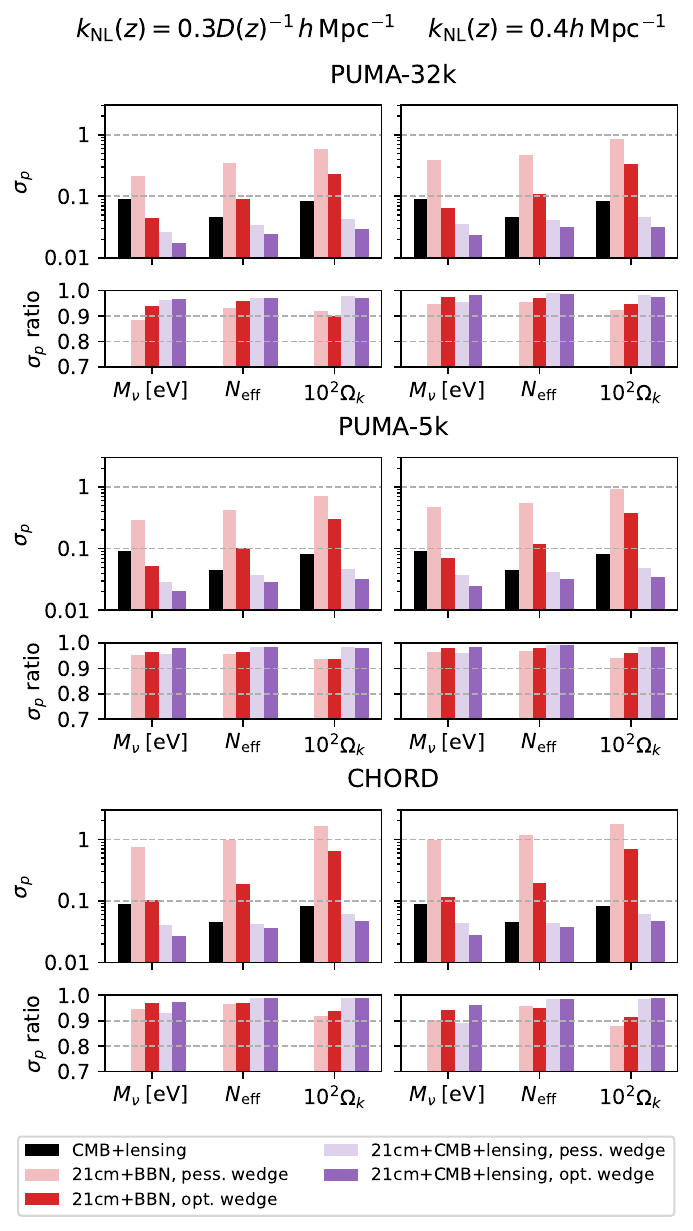}
\caption{%
As Fig.~\ref{fig:lcdm}, but for three one-parameter extensions of $\Lambda$CDM, which vary the 
sum of neutrino masses $M_\nu$, the effective number of relativistic species $N_{\rm eff}$, or 
the curvature parameter $\Omega_k$.
For these parameters, we plot the constraints on the parameter itself in the upper panels, as opposed
to the relative constraints in the upper panels of Fig.~\ref{fig:lcdm}.
 In some cases where the constraining power is competitive with 
the reference CMB+lensing constraints, a more optimal analysis enhances the constraining power by 
up to $\sim$10\%.
}
\label{fig:lcdm_extensions}
\end{figure}

Finally, we perform forecasts for the standard $\Lambda$CDM cosmological parameters, along 
with one-parameter extensions where the sum of neutrino masses $M_\nu$, the effective number 
of relativistic species $N_{\rm eff}$, or the curvature parameter $\Omega_k$ are allowed to vary. 
The redshift-dependent parameters listed in Sec.~\ref{sec:forecasts:growth} are also allowed to 
vary, and our results are marginalized over these parameters.

These results are shown in Figures~\ref{fig:lcdm} and~\ref{fig:lcdm_extensions}. As in 
Figures~\ref{fig:bao_aperp},~\ref{fig:bao_apar}, and~\ref{fig:sigma8_fixedshape_freeTb}, the 
upper panels for each survey show 1$\sigma$ parameter uncertainties for a 
power spectrum
analysis, while the lower panels show the relative improvement incurred by a more optimal
analysis. The black bars are for a reference forecast that combines CMB temperature, 
polarization, and lensing; specifically, we assume that $\ell<30$ multipoles from Planck have 
been combined with $30<\ell<3000$ in temperature and $30<\ell<5000$ in polarization from the 
Simons Observatory~\cite{SimonsObservatory:2018koc}, and also use $30<\ell<500$ from the 
lensing power spectrum from the Simons Observatory, assuming $f_{\rm sky}=0.4$. The pink and 
red bars include the indicated \tcm survey along with a big bang nucleosynthesis prior of 
$\sigma(\omega_{\rm b})=0.0005$ (2\% of the fiducial value)~\cite{Planck:2018vyg}. 
The light and 
dark purple bars combine the \tcm survey with the CMB and lensing information\footnote{We do 
not consider cosmic shear in this work, and therefore we always use ``lensing" to refer to CMB 
lensing only.} from the black bars.\footnote{The difference between power spectrum and 
more optimal parameter constraints is unchanged if lensing information is omitted.} 

For each of the standard $\Lambda$CDM parameters in Fig.~\ref{fig:lcdm}, the CMB+lensing 
constraints are stronger than the {\tcm}+BBN constraints. When combining \tcm, CMB, and 
lensing, constraints on $h$ improve by as much as a factor of 4 (for optimistic PUMA-32k) over 
CMB+lensing, thanks to the addition of low-redshift BAO information. These constraints improve 
further, by $\mathcal{O}(15\%)$ for PUMA-32k, when the \tcm analysis is done more optimally. The 
addition of \tcm to CMB+lensing also has a notable impact on $\omega_{\rm c}$, arising both 
from geometric information (the observed BAO scale) and the shapes of the BAO wiggles and 
broad-band power spectrum~\cite{DAmico:2019fhj,Ivanov:2019pdj}: the constraints tighten by a 
factor of 2-3 for PUMA-32k, and a more optimal \tcm analysis brings a further $\mathcal{O}(10\%)$ 
improvement. For the other parameters ($A_{\rm s}$, $n_{\rm s}$, and $\omega_{\rm b}$), the 
addition of \tcm to CMB+lensing is less beneficial, and a more optimal \tcm analysis improves the 
results by less than 10\%.

In Figure~\ref{fig:lcdm_extensions}, we find that {\tcm}+BBN constraints on $M_\nu$ can be 
tighter than those from CMB+lensing when considering PUMA (either 32k or 5k) with optimistic 
assumptions about foregrounds, due to the longer lever-arm in $k$ that allows for a better 
measurement of neutrino-induced power spectrum suppression. Either with or without adding 
CMB+lensing to this, a more optimal \tcm analysis brings a further 5-10\% improvement. On the 
other hand, combining \tcm measurements with CMB+lensing is needed to improve upon 
CMB+lensing constraints on $N_{\rm eff}$ or $\Omega_k$. The addition of \tcm information 
tightens each constraint by a up to a factor of 3, but in those cases, a more optimal \tcm analysis 
only alters the constraining power by a few percent.

By re-running the forecasts and restricting PUMA's redshift range to either $z<3$ or $z>3$, we 
have found that both ranges contribute non-negligibly to the parameter constraints, but the 
lower-redshift 
range provides more information than the high-redshift range. Evidently, for these 
constraints, the lower instrumental noise and larger range of signal-dominated scales at lower 
redshift is a larger advantage than the higher $k_{\rm NL}$ at higher redshift. However, for both 
redshift ranges, the impact of a more optimal \tcm analysis remains minor.

\section{The Role of Thermal Noise}
\label{sec:thermal_noise}

\subsection{Effect on forecasts}

\begin{figure*}[t]
\includegraphics[width=\textwidth, trim=0 20 0 0]{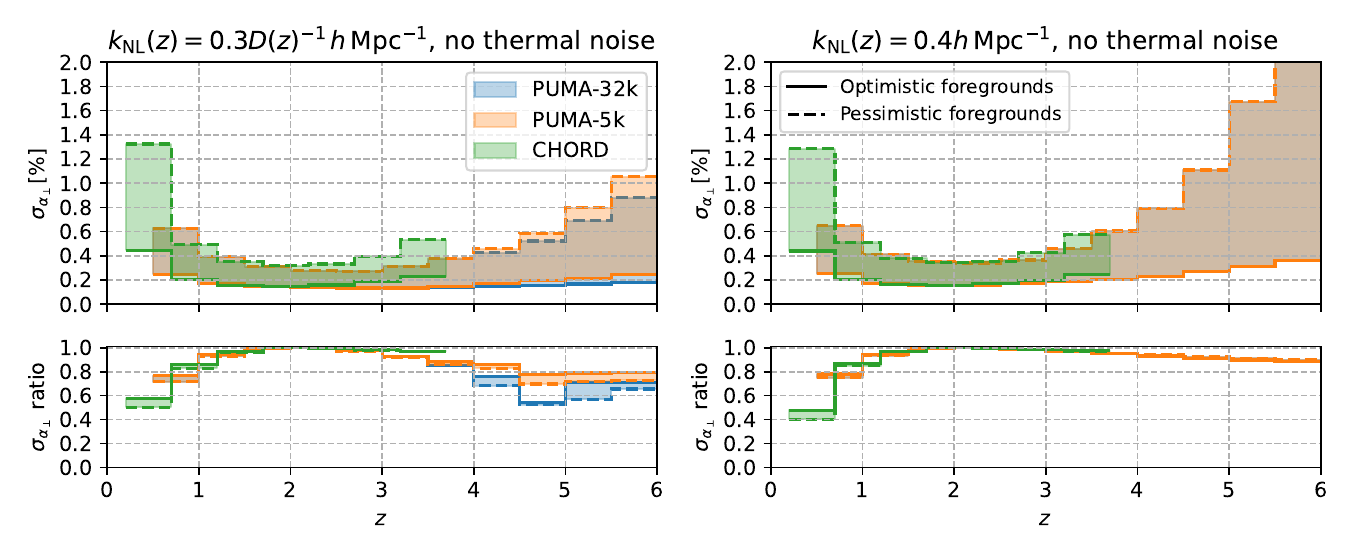}
\caption{%
Forecasts for the Alcock-Paczynski parameter $\alpha_\perp$, analogous to Fig.~\ref{fig:bao_aperp} but in 
the limit of no thermal noise.
Constraints from the power spectrum ({\em upper panels}) improve substantially compared to the baseline 
case. A more optimal analysis ({\em lower panels}) further improves upon power spectrum constraints by 
$40$-$60\%$ 
in the 
lowest-$z$ 
bin of CHORD and $\sim$$20$-$30\%$ in the lowest-$z$ bin of PUMA. As $z\gtrsim 3.5$ and for the 
more optimistic assumption about $\knl(z)$, constraints improve by $20$-$30\%$ for PUMA-5k and as 
much 
as 50\% for PUMA-32k with optimistic foregrounds.
}
\label{fig:bao_aperp_nothermalnoise}
\end{figure*}

In Sec.~\ref{sec:forecasts}, we saw that thermal noise plays a major role in determining whether a more optimal
analysis is significantly better than a power spectrum analysis: if the thermal noise is greater than the 
nonlinearity-induced stochasticity in the power spectrum, both types of analysis will be limited by thermal 
noise, and we will not see huge gains from a more optimal analysis. This is approximately true for both PUMA 
and CHORD at $z\gtrsim 1.5$ (recall Fig.~\ref{fig:power_spectra}). However, it's worth asking how the 
situation changes if the thermal noise level could be pushed much lower than the default expectations.

\begin{figure*}[t]
\includegraphics[width=\textwidth, trim=0 20 0 0]{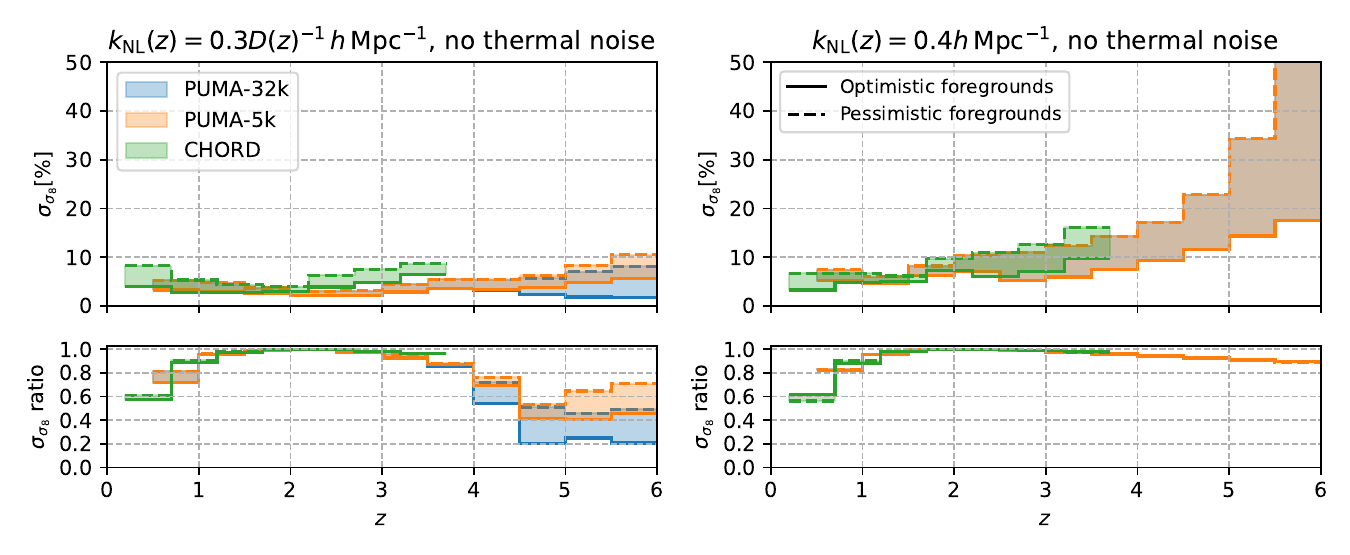}
\caption{
Forecasts for $\sigma_8(z)$, analogous to Fig.~\ref{fig:sigma8_fixedshape_freeTb} but in the limit of no 
thermal noise. As with $\alpha_\perp(z)$, we find that the absolute constraints substantially improve over the 
baseline forecasts, while further large gains are possible with a more optimal analysis at low and high redshift in 
certain cases. Notably, in the most ambitious case (PUMA-32k with optimistic foregrounds), constraints on 
$\sigma_8(z)$ 
from a more optimal analysis can improve upon those from a power spectrum analysis 
by 80\% (i.e.\ a factor of~5) at high redshift.
}
\label{fig:sigma8_fixedshape_freeTb_nothermalnoise}
\end{figure*}

\begin{figure}[t]
\includegraphics[width=\columnwidth, trim=0 10 0 0]{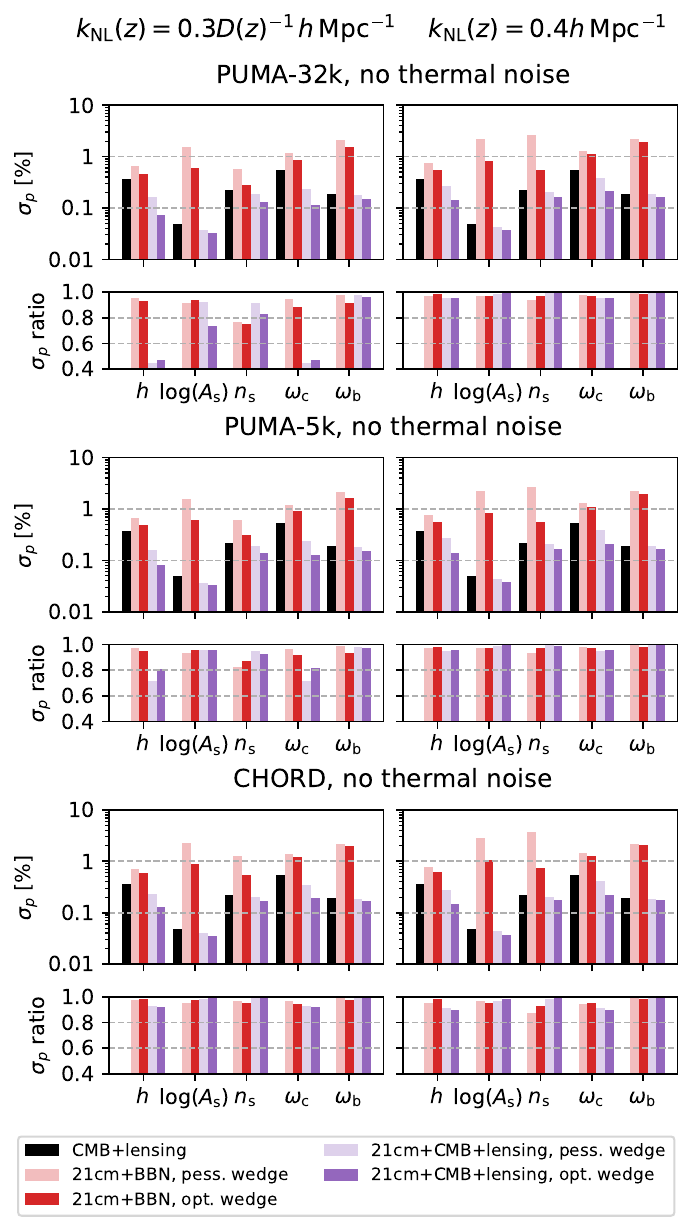}
\caption{%
Forecasts for standard $\Lambda$CDM parameters, analogous to Fig.~\ref{fig:lcdm} but in the limit of no 
thermal noise. Compared to our baseline forecasts, constraints on several parameters improve by a factor 
of 2. In the no-noise limit, a more optimal analysis incurs an improvement of up to 20\% in several cases, and in 
the case of PUMA-32k with $\knl(z)=0.3D(z)^{-1}\hinvMpc$, this improvement can reach 40-50\%.
}
\label{fig:lcdm_nothermalnoise}
\end{figure}

To answer this question, we take the limit of zero thermal noise for each experiment, while maintaining the 
same high-$k_\perp$ cutoff arising from the longest baseline in each array, and also preserving the 
assumptions about foregrounds. We show a representative set of results, for the transverse 
Alcock-Paczynski parameter $\alpha_\perp$ (Fig.~\ref{fig:bao_aperp_nothermalnoise}; compare with Fig.~\ref{fig:bao_aperp}), 
$\sigma_8(z)$ (Fig.~\ref{fig:sigma8_fixedshape_freeTb_nothermalnoise}; compare with 
Fig.~\ref{fig:sigma8_fixedshape_freeTb}), and the base $\Lambda$CDM parameters 
(Fig.~\ref{fig:lcdm_nothermalnoise}; compare with Fig.~\ref{fig:lcdm}). Note that the $y$-axis ranges are 
generally different for the default-noise and no-noise figures. Results for $\alpha_\parallel$ are similar to those 
for 
$\alpha_\perp$.

In Fig.~\ref{fig:bao_aperp_nothermalnoise}, we see that the power-spectrum-derived constraints on 
$\alpha_\perp$ 
(upper panels) for PUMA-32k change by a minor amount, but constraints from PUMA-5k and CHORD
become quite similar to those for PUMA-32k
in the redshift range where 
they overlap. Therefore, if HI stochasticity is the limiting factor instead of thermal noise, the additional scales 
resolvable by PUMA (due to its longer baselines) provide no advantage over CHORD in a 
power spectrum
BAO analysis (except at the lowest redshifts probed by CHORD).

Comparing the lower panels of Figs.~\ref{fig:bao_aperp_nothermalnoise} and~\ref{fig:bao_aperp}, we find 
that if there is no thermal noise, $\alpha_\perp$ in the lowest-$z$ CHORD bin can be measured 
$\sim$$50\%$ 
more precisely using a more optimal analysis compared to a power spectrum analysis, while the improvement 
for PUMA-5k and PUMA-32k is closer to $20$-$30\%$ at low $z$, depending on the assumed $\knl$.
At higher redshifts, the results depend more strongly on $\knl(z)$. If $\knl(z)=0.4\hinvMpc$, more optimal
constraints are around 10\% better than power spectrum constraints. However, if 
$\knl(z)=0.3D(z)^{-1}\hinvMpc$, 
the difference is around 20-30\% for PUMA-5k, and can be as much as 50\% for 
PUMA-32k with pessimistic foregrounds. 

For $\sigma_8$ (Fig.~\ref{fig:sigma8_fixedshape_freeTb_nothermalnoise}), in the limit of zero thermal 
noise, the absolute constraints on $\sigma_8(z)$ 
tighten by a factor of a few in several redshift bins compared to the default-noise forecasts, 
particularly for the more optimistic assumption about $\knl(z)$. 
At high $z$, for PUMA with optimistic $\knl(z)$, significant gains are possible using a more optimal analysis: 
40-60\% for PUMA-5k, and up to 80\% (i.e.\ a factor of 5 improvement in constraining power) for PUMA-32k.

Finally, examining the $\Lambda$CDM parameter constraints in Fig.~\ref{fig:lcdm_nothermalnoise}, we find 
that absolute constraints on several parameters improve by a factor of $\sim$2 for CHORD and PUMA-5k 
compared to our baseline forecasts. In the {\tcm}+CMB+lensing case and with 
$\knl(z)=0.3D(z)^{-1}\hinvMpc$, 
a more optimal analysis further improves constraints on $h$ and $\omega_{\rm c}$ 
by 20-30\% for CHORD and PUMA-5k, and $\sim$$50\%$ for PUMA-32k.

These results reveal a few important points:
\begin{enumerate}
\item With all other assumptions and survey properties held fixed, statistical uncertainties on a variety of 
cosmological quantities can be substantially improved if the thermal noise can be lowered.
\item If a lower level of thermal noise is achievable, even further improvements in statistical precision can be 
obtained by performing a more optimal analysis.
\end{enumerate}
The next subsection discusses possible pathways to lowering the thermal noise and enabling these 
improvements.

\subsection{Instrument/survey design}

The thermal noise power spectrum $P_{\rm N}(\vk, z)$ depends on several properties of the instrument and 
survey (see Appendix~\ref{app:thermal_noise} for a review). Below, we discuss the most important of these 
properties.

\subsubsection{Observing time} 

The thermal noise power spectrum depends inversely on the total usable observing time, where ``usable" 
excludes instrument downtime and periods of heavily contaminated data (e.g.\ strong RFI events and, in 
analyses of current surveys, daytime observations). Therefore, increasing the duration of the survey is one way to 
decrease the thermal noise, assuming that long-term drifts in observing conditions or instrument properties can be 
corrected for. However, the lifespan of a given survey is typically constrained by practicalities such as 
operating costs and availability of personnel, so there is no guarantee that drastic increases over our 
baseline assumption of $t_{\rm survey}=5\,{\rm yr}$ would be possible.

\subsubsection{System temperature} 

The thermal noise power spectrum is proportional to $T_{\rm sys}(z)^2$. The contribution of a given patch of 
sky to $T_{\rm sys}(z)$ is irreducible; one could consider an observing strategy that only concentrates on 
dimmer patches, which have brightness temperatures $\sim$$10\,{\rm K}$ lower than brighter patches at 
the 
relevant frequencies~\cite{CHIME:2022dwe}, but this would necessarily decrease the sky fraction surveyed 
and increase the thermal noise, likely negating any improvement from restricting to
dimmer patches.

However, the non-sky contributions to $T_{\rm sys}(z)$ can in principle be reduced by careful engineering of 
the electronics and dish structure. At higher redshifts, these contributions become subdominant to $T_{\rm 
sky}(z)$ (see Fig.~\ref{fig:Tsys} in Appendix~\ref{app:alternative_forecasts:Tsys}), but at lower redshifts a 
reduction can have a larger impact. We have carried out forecasts with $T_{\rm sys}(z)=T_{\rm sky}(z)$, and 
they generally exhibit only minor differences from our main forecasts. (An exception is BAO forecasts at the 
higher redshifts accessible by CHORD, where constraints are improved by $\mathcal{O}(30\%)$: at these 
redshifts, the reduction of $T_{\rm sys}$ increases the S/N on crucial BAO-scale modes close to the 
longest-baseline limit of CHORD.)

\subsubsection{Number of dishes}

For a close-packed array of $N_{\rm s}^2$ dishes, the thermal noise power spectrum scales like $N_{\rm 
s}^{-2}$, 
such that increasing the number of dishes can substantially decrease the thermal noise for a fixed 
survey duration. This increase is still realized if the total collecting area is held fixed by decreasing  $D_{\rm 
dish}$ by the same factor that $N_{\rm s}$ is increased (assuming a close-packed configuration). 
Assuming the dishes are correlated via fast Fourier transform correlation~\cite{Tegmark:2008au} or direct 
imaging~\cite{Morales:2008wt}, the computational cost of the correlations scales like $N_{\rm s}^2 \log 
N_{\rm s}$, while the total cost of per-dish instrumentation will scale no more favorably than $N_{\rm s}^2$. However, 
more detailed forecasts and costing would be required to determine the optimal array size for a given set of 
goals and practical constraints.

\section{Relation to previous \tcm forecasts}
\label{sec:discussion}

Numerous forecasts for \tcm intensity mapping surveys exist in the literature, and here we briefly comment on their assumptions about model error in the \tcm power spectrum. 
In many cases, these forecasts have not properly accounted for bias-induced stochasticity in the power spectrum covariance. Including this contribution would change their results at some level, but this change is likely far smaller than the uncertainty in their results due to the unknown level of foreground cleaning that is practically achievable. Nevertheless, as foreground cleaning strategies improve, it will be important to account for bias-induced stochasticity in future forecasts, particularly when considering low redshifts or surveys with low levels of thermal noise.

Many existing forecasts use a linear bias model for the HI power spectrum, and therefore do not include bias-induced stochasticity in the assumed covariance. Some forecasts in this class include HI sampling noise (e.g.~\cite{Chang:2007xk,Wyithe:2008mv,Visbal:2008rg,Masui:2009cj,Masui:2010mp,Seo:2009fq,Pober:2012zz,Bull:2014rha,Obuljen:2017jiy,Chen:2018qiu,CosmicVisions21cm:2018rfq,Karagiannis:2019jjx,Karagiannis:2020dpq,Costa:2021jsk,Karagiannis:2022ylq,Xiao:2021nmk,Jolicoeur:2023tcu,Byrne:2023rza}), while others omit it on the principle that is subdominant to thermal noise (e.g.~\cite{CHIME:2022dwe,SKA:2018ckk,Foreman:2018gnv}), or do not mention it explicitly in their descriptions (e.g.~\cite{Crichton:2021hlc,Wyithe:2007rq,Ali:2013rfa,Bharadwaj:2015vwa,Pourtsidou:2015qaa,Pourtsidou:2015mia,Pourtsidou:2016dzn,Yohana:2019ahg}). In many of these cases, the HI shot noise is indeed likely to be far lower than the thermal noise, but the total HI stochasticity may be of the same order as the thermal noise; this is what we found for 5 years of CHORD operations in Figure~\ref{fig:power_spectra}.

A smaller number of existing forecasts have incorporated nonlinear biasing. Ref.~\cite{Sailer:2021yzm}, whose forecasting code we have adapted in this work, computed the HI power spectrum using one-loop EFT using \texttt{velocileptors}~\cite{Chen:2020fxs}, which includes bias-induced stochasticity in its outputs by default. Therefore, the forecasts in Ref.~\cite{Sailer:2021yzm} incorporated this contribution in both the power spectrum covariance and the power spectrum itself. This is the correct approach for the covariance, but one technically must impose non-standard renormalization conditions on the bias parameters in order for this approach to the power spectrum to be correct (e.g.~\cite{Cabass:2023nyo,Rubira:2023vzw}). Ref.~\cite{Randrianjanahary:2023rgp} also used the formalism from Ref.~\cite{Sailer:2021yzm} for their own forecasts.

Ref.~\cite{Pourtsidou:2022gsb} used \texttt{PyBird}~\cite{DAmico:2020kxu} to produce synthetic measurements of the monopole and quadrupole of the HI power spectrum, and carried out forecasts by fitting \texttt{PyBird} predictions to these simulations. By default, \texttt{PyBird} subtracts bias-induced stochasticity from the power spectrum, and therefore the stochasticity in the synthetic measurements does not include this contribution.

\section{Conclusion}
\label{sec:conclusion}

In this paper, we have investigated the issue of stochastic noise in the clustering of neutral hydrogen, and its impact on cosmological analyses of \tcm intensity mapping surveys. We first reviewed the distinction between sampling noise and stochasticity of biased tracers in the context of large-scale structure perturbation theory.
We have argued that, in certain cases, the stochasticity induced by quadratic bias can far exceed the sampling noise that is commonly assumed to dominate the total stochasticity.

Using perturbative field-level modelling applied to two suites of hydrodynamical simulations, we have extended the results of Ref.~\cite{Obuljen:2022cjo} to higher redshift. Our simulation measurements demonstrate that stochasticity in the distribution of HI always exceeds the model error power spectrum.
We have further demonstrated that the redshift-dependence of this behavior can be explained (within reasonable uncertainties) by the contribution of the quadratic bias term (with coefficient $b_2$) to the HI power spectrum at low $k$.
We have presented fitting functions for linear, quadratic, and tidal bias of HI (Eqs.~\ref{eq:fitting_functions}-\ref{eq:fitting_functions_Eul}), along with the total HI stochasticity (Eq.~\ref{eq:stoch_fitting_function}), as a function of redshift. 
The bias fitting functions can be used in future forecasts and simulations of HI clustering, and we advocate that the HI stochasticity also be included in future HI power spectrum forecasts.

We have also performed forecasts for \tcm intensity mapping with the CHORD telescope and the PUMA telescope concept. The goal of these forecasts is to estimate the improvement in constraining power that would follow from a (beyond-power-spectrum) cosmological analysis in which the bias-induced stochasticity does not contribute to the noise.\footnote{Our conclusions do not depend on a specific choice of beyond-power-spectrum analysis method. This implies that there are several potential options for harnessing the improved constraining power in practice, including simple beyond-2-point statistics~\cite{Cabass:2023nyo}, field-level likelihood analyses~\cite{Nguyen:2024yth}, and simulation-based inference~\cite{SimBIG:2023ywd}.}
 (A byproduct of this exercise is a set of forecasts for power spectrum analyses of CHORD and PUMA, but we remind the reader that the results of those forecasts are highly sensitive to the assumed foreground cleaning efficiency, and should therefore not be interpreted as definitive claims about the absolute constraining power of these telescopes.) In these forecasts, we considered both optimistic and pessimistic assumptions about foreground cleaning and the nonlinear scale $\knl(z)$ up to which perturbative modelling is valid.

We found that elimination of bias-induced stochastic noise can improve constraints on the BAO scale or $\sigma_8(z)$ by several tens of percents, up to a factor of $\sim$2 for the lowest-redshift bin we consider for each survey ($0.2<z<0.7$ for CHORD, $0.5<z<1$ for PUMA). The large improvement at low redshift is due to the combination of higher stochastic noise and lower thermal noise than at higher redshifts. The improvement is particularly low around $z\sim 2$, where $b_2(z)$, and therefore bias-induced stochasticity, reaches a minimum for HI.
Improvements in constraints on standard $\Lambda$CDM parameters, or common one-parameter $\Lambda$CDM extensions, are less impressive, staying below 10\% in most cases.

The benefits of a more optimal analysis over a power spectrum analysis depend on the thermal noise level of the \tcm survey, because reducing the stochastic noise will have minimal impact if the stochastic noise is well below the thermal noise. However, thermal noise can in principle be lowered by altering the telescope/survey design, and we also explored the impact of lowering the thermal noise using forecasts that set this noise to zero. We can draw the following conclusions from these forecasts:
\begin{enumerate}
\item In the zero-thermal-noise limit, the CHORD forecasts for BAO and~$\sigma_8(z)$ become quite similar to the PUMA forecasts, with almost identical results over $0.5\lesssim z \lesssim 2$ and agreement within a factor of~2 for $2\lesssim z \lesssim 3.5$.
\item A more optimal analysis is definitively more beneficial than a power spectrum analysis in this limit. For example, BAO in the $0.2<z<0.7$ CHORD bin can be measured 50\% more precisely with a more optimal analysis, and PUMA-32k's constraints on $h$ and $\omega_{\rm c}$ can improve by 20-30\%.
\end{enumerate}

Under the assumption that foreground and systematics can be controlled at the appropriate level, these results motivate the design of \tcm intensity mapping surveys with thermal noise as close as possible to the expected HI sampling noise. Maximizing the number of baselines that are sensitive to quasi-linear scales is a conceptually straightforward way to achieve this goal, but must of course be weighed against other practical concerns.

We conclude by mentioning a few related points that could be worth further investigation:
\begin{itemize}
\item We have focused on HI in this work, but the issue of bias-induced stochasticity could be relevant for other tracers of LSS, particularly very dense samples of galaxies that could be collected at low redshift (such as the DESI Bright Galaxy Survey~\cite{Hahn:2022dnf}) or the Lyman-$\alpha$ forest on perturbative scales~\cite{Ivanov:2023yla}.
\item We have focused on quasi-linear scales around where BAO occur, but bias-induced stochasticity could in principle also play a role at much larger scales that could be used to measure the matter-radiation equality scale or scale-dependent bias from local primordial non-Gaussianity. These scales can potentially be accessed by antenna auto-correlations in \tcm surveys, or by galaxy surveys with sufficient control of large-scale systematics.
\end{itemize}

\acknowledgments

We thank Giovanni Cabass, Emanuele Castorina, and Neal Dalal for useful discussions.
We thank Noah Sailer and Stephen Chen for making the \texttt{FishLSS} and \texttt{velocileptors} codes publicly available. We thank Klaus Dolag for providing access to Magneticum simulation data. AO acknowledges financial support from the Swiss National Science Foundation (grant no CRSII5{\_}193826).
This material is based upon work supported by the U.S.\ Department of Energy, Office of Science, Office of High Energy Physics under Award Number DE-SC0024309.

\appendix
\section{Bias parameter transformations} 
\label{app:bias_params}

In our forecasts in Sec.~\ref{sec:forecasts}, we use the LPT formalism from 
Refs.~\cite{Chen:2020fxs,Chen:2020zjt}. On the other hand, the bias parameters we measure 
from simulations in Sec.~\ref{sec:simulations:biascoefficients} 
correspond to the shifted-operator framework described in 
Sec.~\ref{sec:simulations:fieldlevel}. 
This appendix describes how to translate these parameters from one 
formalism to the other, such that our simulation measurements can be used as fiducial 
parameters in our forecasts.

Recall that the shifted-operator framework begins with the following bias expansion in Lagrangian 
space,
\beq
\label{eq:deltag_Lag_G2}
\deltag^\Lag(\vq) 
	= b_1^\Lag \delta_1(\vq) 
	+ b_2^\Lag\lp \delta_1(\vq)^2 - \langle \delta_1^2 \rangle \rp
	+ b_{\Gtwo}^\Lag \Gtwo(\vq)
	+ \cdots\ ,
\eeq
and shifts each operator by the Zel'dovich displacements to form a prediction in Eulerian space. If 
we denote the bias coefficients in the shifted basis by an ``S" superscript, they are related to the 
Lagrangian coefficients via~\cite{Schmittfull:2018yuk}
\beq
b_1^\Shi = 1 + b_1^\Lag\ , \quad
b_2^\Shi = b_2^\Lag\ , \quad
b_{\Gtwo}^\Shi = \frac{2}{7} + b_{\Gtwo}^\Lag\ .
\eeq

The LPT formalism from Refs.~\cite{Chen:2020fxs,Chen:2020zjt} uses a different expansion in 
Lagrangian space,
\begin{align*}
\deltag^\Lag(\vq) 
	&= \tilde{b}_1^{\Lag} \delta_1(\vq) 
	+ \frac{1}{2} \tilde{b}_2^\Lag \lp \delta_1(\vq)^2 - \langle \delta_1^2 \rangle \rp \\
&\quad + \tilde{b}_{s}^\Lag \lp s^2(\vq) - \langle s^2 \rangle \rp
	+ \cdots\ ,
	\numberthis
	\label{eq:deltag_Lag_s2}
\end{align*}
where the corresponding coefficients are denoted by tildes and the squared tidal operator $s^2$ 
is defined as
\beq
s^2 \equiv \lp \frac{\d_i \d_j}{\d^2} \delta_1\rp^2 - \frac{1}{3} \delta_1^2\ ,
\eeq
implying that
\beq
\label{eq:G2_s2}
\Gtwo = s^2 - \frac{2}{3} \delta_1^2\ .
\eeq
By equating Eqs.~\eqref{eq:deltag_Lag_G2} and~\eqref{eq:deltag_Lag_s2}, and using 
Eq.~\eqref{eq:G2_s2}, we find the following relationships between the two sets of Lagrangian 
biases:
\beq
\tilde{b}_1^\Lag = b_1^\Lag\ , \quad
\tilde{b}_2^\Lag = 2b_2^\Lag - \frac{4}{3} b_{\Gtwo}^\Lag\ , \quad
\tilde{b}_s^\Lag = b_{\Gtwo}^\Lag\ .
\eeq

Finally, the forecasting code used in this work accepts Eulerian biases as inputs, and these are 
related to the Lagrangian biases via (e.g.~\cite{Desjacques:2016bnm}):
\beq
\tilde{b}_1^\Eul = 1 + \tilde{b}_1^\Lag\ , \quad
\tilde{b}_2^\Eul = \tilde{b}_2^\Lag + \frac{8}{21} \tilde{b}_1^\Lag\ , \quad
\tilde{b}_s^\Eul = \tilde{b}_s^\Lag - \frac{2}{7} \tilde{b}_1^\Lag\ .
\eeq
Thus, putting everything together, the Eulerian biases in the basis of our forecasts are related to 
the shifted-operator biases reported in Sec.~\ref{sec:simulations:biascoefficients} by
\begin{align}
\tilde{b}_1^\Eul &= b_1^\Shi\ , \\
\tilde{b}_2^\Eul &= 2b_2^\Shi - \frac{4}{3} b_{\Gtwo}^\Shi + \frac{8}{21} b_1^\Shi\ , \\
\tilde{b}_s^\Eul &= b_{\Gtwo}^\Shi - \frac{2}{7} b_1^\Shi\ .
\end{align}

\section{Alternative versions of forecasts} 
\label{app:alternative_forecasts}

In this appendix, we show a selection of forecasts with different assumptions than used in the 
main text.

\subsection{Alternative assumptions for Finger-of-God damping scale $\sigma_v(z)$}
\label{app:alternative_forecasts:sigmav}

In Sec.~\ref{sec:forecasts:setup:predictions}, we discussed how our baseline forecasts use a 
Finger-of-God 
damping scale $\sigma_v(z)$ based on the velocity dispersion of HI in halos, but an inference of 
this scale from the error power spectrum in field-level modelling~\cite{Obuljen:2022cjo} gives a much higher value ($285\,{\rm 
km}\,{\rm s}^{-1}$ instead of $10\,{\rm km}\,{\rm s}^{-1}$, or $2.85\Mpcinvh$ instead of $0.1\Mpcinvh$, at 
$z=0$). 

We have run forecasts that assume the higher value, and have found that this does not qualitatively 
change our conclusions. When quantifying the improved constraining power of a more optimal analysis over 
a power spectrum analysis, we found that the higher $\sigma_v(z)$ normalization causes no more than 
a 10\% change in the improvement in BAO constraints, and no more than a 10\% change in 
$\sigma_8(z)$ 
constraints except in a few redshift bins when assuming $\knl(z)=0.3D(z)\Mpcinvh$. For 
cosmological parameter constraints, we found less than 5\% changes in the improvement, except for 
$h$ and $\omega_{\rm c}$ for PUMA-32k and $\knl(z)=0.3D(z)\Mpcinvh$. In those cases, the 
improvement in constraining power was $\sim$$15\%$ for the lower $\sigma_v(z)$ assumption but 
$\sim$$50\%$ 
for the higher assumption.

In all cases, the expected constraints themselves were slightly weaker with the higher $\sigma_v(z)$ 
assumption, reflecting the fact that fewer high-$k$ modes are above the noise in that case. The modes 
that remain above the noise are more important to the constraints than they are with the lower 
$\sigma_v(z)$ 
assumption, and this is reflected in the slight shifts in the impact of a more optimal analysis.

\subsection{Alternative assumptions for $\knl(z)$}
\label{app:alternative_forecasts:kNL}

\begin{figure}[t]
\includegraphics[width=\columnwidth, trim=0 20 0 0]{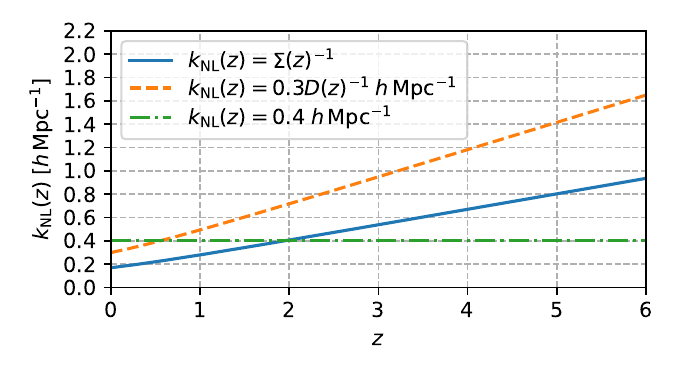}
\caption{%
Redshift-dependence of the nonlinear scale $k_{\rm NL}(z)$ when it is assumed to be set by the 
inverse of the root-mean-square displacements in the Zel'dovich approximation ({\em blue solid 
line}), compared with the two assumptions used in the main text ({\em orange dashed and green 
dot-dashed lines}).
}
\label{fig:knl_z}
\end{figure}

Our forecasts in the main text assume that either $\knl(z)=0.3D(z)^{-1}\hinvMpc$ or 
$\knl(z)=0.4\hinvMpc$. 
Another common assumption about the nonlinear scale is that it is given by 
the inverse of the root-mean-square displacements $\Sigma(z)$, computed in the Zel'dovich 
approximation~\cite{CosmicVisions21cm:2018rfq,Sailer:2021yzm}. The resulting $k_{\rm NL}(z)$ 
is shown in Fig.~\ref{fig:knl_z}, along with our two baseline assumptions.

\begin{figure*}[t]
\includegraphics[width=\textwidth, trim=0 20 0 0]{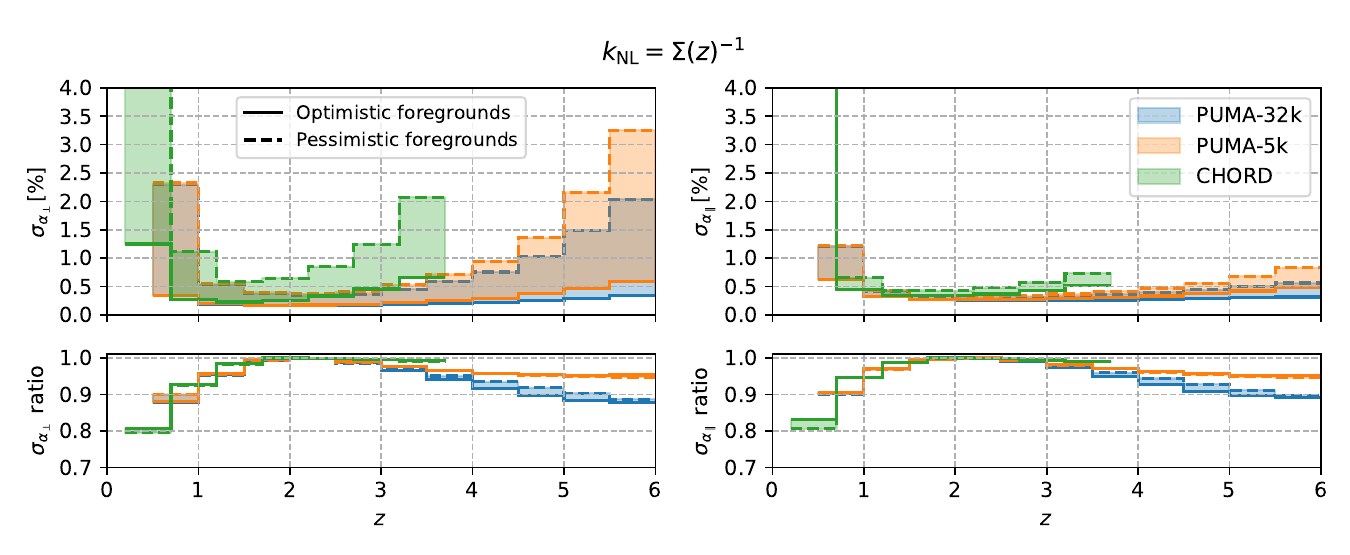}
\caption{%
BAO constraints in the same format as Figs.~\ref{fig:bao_aperp}-\ref{fig:bao_apar}, but assuming 
that $k_{\rm NL}(z)=\Sigma^{-1}(z)$. In this case $k_{\rm NL}(z)$ at low $z$ is lower than either 
of the assumptions in the main text, and nonlinearity-induced stochastic noise is more important 
at high $k$, so the advantage of a more optimal \tcm analysis is reduced.
}
\label{fig:bao_knlZel}
\end{figure*}

\begin{figure}[t]
\includegraphics[width=\columnwidth, trim=0 20 0 
0]{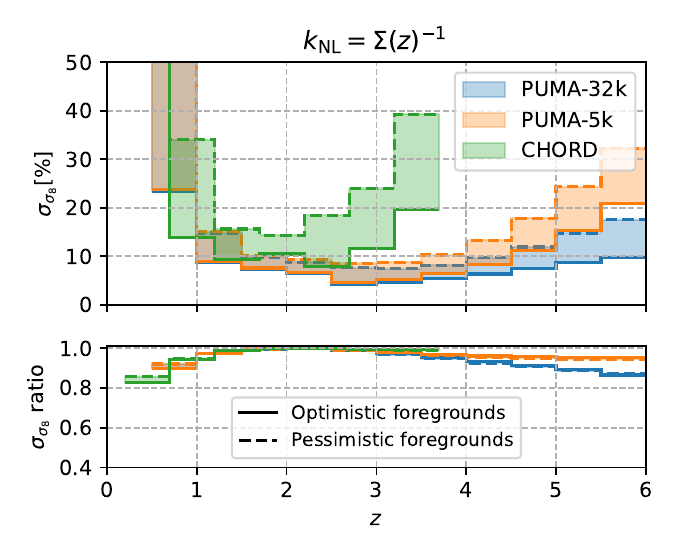}
\caption{%
Constraints 
on $\sigma_8(z)$ in the same format as Fig.~\ref{fig:sigma8_fixedshape_freeTb}, but 
with $k_{\rm NL}(z)=\Sigma(z)^{-1}$. As with BAO, the reduced nonlinear scale implies that a 
more optimal analysis does not improve the constraints as much as for our other assumptions about 
$k_{\rm NL}(z)$.
}
\label{fig:sigma8_fixedshape_freeTb_knlZel}
\end{figure}

At $z\lesssim 1.5$, the BAO constraints noticeably weaken if $k_{\rm NL}(z)=\Sigma(z)^{-1}$ 
(compare Fig.~\ref{fig:bao_knlZel} with Figs.~\ref{fig:bao_aperp} and~\ref{fig:bao_apar}): at these 
redshifts, the nonlinear scale sets the maximum usable wavenumber, rather than instrumental 
noise. The impact of a more optimal \tcm analysis is also reduced, because the stochastic noise that 
is absent in more optimal predictions is most important at higher $k$. At $z\gtrsim 1.5$, however, the 
constraints lie in between the $\knl(z)=0.3D(z)^{-1}\hinvMpc$ and $\knl(z)=0.4\invMpc$ results, as we'd 
expect from Fig.~\ref{fig:knl_z}.
The same conclusions hold for constraints 
on $\sigma_8(z)$ (compare Fig.~\ref{fig:sigma8_fixedshape_freeTb_knlZel} with 
Fig.~\ref{fig:sigma8_fixedshape_freeTb}).

\begin{figure}[t]
\includegraphics[width=\columnwidth, trim=0 10 0 0]{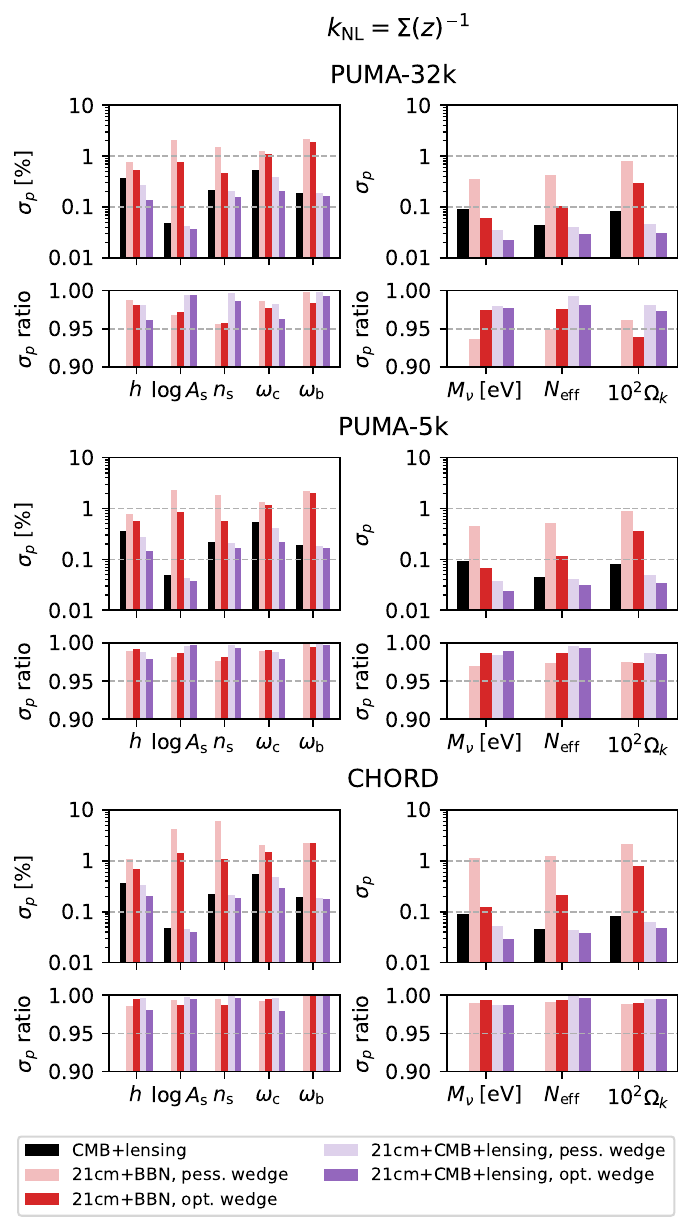}
\caption{%
Constraints on $\Lambda$CDM parameters or one-parameter extensions, in the same format as 
Figs.~\ref{fig:lcdm}-\ref{fig:lcdm_extensions} but with $k_{\rm NL}(z)=\Sigma(z)^{-1}$. A field-level 
\tcm analysis impacts the constraining power by less than 5\% for this form of $k_{\rm NL}(z)$.
}
\label{fig:lcdm_plus_extensions_knlZel}
\end{figure}

In forecasts for individual cosmological parameters (compare 
Fig.~\ref{fig:lcdm_plus_extensions_knlZel} with Figs.~\ref{fig:lcdm} 
and~\ref{fig:lcdm_extensions}), constraints in the {\tcm}+CMB+lensing analysis worsen by up to 
50\% if $k_{\rm NL}(z)=\Sigma(z)^{-1}$ instead of $\knl(z)=0.3D(z)^{-1}\hinvMpc$, because the 
amount of usable low-$z$ information is reduced. If $k_{\rm NL}(z)=\Sigma(z)^{-1}$ is used 
instead of $\knl(z)=0.4\hinvMpc$, some constraints improve by up to 10\% while others worsen 
by up to 5\%, depending on whether the higher $k_{\rm NL}$ at high redshift or the lower $k_{\rm 
NL}$ at lower redshift is dominating the constraining power. In all cases, though, using $k_{\rm 
NL}(z)=\Sigma(z)^{-1}$ reduces the advantage of a more optimal \tcm analysis, since 
nonlinearity-induced 
stochastic noise is most important at low $z$.

In summary, for all forecasts we consider, if $k_{\rm NL}(z)=\Sigma(z)^{-1}$, a more optimal \tcm 
analysis will not increase the overall constraining power as much as if $k_{\rm NL}(z)$ follows one 
of our baseline assumptions.

\subsection{Alternative system temperature}
\label{app:alternative_forecasts:Tsys}

\begin{figure}[t]
\includegraphics[width=\columnwidth, trim=0 20 0 0]{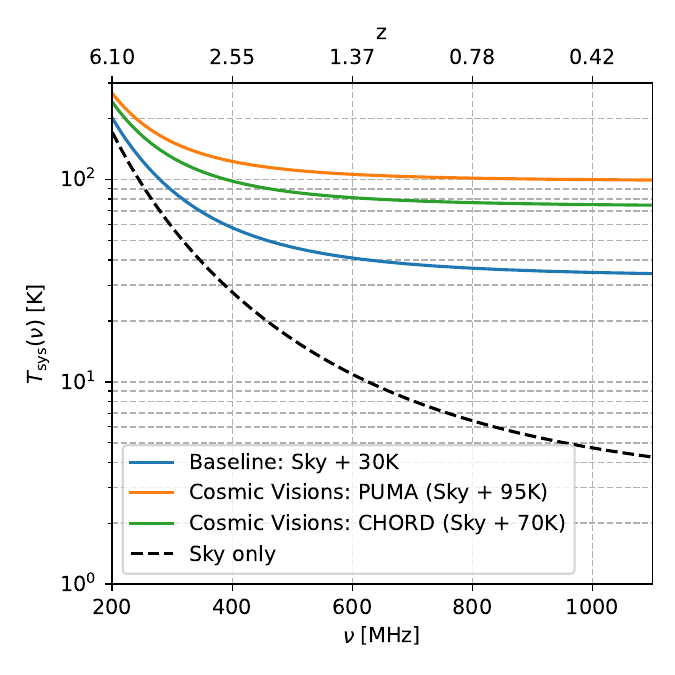}
\caption{%
Assumptions about system temperature $T_{\rm sys}$ as a function of frequency ({\em lower axis}) or 
redshift ({\em upper axis}) used in our forecasts. Our baseline assumption for PUMA and CHORD ({\em blue 
solid}) is based on the specification that has been met by prototype CHORD 
components~\cite{CHORD-feed}, 
while the ``Cosmic Visions" assumptions ({\em orange and green solid}) are based on 
Ref.~\cite{CosmicVisions21cm:2018rfq}, with $T_{\rm ampl}=50\,{\rm K}$ for PUMA and $30\,{\rm K}$ for 
CHORD.
We also isolate the sky contribution ({\em black dashed}), dominated by Galactic synchrotron and the CMB.
}
\label{fig:Tsys}
\end{figure}

As discussed in Sec.~\ref{sec:forecasts:setup:system_temperature}, our baseline forecasts for PUMA and 
CHORD assume a system temperature of $T_{\rm sys}(\nu) = T_{\rm sky}(\nu) + 30\,{\rm K}$, which is 
based on measurements of CHORD components but is more optimistic than the common assumptions for 
PUMA in the literature, which generally follow Ref.~\cite{CosmicVisions21cm:2018rfq}. In this appendix, we 
present forecasts for PUMA and CHORD that use these more common assumptions: the values from 
Ref.~\cite{CosmicVisions21cm:2018rfq} for all parameters in Eq.~\eqref{eq:Tsys}, except for the amplifier 
temperature $T_{\rm ampl}$ for CHORD, which we set to $30\,{\rm K}$ instead of $50\,{\rm K}$. The 
various assumptions are shown in Fig.~\ref{fig:Tsys}; our baseline assumption is lower than the alternative 
(``Cosmic Visions") assumption for PUMA by a factor of $\sim$2 for $\nu\gtrsim 400\,{\rm MHz}$, and by at 
least 25\% over $200\,{\rm MHz} < \nu < 400\,{\rm MHz}$.

\begin{figure*}[t]
\includegraphics[width=\textwidth, trim=0 20 0 0]{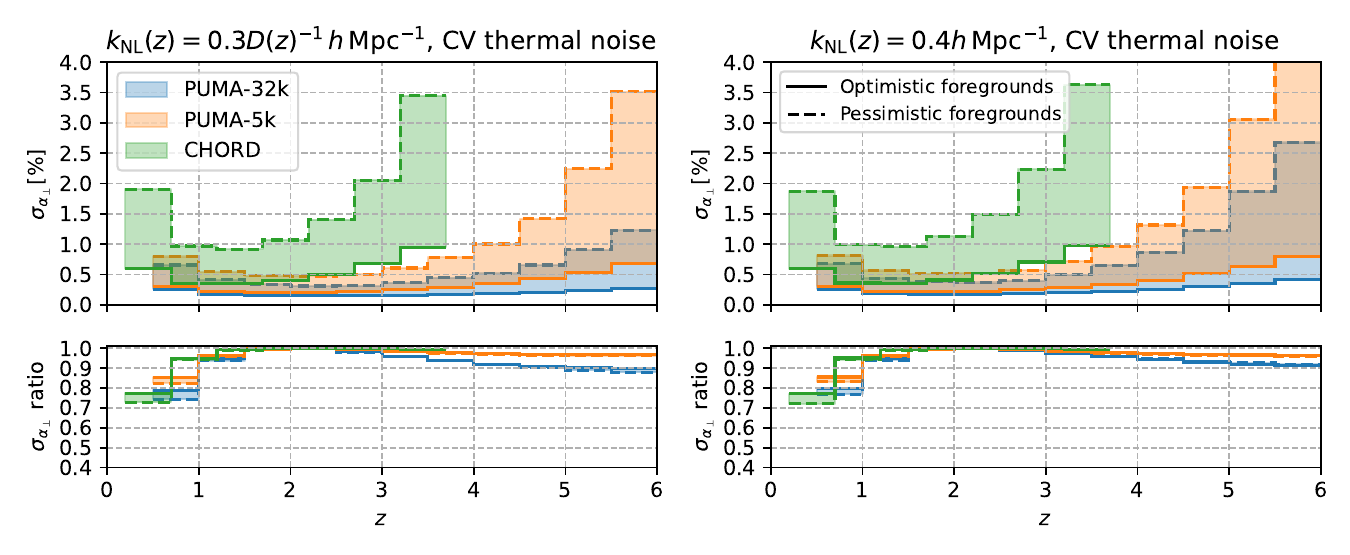}
\caption{%
Forecasts for $\alpha_\perp$, analogous to Fig.~\ref{fig:bao_aperp} but using the thermal noise 
assumptions from Ref.~\cite{CosmicVisions21cm:2018rfq} (labelled by ``CV" in this figure and 
Figs.~\ref{fig:bao_apar_cvnoise} to~\ref{fig:lcdm_extensions_cvnoise}).
}
\label{fig:bao_aperp_cvnoise}
\end{figure*}

\begin{figure*}[t]
\includegraphics[width=\textwidth, trim=0 20 0 0]{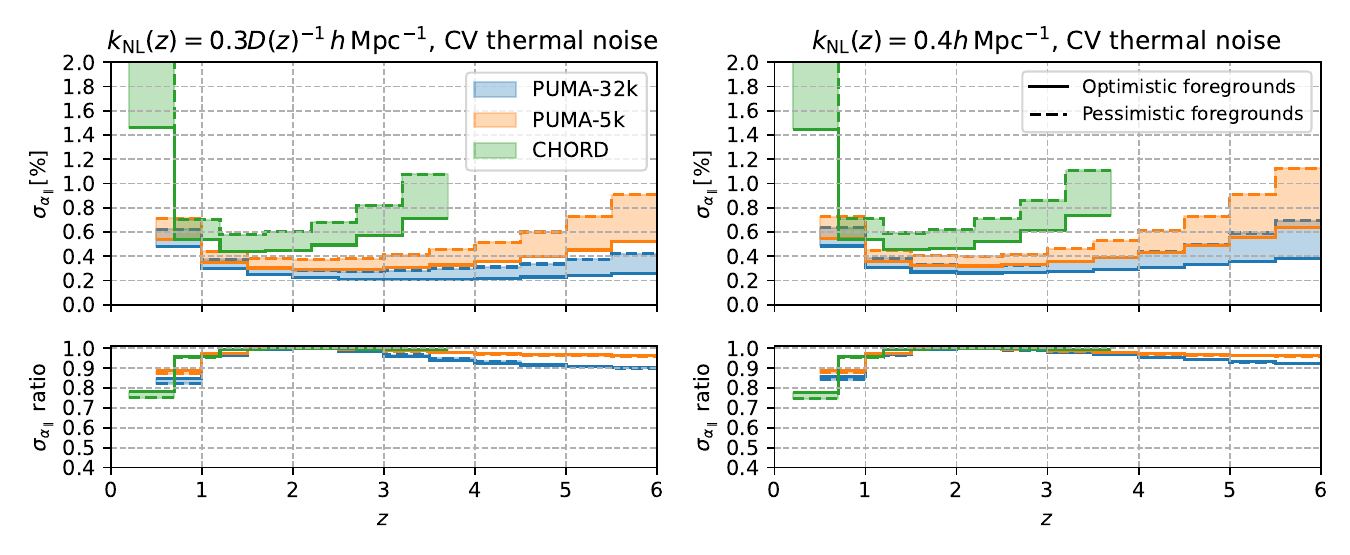}
\caption{%
Forecasts for $\alpha_\parallel$, analogous to Fig.~\ref{fig:bao_apar} but using the thermal noise 
assumptions from Ref.~\cite{CosmicVisions21cm:2018rfq}.
}
\label{fig:bao_apar_cvnoise}
\end{figure*}

\begin{figure*}[t]
\includegraphics[width=\textwidth, trim=0 20 0 0]{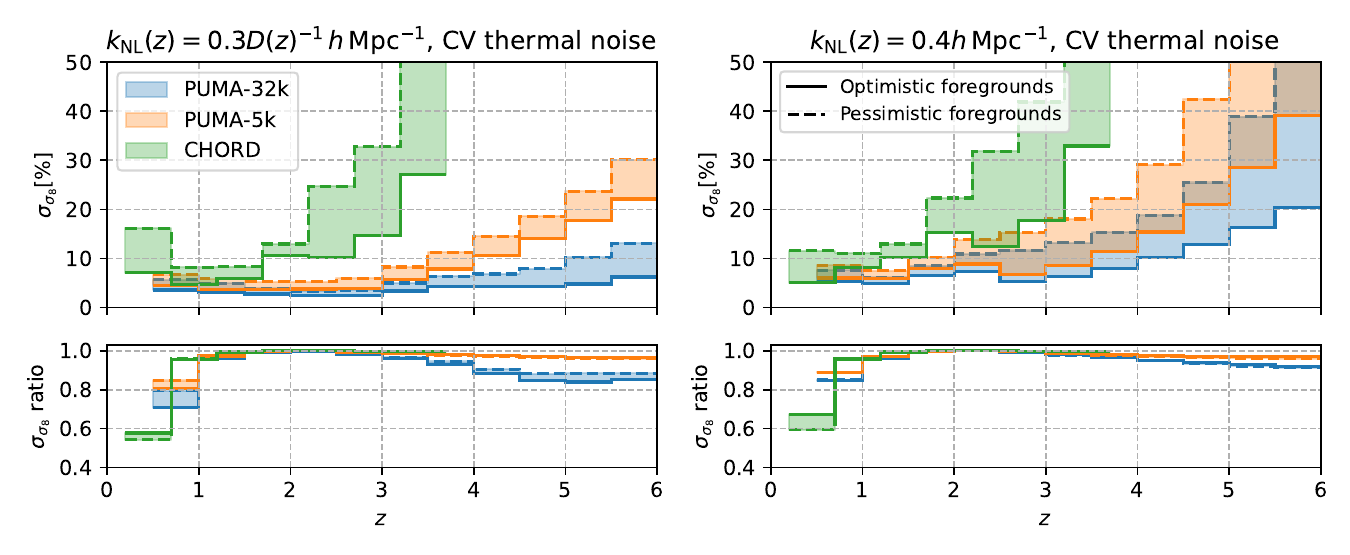}
\caption{%
Forecasts for $\sigma_8(z)$, analogous to Fig.~\ref{fig:sigma8_fixedshape_freeTb} but using the thermal 
noise assumptions from Ref.~\cite{CosmicVisions21cm:2018rfq}.
}
\label{fig:sigma8_fixedshape_freeTb_cvnoise}
\end{figure*}

\begin{figure}[t]
\includegraphics[width=\columnwidth, trim=0 10 0 0]{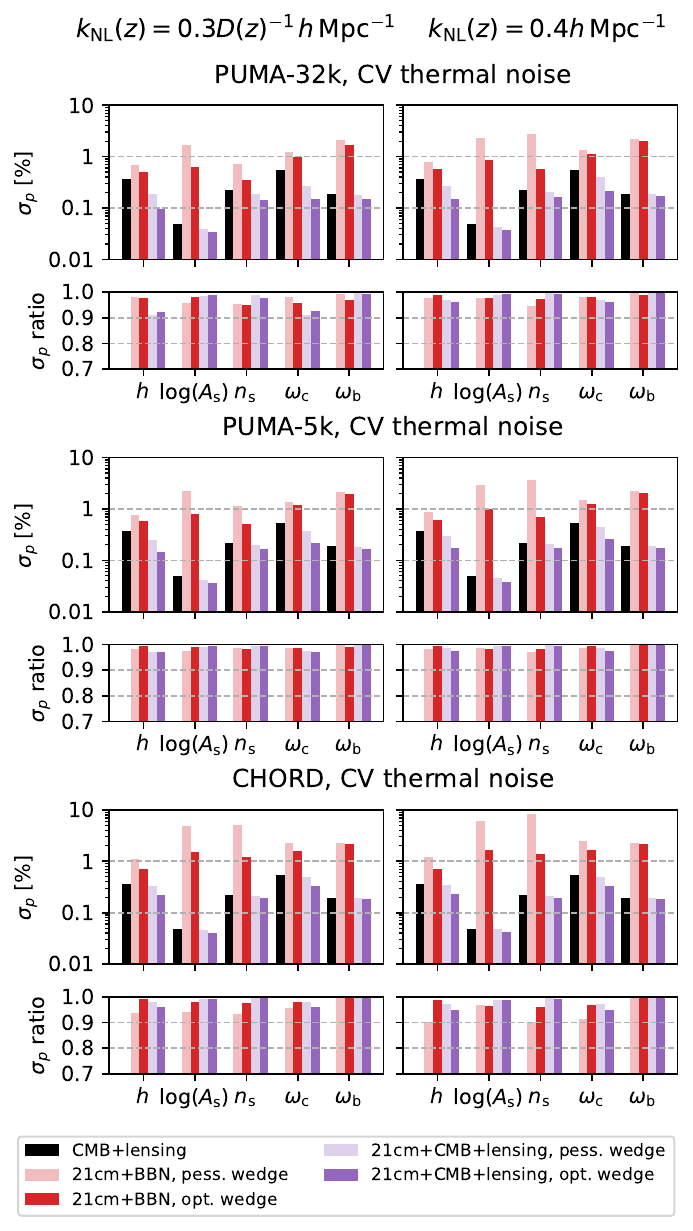}
\caption{%
Forecasts for standard $\Lambda$CDM parameters, analogous to Fig.~\ref{fig:lcdm} but using the thermal 
noise assumptions from Ref.~\cite{CosmicVisions21cm:2018rfq}.
}
\label{fig:lcdm_cvnoise}
\end{figure}

\begin{figure}[h]
\includegraphics[width=\columnwidth, trim=0 10 0 0]{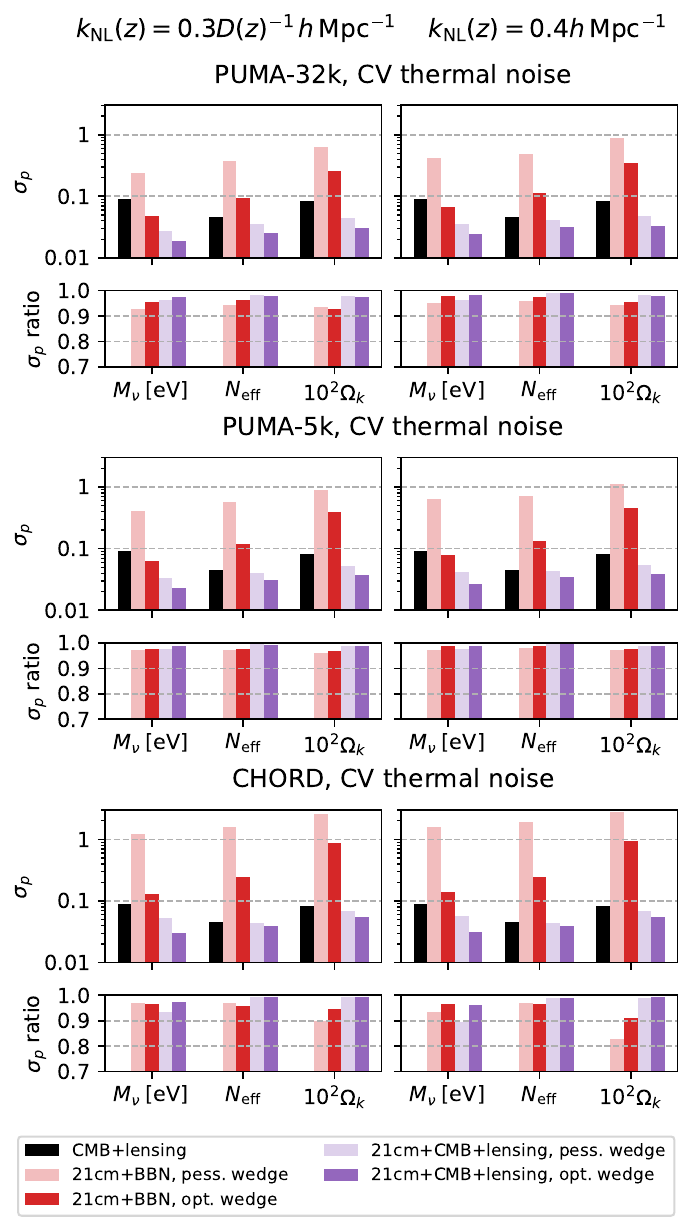}
\caption{%
Forecasts for one-parameter extensions of $\Lambda$CDM, analogous to Fig.~\ref{fig:lcdm_extensions} but using 
the thermal noise assumptions from Ref.~\cite{CosmicVisions21cm:2018rfq}.
}
\label{fig:lcdm_extensions_cvnoise}
\end{figure}

These forecasts are shown in Figures~\ref{fig:bao_aperp_cvnoise} 
($\alpha_\perp$),~\ref{fig:bao_apar_cvnoise} 
($\alpha_\parallel$),~\ref{fig:sigma8_fixedshape_freeTb_cvnoise} 
($\sigma_8$),~\ref{fig:lcdm_cvnoise} 
(standard $\Lambda$CDM parameters), 
and~\ref{fig:lcdm_extensions_cvnoise} (one-parameter $\Lambda$CDM extensions). The absolute 
parameter constraints are worse by up to 30\% compared to our baseline forecasts, while the impact of a 
more optimal analysis is weaker than for our baseline forecasts.

\subsection{Fixed mean brightness temperature}
\label{app:alternative_forecasts:fixed_Tb}

\begin{figure*}[t]
\includegraphics[width=\textwidth, trim=0 20 0 0]{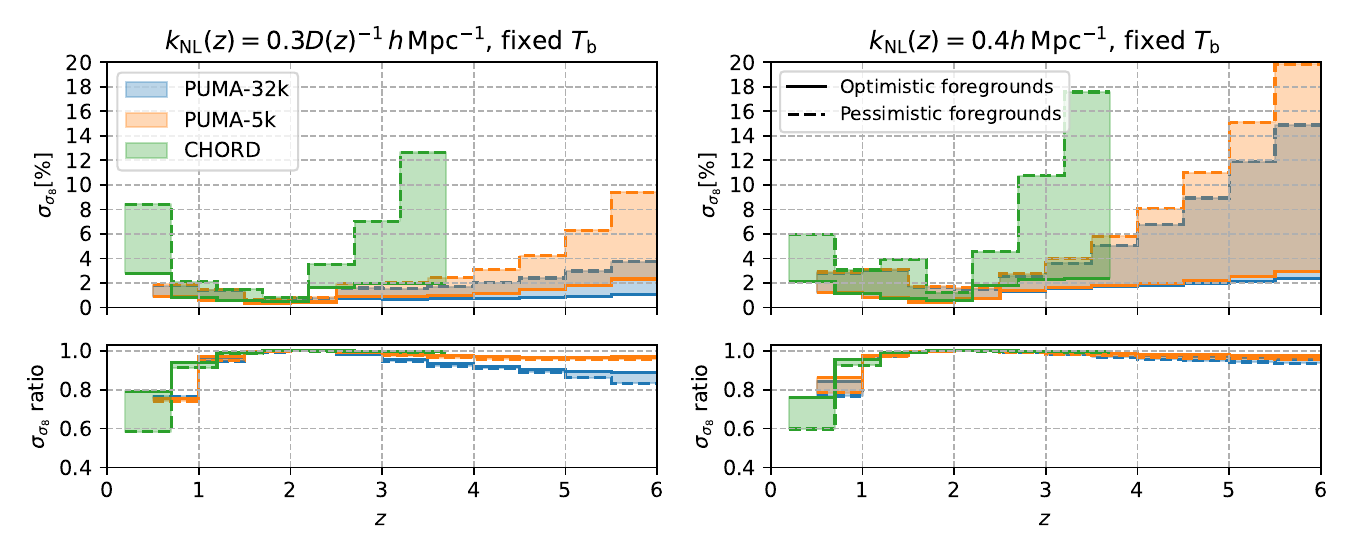}
\caption{%
Constraints on $\sigma_8(z)$ in the same format as Fig.~\ref{fig:sigma8_fixedshape_freeTb}, 
with the mean \tcm brightness temperature $T_{\rm b}(z)$ fixed to its fiducial value, as a proxy 
for external constraints that could come from damped Lyman-$\alpha$ absorbers or direct \tcm 
galaxy surveys. While the absolute $\sigma_8(z)$ constraints tighten considerably, the impact of 
a more optimal \tcm analysis on the constraining power does not change by more than 15\% 
compared to the free-$T_{\rm b}$ case in the main text.}
\label{fig:sigma8_fixedshape_fixedTb}
\end{figure*}

Figure~\ref{fig:sigma8_fixedshape_fixedTb} shows a version of the $\sigma_8(z)$ forecasts from 
Fig.~\ref{fig:sigma8_fixedshape_freeTb} where the mean \tcm brightness temperature~$T_{\rm 
b}(z)$ 
is fixed instead of being allowed to vary. While perfect knowledge of $T_{\rm b}(z)$ will not 
be possible in practice, fixing its value in forecasts provides an indication of how external 
constraints from damped Lyman-$\alpha$ absorbers or \tcm galaxy surveys would affect our 
results.

Comparing Fig.~\ref{fig:sigma8_fixedshape_fixedTb} with 
Fig.~\ref{fig:sigma8_fixedshape_freeTb}, we find that uncertainties on $\sigma_8(z)$ are 
significantly reduced if $T_{\rm b}(z)$ is fixed, due to the $\sigma_8-T_{\rm b}$ degeneracy in 
linear theory. However, the relative improvement due to a more optimal vs.\ power spectrum 
analysis is similar in both cases, with no more than 10\% differences in the improvement in 
constraining power.

When forecasting for individual cosmological parameters, constraints on $\log A_{\rm s}$ are 
tightened by a factor of~2 (for PUMA-32k+CMB+lensing with optimistic foregrounds and $k_{\rm 
NL}(z)=0.3D(z)^{-1}\hinvMpc$) in the fixed-$T_{\rm b}(z)$ case. Constraints on $M_\nu$ also 
improve by 10-25\% depending on the \tcm survey. The improvement from more optimal \tcm analyses can 
also change by 10\% between the fixed- and free-$T_{\rm b}(z)$ cases, but only for {\tcm}+BBN 
analyses where the constraints are much weaker than for CMB+lensing. For full 
{\tcm}+CMB+lensing 
forecasts, the choice of fixing or varying $T_{\rm b}(z)$ only changes the impact 
of a more optimal analysis by a few percent.

\section{Expressions for thermal noise} 
\label{app:thermal_noise}

The power spectrum of thermal noise for a drift-scan interferometer is given by 
(e.g.~\cite{CosmicVisions21cm:2018rfq})
\beq
P_{\rm N}(\vk, z) 
	= \frac{C(z) T_{\rm sys}(z)^2}
	{A_{\rm e}^2 \,t_{\rm survey}\, n_{\rm b}(u=k_\perp \chi(z) / 2\pi; z) \,{\rm FOV}(z)}\ ,
	\label{eq:PN_CV}
\eeq
where factors that are irrelevant to our arguments in this section have been grouped together into a function 
$C(z)$, with dimensions of $[{\rm length}]^7[{\rm time}]^1$. The system temperature $T_{\rm sys}(z)$ has 
contributions from the sky (dominated by smooth-spectrum foregrounds) and instrument (amplifiers, 
ground spill, etc.). The effective collecting area per antenna, $A_{\rm e}$, and effective field of view, ${\rm 
FOV}(z)$, are given by
\beq
A_{\rm e} = \pi \lp \frac{D_{\rm eff}}{2} \rp^2\ ,
	\quad {\rm FOV}(z) = \lp \frac{\lambda_{\rm obs}(z)}{D_{\rm eff}} \rp^2\ ,
\eeq
where the effective dish area $D_{\rm eff}^2$ is related to the physical diameter $D_{\rm phys}$ by an 
aperture efficiency factor $\eta_{\rm a}$ as
\beq
D_{\rm eff}^2 = \eta_{\rm a} D_{\rm phys}^2\ .
\eeq
The $uv$-plane baseline number density $n_{\rm b}(u)$ is given by
\beq
n_{\rm b}(u; z) = \lambda_{\rm obs}(z) \lp \frac{N_{\rm s}}{D_{\rm phys}} \rp^2
	F\!\lp \frac{u\lambda_{\rm obs}(z)}{N_{\rm s} D_{\rm phys}} \rp\ ,
\eeq
where $N_{\rm s}^2$ is the number of antennas and $F(x)$ is well-described by a fitting function given in 
Ref.~\cite{CosmicVisions21cm:2018rfq}. $F(x)$ is a decreasing function of $x$, reflecting the fact that the 
number of baselines in the array is a decreasing function of baseline length.

Combining the above equations, we find that
\beq
P_{\rm N}(\vk, z) \propto T_{\rm sys}(z)^2\, t_{\rm survey}^{-1}\, N_{\rm s}^{-2}\,
	F\!\lp \frac{u\lambda_{\rm obs}(z)}{N_{\rm s} D_{\rm phys}} \rp^{-1}\ .
\eeq
Thus, the thermal noise power spectrum is proportional to the square of the system temperature, and 
inversely proportional to the total survey time, number of antennas, and baseline distribution fitting function 
$F$. Notably, the $D_{\rm phys}$ prefactors cancel completely: for fixed $N_{\rm s}$, as the dish size 
increases, the larger collecting area is perfectly compensated by the smaller instantaneous field of view and 
the lower overall density of baselines. For angular scales where $F(x)$ is not constant, however, the noise 
does decrease with dish size, as larger dish size (assuming a close-packed array) implies a larger number 
of 
longer baselines.

\bibliography{references}

\end{document}